%% file: draft.tex
\documentclass[11pt,letterpaper]{article}
\usepackage[utf8]{inputenc}
\usepackage[T1]{fontenc}
\usepackage{amsmath,amssymb,amsfonts,mathtools,bm}
\usepackage{newpxtext}
\usepackage{newpxmath}
\usepackage{graphicx}
\usepackage{booktabs,threeparttable,tabularx,array,multirow,makecell}
\usepackage{enumitem}
\usepackage{subcaption}
\usepackage[table]{xcolor}
\usepackage[hyphens]{url}
\usepackage{hyperref}
\hypersetup{colorlinks=true,linkcolor=blue,citecolor=blue,urlcolor=blue,
            breaklinks=true}
\usepackage{natbib}
\usepackage{geometry}
\usepackage{setspace}
\usepackage{float}
\usepackage{xr}
\setlist[itemize]{leftmargin=1.55em, itemsep=2pt, topsep=2pt}
\setlist[enumerate]{leftmargin=1.7em, itemsep=2pt, topsep=2pt}

\newcommand{\EV}{\mathrm{EV}}

\begin{document}

\begin{titlepage}
\title{\textbf{Heterogeneous Diffusion of Electric Vehicles in China:
Demand, Learning, Product Entry, and the Incidence of Industrial Policy}}
\author{Yu (Jasmine) Hao\thanks{}\\
Faculty of Business and Economics\\
The University of Hong Kong\\
haoyu@hku.hk \and
Jinge Li\\
Harris School of Public Policy \\
The University of Chicago\\
jingeli@uchicago.edu
}
\date{June 2026}
\maketitle

\begin{abstract}
\noindent
China's electric-vehicle (EV) sales share rose from about $1\%$ in
2015 to roughly $45\%$ in 2024. We evaluate this technology
transition with an equilibrium differentiated-products model of the
Chinese auto market, and quantify both its attribution and its
welfare and reallocation consequences. Every yuan of 2024 EV
subsidy delivered about $3.38$ yuan of private surplus, but this
surplus accrued asymmetrically. Per-capita consumer-surplus loss
from subsidy removal is about five times larger in Tier 1 than in
the Rest tier; about half of the aggregate welfare loss operates
through indirect Wright's-law learning rather than the direct cash
transfer; and EV-native firms (BYD, Tesla, New Forces) retain
$16$--$27\%$ of their 2024 EV business under subsidy removal while
traditional state-owned manufacturers retain only $11\%$. A Shapley
decomposition into six channels --- Quality, Variety, Battery,
Subsidy, Residual, and Market --- attributes the historical
2015--2024 rise primarily to product-quality gains ($+45.49\%$),
choice-set expansion ($+14.81\%$), and battery-cost decline
($+8.20\%$). The Subsidy block is negative ($-13.63\%$) because
direct purchase subsidies were phased down, not because subsidies
reduce demand: a separate counterfactual that removes the 2024
subsidy entirely lowers EV share by $23$--$33\%$.
\\
\end{abstract}

\setcounter{page}{0}
\thispagestyle{empty}
\end{titlepage}
\section{Introduction}

Technological transitions often require more than the arrival of a new
technology. They depend on whether firms embody the technology in
attractive products, whether production costs fall, whether consumers
substitute toward the new products, and whether policy support arrives
at the right stage of diffusion. These forces are especially difficult
to separate in markets with endogenous prices and differentiated
products, because technology, entry, costs, and demand interact in
equilibrium. A new technology may have little effect when product variety
is thin or costs are high, but the same technology may diffuse rapidly
once quality improves, firms enter, and consumers face a broader and
more competitive choice set.

This paper studies these mechanisms in the context of China's electric
vehicle transition. The national EV share rose from $1.03\%$ in 2015
to $44.66\%$ in 2024. This rapid diffusion took place during a period
of major technological improvement, large-scale product entry,
battery-cost decline, changing subsidy policy, and substantial
reallocation of demand across cities and firms.

Given a transition of this magnitude, two questions follow. First,
what fraction of the 2015--2024 EV-share rise is attributable to
each policy and market force? Second, what are the welfare and
reallocation consequences when the policy stack that supported this
transition is removed? We answer both with an equilibrium structural
model of the Chinese auto market.

The attribution question is difficult because the underlying
channels are complementary. Battery-cost reductions matter more when
firms offer attractive EV products. Entry matters more when EV
quality has improved. Subsidies have different effects in an early
market with few EV options than in a mature market with many
high-quality products. A decomposition that adds channels in an
arbitrary order can therefore give misleading answers.

To address this, we estimate an equilibrium differentiated-products
model of the automobile market and use it to decompose the
2015--2024 EV transition. The model links micro-level product and
market data to equilibrium prices, quantities, and substitution
across gasoline vehicles and EVs. We then conduct a Shapley
decomposition that switches groups of economic primitives from their
2015 values to their endpoint-year values. The decomposition
separates six channels: Quality, Variety, Battery, Subsidy,
Residual, and Market. The Quality and Battery blocks correspond to a
product-level productivity gain — longer range and lower marginal
cost per vehicle-service unit. The remaining blocks shape how that
productivity gain diffuses across products, firms, and cities.
Because the Shapley value averages each channel's marginal
contribution over all possible orderings, it provides an
order-invariant allocation of the aggregate EV-share change across
complementary mechanisms.

The first result is that product improvement is the central force
behind China's EV transition. The Quality block contributes $45.49\%$
to the 2015--2024 EV-share change. It captures the within-product
attribute trajectory — driving range, engine power, vehicle size, and
their interactions — for products present in both endpoint sets.
The implication is that the 2024 EV was not simply a cheaper version
of the 2015 EV. It is a substantially different product with longer
range, more competitive power and size, and a shifting body- and
size-segment composition.

The second result is that product-market expansion and cost decline
were important complements to quality improvement. The Variety block
contributes $14.81\%$, reflecting the expansion of the EV choice set.
The Battery block contributes $8.20\%$, reflecting battery-related
cost reductions. These channels are not independent: lower battery
costs matter most when firms can embody them in appealing EV
products, and entry matters most when EV quality is already high.

The third result is that the historical Subsidy contribution is
negative, even though subsidies remain important in counterfactuals.
The Subsidy block compares the 2024 schedule with the 2015 schedule.
Direct purchase subsidies were phased down between these dates, so
this block contributes $-13.63\%$. The negative sign does not imply
that subsidies reduced EV demand. A separate no-subsidy
counterfactual shows that subsidies accelerated adoption through
early demand, cumulative learning, and product availability. The two
exercises measure different counterfactuals.

The fourth result concerns incidence. During the active subsidy years,
EV adoption was concentrated in high-income cities, so historical subsidy
benefits accrued disproportionately to richer markets. By 2024, EV
adoption had diffused much further down the city-income distribution,
but direct purchase subsidies had been phased out. A counterfactual that
applies the 2015 subsidy level to the 2024 market produces a much more
balanced distribution of benefits across city income tiers. This timing
misalignment suggests that the same policy instrument can have very
different distributional effects depending on the stage of technology
diffusion.

The fifth result decomposes the welfare loss from subsidy removal
into a direct cash-transfer channel and an indirect Wright's-law
learning channel. The two channels contribute roughly equal shares:
$52\%$ direct and $48\%$ indirect. The indirect share is higher in
Tier 1 than in the Rest tier because the Wright channel transmits
through the equilibrium EV price, which moves more consumer surplus
where EV adoption is concentrated.

The sixth result is that subsidy removal reallocates EV production
across firm types. EV-native firms --- BYD, Tesla, and the New
Forces --- retain $16$--$27\%$ of their 2024 EV business. Traditional
state-owned OEMs retain only $11\%$. This producer-incidence
asymmetry is a technology-driven reallocation: the same equilibrium
that delivers consumer surplus to high-tier cities also concentrates
producer surplus in the EV-native firm groups whose product
portfolios are most exposed to the EV margin.

The paper contributes to four strands of literature. The first is the
empirical industrial-organization literature on differentiated-product
demand and supply. The demand system follows \citet{Berry1995},
\citet{BerryLevinsohnPakes2004}, and \citet{Nevo2001}. Product-market
shares invert mean utilities; observed and unobserved attributes
shape substitution; markups are recovered from multi-product Bertrand
first-order conditions. \citet{BerryHaile2014} clarify the
market-level identification problem in this class of models. That
problem is especially relevant here because price, range, and product
availability are all equilibrium objects.

The second strand studies new products, product entry, and endogenous
product characteristics. \citet{Petrin2002} shows that a new
differentiated product can generate large welfare gains.
\citet{Fan2013}, \citet{Sweeting2013}, and \citet{Wollmann2018} treat
product characteristics and availability as strategic choices.
\citet{Reynaert2021} shows that firms respond to environmental
regulation by reshaping the product portfolio rather than only
adjusting prices, which makes the choice-set margin first-order in
auto markets. In China's EV market, entry is not a single product
launch but a large shift in the product space. The 2024 market contains many EV models, body
types, and firm groups that did not exist or were commercially minor
in 2015. The Variety block is therefore an accounting measure of
choice-set expansion, not a structural estimate of firms' entry
values. The endogenous-range exercise treats range similarly, as a
strategic quality choice rather than an exogenous engineering
statistic.

The third strand studies durable goods and dynamic adoption.
\citet{GowrisankaranRysman2012} show that durable-good demand is
inherently dynamic because consumers can wait, firms can adjust future
offerings, and current sales affect future market conditions. This
paper is not a fully dynamic durable-good model, but the interpretation
of subsidies and range is dynamic in this reduced-form sense. If
subsidies raise early EV sales, they can affect later battery costs,
product entry, and endogenous range choices.

The fourth strand studies environmental and energy technology policy.
\citet{Ryan2012} and \citet{FowlieReguantRyan2016} show that
environmental policy can reshape industry dynamics, not only static
prices. \citet{AcemogluAghionBursztynHemous2012},
\citet{NewellJaffeStavins1999}, and \citet{Popp2002} provide the
broader logic for directed technical change.
\citet{AghionCaiDewatripontDuHarrisonLegros2015} show that
industrial-policy subsidies in China raise productivity more when
they target more-competitive sectors, which is the empirical
condition our Variety + Subsidy interaction speaks to. In the EV
context, \citet{Li2017} and \citet{Springel2021} emphasize indirect
effects through complementary infrastructure, installed base, and
expectations. On EV subsidy incidence specifically,
\citet{Sallee2011} documents that hybrid-vehicle tax credits accrue
disproportionately to high-income early adopters, a finding our
tier-incidence accounting echoes for China's phase-down schedule.
\citet{MuehleggerRapson2022} use quasi-experimental variation to
estimate the marginal effect of EV subsidies on low- and
middle-income adoption, the closest reduced-form analogue to our
no-subsidy counterfactual.

We place direct purchase subsidies inside a broader transition that
also includes battery-cost decline and product quality. That is why
the sign of the Subsidy block depends on the counterfactual: the
historical decomposition captures policy phase-out; the 2024 removal
exercise captures the value of the remaining subsidy in a mature
market. The industrial-policy scope is narrow: we evaluate the
2015--2022 cash schedule plus the 2014--2025 purchase-tax exemption
against a no-instrument counterfactual. We do not compare cash
subsidies to fuel-economy mandates, emission caps, or R\&D credits,
and we do not compute cross-country welfare benchmarks.

Finally, the paper connects the industrial-organization approach to
learning-by-doing and decomposition methods. The learning-curve
framework dates to \citet{Wright1936} and remains the standard tool
for projecting battery costs. Battery-cost decline is motivated by
the broader evidence on learning and scale in industrial production
\citep{Benkard2004,Thompson2012,BollingerGillingham2019}. For the EV
battery itself, \citet{NykvistNilsson2015} and \citet{ZieglerTrancik2021}
estimate long-run lithium-ion learning rates; the latter, using the
widest price dataset, is the benchmark against which we report our
$22.8\% \pm 2.3\%$ Wright fit to BNEF + IEA data.
The Shapley framework \citep{Shapley1953,Shorrocks2013} is used not
because mechanisms are separable, but precisely because they are not:
it provides a transparent average over possible orders and makes
order-sensitivity observable. In that sense, the wide Shapley ranges
are an empirical finding about complementarity, not simply an
estimation inconvenience.

The rest of the paper follows this logic. Section 2 describes the
policy background and data. Sections 3 and 4 present the demand and
supply models. Section 5 uses the estimated model to decompose the
2015--2024 EV diffusion and reports diagnostics for the decomposition.
Section 6 studies policy counterfactuals and incidence. Section 7
concludes.

\section{Policy Background and Data}

\subsection{Policy Background}
\label{sec:policy_background}

\begin{figure}[!ht]
\caption{Rebuilt national NEV subsidy path.}
\label{fig:subsidy}
\centering
\includegraphics[width=0.93\linewidth]{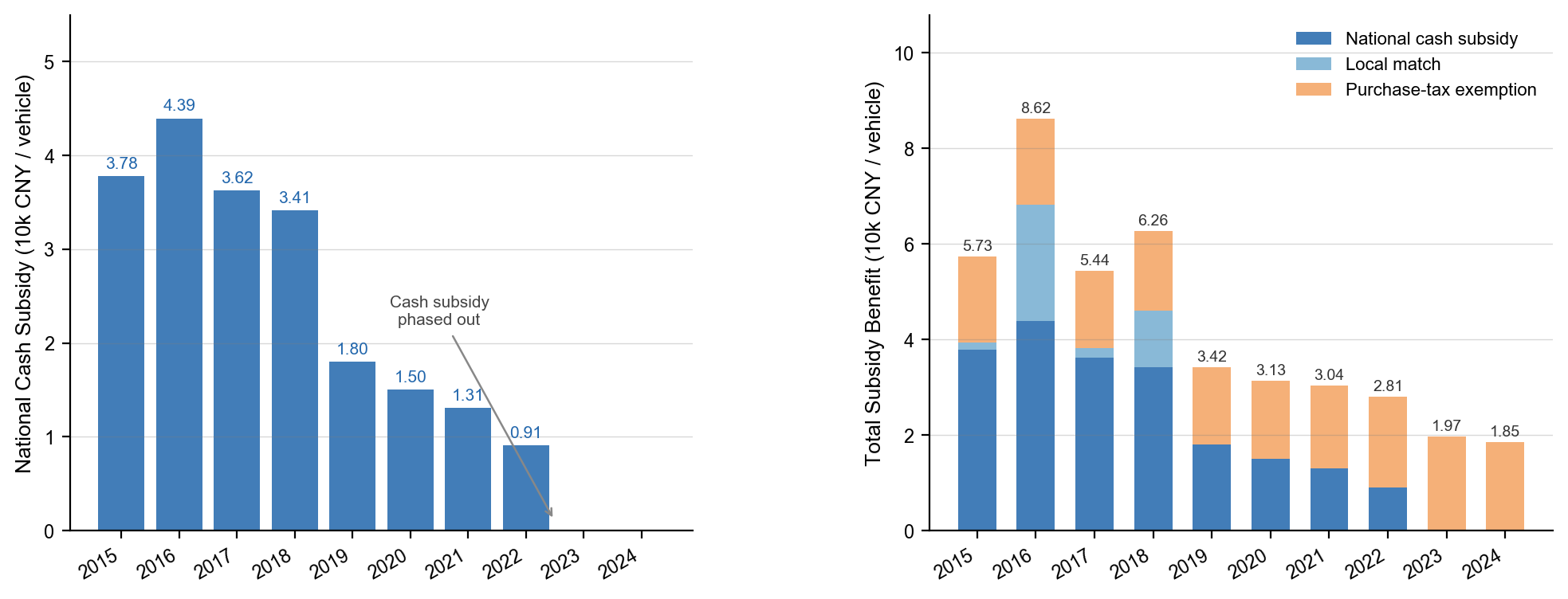}
\par\smallskip
{\footnotesize\flushleft\textit{Notes:} Cash subsidies decline in steps and end after 2022; the 2023--2024 direct subsidy is mainly the purchase-tax exemption.}
\end{figure}

China's EV policy combined demand subsidies, industrial policy,
charging infrastructure, and local-market restrictions on internal
combustion vehicles. The direct purchase subsidy changed substantially
over time. In the early sample, national cash subsidies and local
matches made EVs considerably cheaper at the point of purchase. By
2023 and 2024, national cash subsidies had ended, and the main direct
consumer subsidy in the data was the purchase-tax exemption. Figure
\ref{fig:subsidy} summarizes the rebuilt direct subsidy path used in
the counterfactuals.

\paragraph{License-plate-priority and other in-kind subsidies.}
Beijing, Shanghai, Guangzhou, Shenzhen, Tianjin, and Hangzhou restrict
internal-combustion license-plate issuance through lottery or auction.
In Shanghai, an auctioned ICE plate trades for roughly $90{,}000$
yuan, and EVs are exempt from the auction. The implicit value of a
free EV plate is comparable to the early-period cash subsidy in those
cities. We do not include this in-kind subsidy in the rebuilt subsidy
path, because the implicit value varies with auction prices and
reliable city-year auction data are not available for the full sample.
Instead we treat the EV plate priority as a city-level fixed factor,
absorbed in the demand fixed effects and the EV-$\times$-city
interactions. The incidence accounting in
Section~\ref{sec:subsidy_incidence} therefore understates the implicit
subsidy that accrued to Tier 1 buyers. A back-of-the-envelope value of
$90{,}000$ yuan times 2024 Tier 1 EV sales of $0.045$ million gives
about $4$ bn yuan per year. That is on the same order as the per-tier
CS values in Table~\ref{tab:channels} and would widen the per-capita
gradient between Tier 1 and the Rest tier.

The policy environment matters for interpretation. In a static
cross-section, a subsidy lowers the consumer net price and raises EV
demand. In the time-series decomposition, however, the subsidy block
asks a different question: what is the marginal contribution of moving
from the 2015 policy schedule to the 2024 policy schedule, holding
other blocks at different coalition states? Since the direct schedule
became less generous over this period, the subsidy block can be
negative even though subsidies remain economically important.

\subsection{EV growth}
\begin{figure}[!ht]
\caption{EV and GV attribute progress, 2015--2024.}
\label{fig:ev_progress}
\centering
\begin{subfigure}[t]{0.48\linewidth}
  \includegraphics[width=\linewidth]{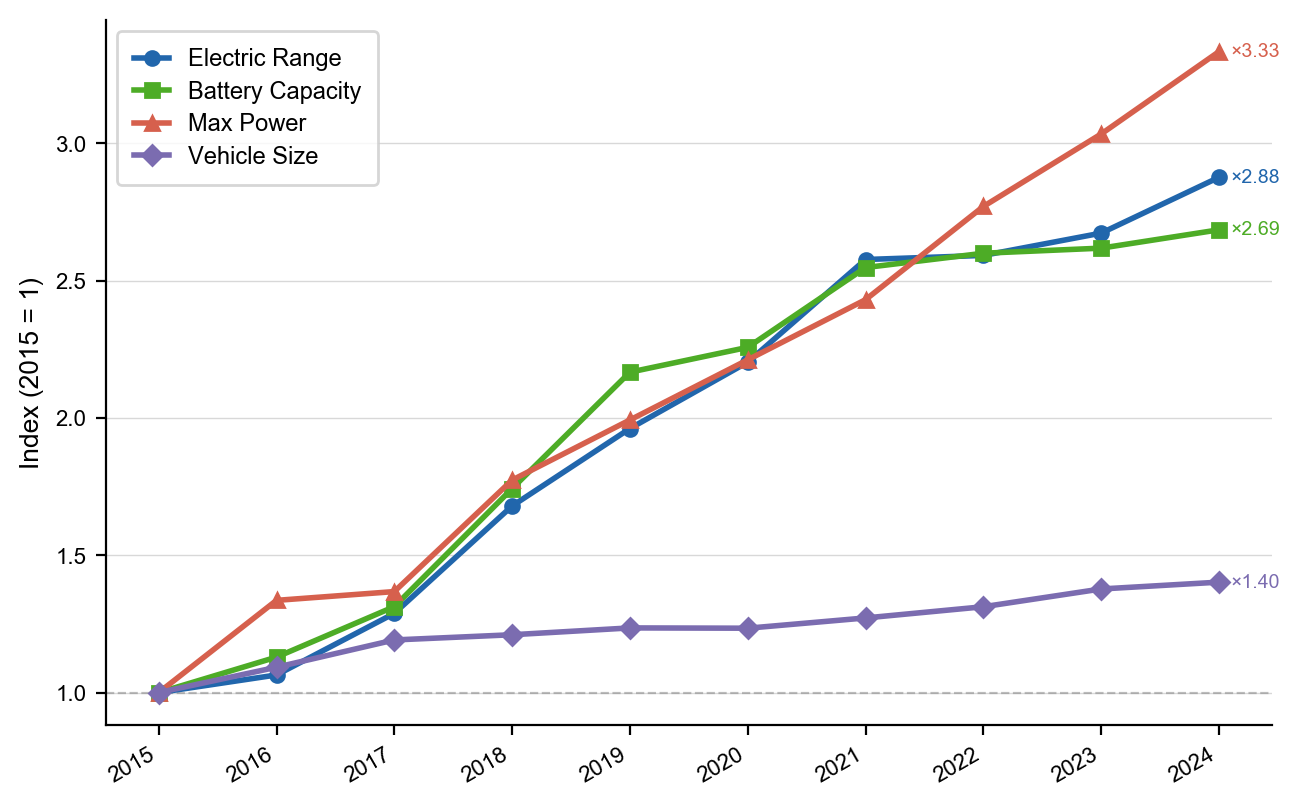}
  \caption{Electric vehicles}
\end{subfigure}
\hfill
\begin{subfigure}[t]{0.48\linewidth}
  \includegraphics[width=\linewidth]{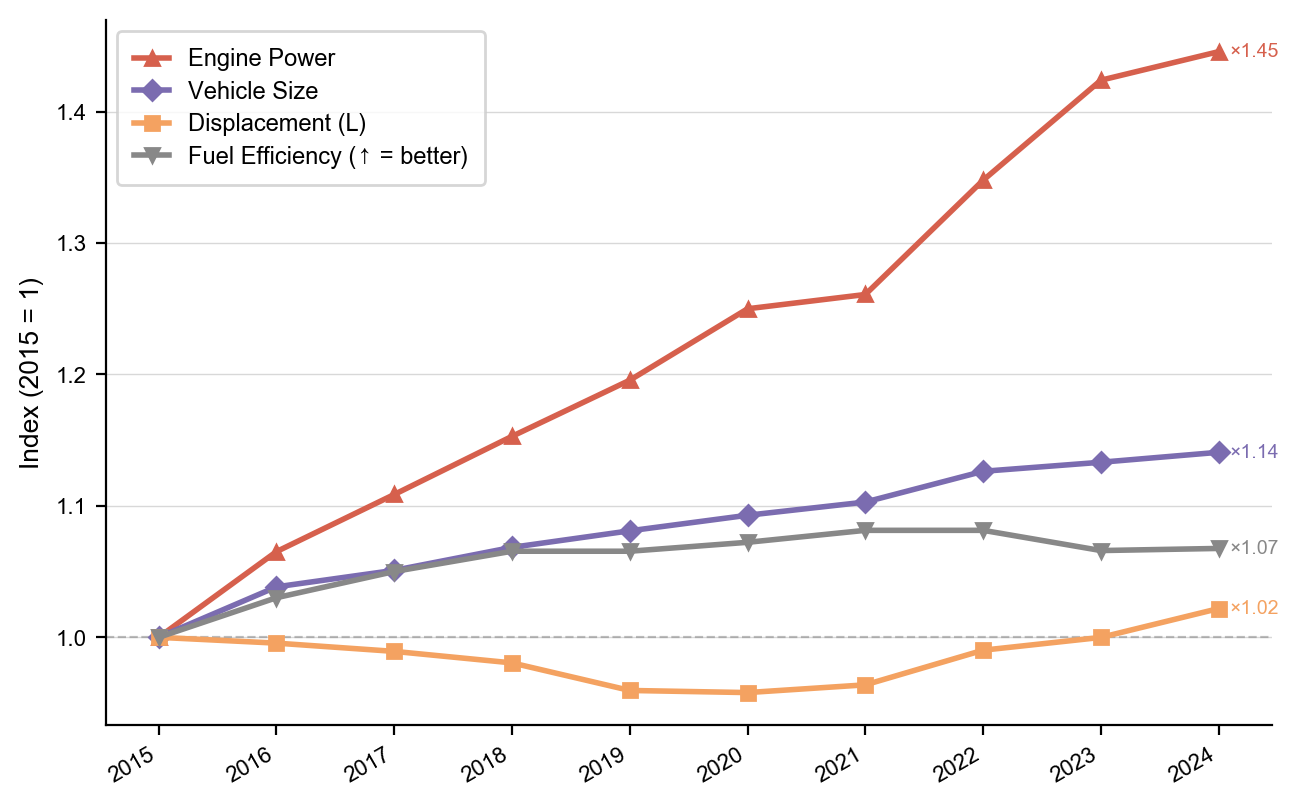}
  \caption{Gasoline vehicles}
\end{subfigure}
\par\smallskip
\flushleft{\footnotesize\textit{Notes:} Each series indexes the within-year product-level median to 2015~$= 1$. Engine Power and Vehicle Size use the same color and marker across panels. Annotations show cumulative growth by 2024.}
\end{figure}

\subsection{Product Panel and Key Variables}

The estimation sample is a product-by-city-by-year panel of passenger
vehicles from 2015 through 2024. Products are model and fuel-type
combinations. Fuel types include battery electric vehicles, plug-in
hybrids, range-extended electric vehicles, internal-combustion
vehicles, and hybrids. The main estimation sample contains about
476,787 product-market-year observations across 790 city-year markets.

The demand system uses consumer net prices, observed vehicle
attributes, city demographics, and policy variables. Consumer net price
equals sticker price less the relevant direct subsidy. Product
attributes include EV range, power, body type, fuel type, size segment,
and firm group. City variables include income, GDP per capita, density,
urbanization, population, and policy indicators. The market-size and
outside-good construction follows the existing code base.

Battery costs are measured using the BNEF annual battery-pack cost
series. The pack cost falls from 373 USD/kWh in 2015 to 115 USD/kWh in
2024. The cost series is used directly in the supply-side marginal-cost
regression rather than through an estimated learning curve. The paper
uses the term ``battery learning'' because the battery-cost path is the
empirical reduced-form outcome of industry-wide learning, scale, and
input-cost dynamics, but the current decomposition does not separately
estimate a structural Wright's-law learning elasticity.

\begin{table}[H]
\centering
\caption{Summary Statistics}
\label{tab:summary}
\footnotesize
\begin{tabularx}{\linewidth}{Xrrrrrr}
\toprule
 & \multicolumn{3}{c}{GVs} & \multicolumn{3}{c}{EVs} \\
\cmidrule(lr){2-4} \cmidrule(lr){5-7}
 & \# Obs. & Mean & Std.\ Dev. & \# Obs. & Mean & Std.\ Dev. \\
\midrule
\multicolumn{7}{l}{\textbf{Panel A: Product--year--city observations (demand estimation)}} \\[0.4em]
Sales (cars) & 366,706 & 21.25 & 52.80 & 110,081 & 10.90 & 45.27 \\
MSRP (10k RMB) & 366,706 & 15.86 & 10.29 & 110,081 & 24.13 & 13.70 \\
Net price (10k RMB) & 366,706 & 15.74 & 10.18 & 110,081 & 20.79 & 12.39 \\
Vehicle size ($\mathrm{m}^3$) & 366,706 & 13.72 & 2.00 & 110,081 & 13.45 & 2.64 \\
Horsepower & 366,706 & 117.00 & 34.60 & 110,059 & 118.70 & 55.63 \\
L/100km & 366,706 & 6.92 & 1.25 & --- & --- & --- \\
kWh/100km & --- & --- & --- & 110,081 & 14.29 & 5.73 \\
Driving range (km) & --- & --- & --- & 106,354 & 337.97 & 192.01 \\
Battery capacity (kWh) & --- & --- & --- & 106,354 & 46.91 & 25.27 \\
National cash subsidy (10k RMB) & --- & --- & --- & 110,081 & 0.92 & 1.22 \\
Local cash subsidy (10k RMB) & --- & --- & --- & 110,081 & 0.07 & 0.41 \\
Purchase-tax exemption (10k RMB) & --- & --- & --- & 110,081 & 1.84 & 1.10 \\
\midrule
\multicolumn{7}{l}{\textbf{Panel B: Model--year observations (supply estimation)}} \\[0.4em]
Sales (cars, national) & 4,762 & 1,636.32 & 2,721.50 & 1,782 & 673.60 & 1,861.90 \\
MSRP (10k RMB) & 4,762 & 15.79 & 10.43 & 1,782 & 24.08 & 13.21 \\
Net price (10k RMB) & 4,762 & 15.68 & 10.32 & 1,782 & 20.30 & 11.90 \\
Vehicle size ($\mathrm{m}^3$) & 4,762 & 13.75 & 2.02 & 1,782 & 13.17 & 2.71 \\
Horsepower & 4,762 & 116.84 & 34.86 & 1,781 & 112.14 & 55.83 \\
L/100km & 4,762 & 6.96 & 1.26 & --- & --- & --- \\
kWh/100km & --- & --- & --- & 1,782 & 14.49 & 6.98 \\
Driving range (km) & --- & --- & --- & 1,733 & 325.73 & 184.27 \\
Battery capacity (kWh) & --- & --- & --- & 1,733 & 45.51 & 24.55 \\
\midrule
\multicolumn{7}{l}{\textbf{Panel C: City--year observations (market structure)}} \\[0.4em]
 & & \# Obs. & Mean & Std.\ Dev. & Min & Max  \\
\cmidrule(lr){3-7}
Income per capita (10k RMB) && 790 & 4.62 & 1.44 & 2.17 & 9.31  \\
GDP per capita (10k RMB) && 790 & 8.93 & 4.31 & 2.22 & 26.92  \\
Population density (10k/km$^2$) && 790 & 0.09 & 0.12 & 0.00 & 0.90  \\
Urbanization rate && 790 & 0.68 & 0.18 & 0.00 & 1.00  \\
EV license priority (dummy) && 790 & 0.09 & 0.29 & 0.00 & 1.00 \\
Pilot city (dummy) && 790 & 0.47 & 0.50 & 0.00 & 1.00  \\
EV policy strength && 790 & 9.17 & 26.48 & {$-$}23.01 & 93.91  \\
Oil price (year-level, USD/bbl) && 790 & 65.90 & 17.32 & 42.00 & 99.00  \\
\# of firms && 790 & 76.29 & 6.32 & 57.00 & 92.00  \\
\# of GV models && 790 & 418.94 & 53.96 & 228.00 & 533.00  \\
\# of BEV models && 790 & 94.85 & 74.60 & 0.00 & 261.00  \\
\# of PHEV models && 790 & 34.77 & 30.01 & 0.00 & 102.00 \\
\# of REEV models && 790 & 4.61 & 7.69 & 0.00 & 25.00  \\
\bottomrule
\end{tabularx}
\vspace{0.5em}
\begin{flushleft}
\footnotesize
\textbf{Notes:} Panel A: product$\times$year$\times$city observations for BLP demand estimation (2015--2024, 79 cities). Panel B: model$\times$year observations for marginal-cost OLS. Panel C: year$\times$city observations for market-structure covariates ($N = 790$ cells). ``GVs'' includes ICE and HEV; ``EVs'' includes BEV, PHEV, and REEV. Vehicle size is exterior length$\times$width$\times$height in $\mathrm{m}^3$. ``L/100km'' is GV fuel economy; ``kWh/100km'' is EV electricity consumption. Subsidy rows split the cash schedule (national + local) from the 10\% purchase-tax exemption (active for NEVs 2014--2025). Driving range and battery capacity are EVs only. Population density is in 10{,}000 persons per $\mathrm{km}^2$. ``\# of firms'' counts unique manufacturers selling at least one model in a city$\times$year cell.
\end{flushleft}
\end{table}

\section{Demand Model}
The demand side follows a differentiated-products discrete-choice model in the spirit of \citet{Berry1995}. A market is defined as a city-by-year pair, indexed by $m=(c,t)$. Consumer $i$ chooses among the set of vehicle products $J_m$ available in market $m$ and an outside option.

The utility of consumer i for product j in market m is given as consists of three components: (i) a consumer-invariant mean utility component that captures the average attractiveness of a product, (ii) a consumer-specific component that generates heterogeneous preferences across consumers, and (iii) an idiosyncratic taste shock. Formally,

\begin{align}
u_{ijm}
&= \delta_{jm}+\mu_{ijm}+\varepsilon_{ijm}, \label{eq:utility}\\
\delta_{jm}
&= \beta_p \log p_{jm}
  + \beta_r \log r_{jm}^{\EV}
  + \beta_{rd}\log r_{jm}^{\EV}D_{cm}
  + x_{jm}'\beta
  + \eta^B_{j}+\eta^S_{j}
  + \eta^F_{j}+\eta^G_{j}
  + \eta_t+\xi_{jm}, \label{eq:mean_utility}\\
\mu_{ijm}
&= \pi_p \frac{1}{Y_{im}}\log p_{jm}. \label{eq:heterogeneity}
\end{align}

Equation~(\ref{eq:utility}) decomposes utility into a mean utility component $\delta_{jm}$, a consumer-specific deviation $\mu_{ijm}$, and an idiosyncratic taste shock $\varepsilon_{ijm}$. The idiosyncratic shock is assumed to follow a Type I extreme-value distribution.

Equation~(\ref{eq:mean_utility}) specifies the mean utility component. The variable $p_{jm}$ denotes the consumer net purchase price after direct subsidies. The variable $r_{jm}^{\EV}$ denotes EV driving range (measured in kilometers), while $D_{cm}$ is the log population density of city $c$. The coefficient $\beta_r$ captures the average valuation of EV driving range, and $\beta_{rd}$ allows the value of range to vary with local population density. The vector $x_{jm}$ contains additional observed product and market characteristics including engine power for gasoline vehicles, interactions between EV status and local policy variables (license-plate restrictions, EV pilot-city designation, and policy strength), interactions between EV status and city demographics (log household income, GDP per capita, log population density, and urbanization rate), as well as oil prices, log population density, GDP per capita, urbanization rate, and log city income. The fixed effects $\eta^B_j$, $\eta^S_j$, $\eta^F_j$, $\eta^G_j$, and $\eta_t$ denote body-type, size-segment, fuel-type, firm-group, and year fixed effects, respectively. The term $\xi_{jm}$ captures product-market specific demand shocks observed by consumers and firms but unobserved by the econometrician.

Equation~(\ref{eq:heterogeneity}) introduces consumer heterogeneity through income-dependent price sensitivity. Household income is denoted by $Y_{im}$. The parameter $\pi_p$ shows how the marginal utility of price varies across consumers. Because the price coefficient is scaled by $1/Y_{im}$, lower-income households are more sensitive to vehicle prices than higher-income households. A negative estimated value of $\pi_p$ implies that a given increase in vehicle prices reduces utility more strongly for lower-income households than for higher-income households.

In summary, the model-specific component of utility is determined by vehicle prices,
vehicle attributes, local market conditions, policy variables, and unobserved product characteristics. Consumers differ only in their
sensitivity to prices. Specifically, the consumer-specific component
$\mu_{ijm}$ allows the marginal disutility of price to vary with
household income $Y_{im}$. This specification is motivated by the idea that vehicle purchases represent a larger share of the budget for
lower-income households, making them more sensitive to price changes
than higher-income households. 

In contrast, the model does not include an unobserved random coefficient on EV status. Instead, the vector $x_{jm}$ includes interactions between the EV indicator and a vector of market demographics $Z_{m}$, consisting of household income, GDP per capita, population density, and urbanization. We also drop the random coefficient on log driving range. In preliminary estimates that allowed both an unobserved EV taste random coefficient $\sigma_{\mathrm{EV}}$ and a range random coefficient $\sigma_r$, the optimizer drove both toward zero whenever the parameter bounds were relaxed, indicating that the data identify mean-utility effects rather than additional unobserved heterogeneity once the rich observable interactions and fixed effects are in the specification. We therefore report a specification that drops both unobserved random coefficients and identifies EV-related heterogeneity through observable interactions and the random coefficient on the income-scaled price term in Equation~(\ref{eq:heterogeneity}). The estimated income-price coefficient $\pi_p$ in the canonical specification is interior to the optimizer bounds, but earlier specifications that imposed tighter bounds delivered estimates at the boundary; the heterogeneity in price sensitivity across income groups should therefore be read as a quantitative property of the IV moment system at the chosen instrument set and bound configuration, not as a tightly identified data feature. This is the basis on which all subsequent cross-tier welfare results in Section~\ref{sec:subsidy_incidence} rest, and we report a calibrated demand-elasticity sensitivity exercise in Section~\ref{sec:no_subsidy_dynamic} that brackets the no-subsidy effects under $|\alpha|=2.5$; readers who view the heterogeneity in price sensitivity as a calibration rather than an estimate should read the cross-tier welfare gradient in the same calibrated spirit. This specification allows EV demand to vary systematically across markets with different demographic and economic characteristics. For example, EV adoption may differ between richer and poorer cities because of differences in purchasing power, between denser
and less dense cities because of differences in driving patterns and
charging convenience, and between more and less urbanized areas because
of differences in infrastructure and consumer preferences.

Given the Type I extreme-value assumption, the predicted market share of product $j$ is

\begin{equation*}
s_{jm}(\theta)
=
\int
\frac{e^{\delta_{jm}+\mu_{ijm}}}
{1+\sum_{k\in J_m} e^{\delta_{km}+\mu_{ikm}}}
\, dF_m(Y_i).
\label{eq:share}
\end{equation*}

where $\theta$ denotes the vector of demand parameters and $F_m(Y_i)$ denotes the market-specific distribution of household
income used to simulate consumer heterogeneity. The outside option is
normalized to have mean utility equal to zero. For a candidate value of
the nonlinear parameter $\pi_p$, the vector of mean utilities
$\delta_m$ is recovered by matching predicted and observed market shares
using the BLP contraction mapping.

\subsection{Identification and Estimation}
Following \citet{whitefoot2017compliance}, we view vehicle design as a multistage
process in which firms choose some product attributes well before observing market-specific demand shocks. Characteristics such as body
type, size segment, and engine power reflect fundamental design choices are costly to modify once a model enters production. We therefore assume that these attributes are
chosen prior to the realization of market-specific demand shocks and are uncorrelated with $\xi_{jm}$. However, in contrast, EV manufacturers can adjust driving range more
easily than fundamental vehicle characteristics by varying battery
capacity or adopting improvements in battery technology. Consistent with
this idea, the data show that manufacturers frequently update the driving range of an existing vehicle models over their lifecycle while leaving its fundamental design characteristics unchanged, as illustrated in Figure~\ref{fig:range_traj}. Firms may therefore choose driving range in response to demand factors that are observed by manufacturers but not by the econometrician. As a result,
driving range may be correlated with the unobserved demand shock
$\xi_{jm}$. 

Vehicle prices are also likely to be endogenous because
products with favorable unobserved characteristics command higher
willingness to pay and therefore higher equilibrium prices. To address
these concerns, we treat both vehicle prices and EV driving range as
endogenous product characteristics and estimate the model using
instrumental variables.

To address the endogeneity of vehicle prices and EV driving range, we
follow the instrumental-variable approach developed in
\citet{Berry1995} and subsequently applied in the automobile demand literature. The estimation relies on moment conditions of the form $E[\xi_{jm}\mid Z_{jm}] = 0$, where $\xi_{jm}$ is the unobserved demand shock and $Z_{jm}$ denotes a
vector of exogenous product characteristics, market controls, fixed
effects, and excluded instruments.

We utilize three sets of excluded instruments. First, following the
BLP differentiation-instrument approach, we construct, for each
product attribute $x_{jm}^{(k)}$ in $\{\log r, \log\mathrm{power}^{GV},
\log r \times \mathrm{density}\}$, the leave-one-out rivals' mean
within the market: $z_{jm}^{\mathrm{diff},(k)} = x_{jm}^{(k)} -
\bar x_{-j,m}^{(k)}$ where $\bar x_{-j,m}^{(k)}$ averages over all
products $j' \neq j$ in market $m$. These instruments shift equilibrium
markups and prices through competitive interactions while remaining
orthogonal to product-specific demand shocks. Second, following the
nested differentiation approach used in the automobile literature, we
construct the same leave-one-out rivals' mean within the
product's fuel-type segment (BEV, PHEV, REEV, HEV, ICE separately). These instruments
capture variation in local competitive pressure arising from the
composition of products within each fuel category. Third, we exploit variation in battery production costs \footnote{The battery-cost instruments include battery costs interacted
with BEV, PHEV, and REEV indicators, battery capacity, and interactions
between battery costs and battery capacity.}. Because batteries constitute a major component of EV production costs, changes in battery prices affect both vehicle prices and the cost of providing longer driving range. We therefore construct instruments based on battery costs, battery capacity, and EV technology type. Conditional on observed vehicle characteristics, market controls, and fixed effects, the identifying assumption is that battery-cost shocks influence demand only through their effects on prices and range.\footnote{We
explored discrete range-tier fixed effects and a Hausman-style leave-out
instrument; neither delivered a well-conditioned moment system. The
canonical IV set is bracketed with a calibrated demand-elasticity
robustness exercise in Section~\ref{sec:no_subsidy_dynamic}.}

The market-specific income distribution $F_m(Y_i)$ used to integrate
Equation~(\ref{eq:share}) is taken from the corresponding year's
\emph{China Statistical Yearbook} city-level household-income panel.
We discretize $F_m$ into $N_S = 25$ quasi-Monte Carlo draws per
market using a lognormal calibration matched to the mean and Gini
coefficient of the city-year cell, and we use the same set of draws
throughout the contraction mapping, the BLP inversion, and all
counterfactual share computations. The random coefficient parameter
governing income-dependent price sensitivity is identified from
variation in household income distributions across markets. Because
the consumer-specific utility component depends on household income,
markets with similar product
offerings but different income distributions imply different
substitution patterns and price responses. This variation provides
identifying information for the nonlinear parameter governing
heterogeneous price sensitivity.

The model is estimated using the generalized method of moments (GMM).
For a candidate value of the nonlinear parameter $\pi_p$, the BLP contraction mapping is used to recover the mean utility levels $\delta_{jm}$ that match observed and predicted market shares. Conditional on the recovered mean utilities, the linear parameters are
estimated using the moment conditions described above. The nonlinear parameter is then chosen to minimize the resulting GMM objective function. Standard errors are clustered at the market level. The
estimation sample contains 476,787 product-market observations.

\paragraph{Instrument strength.}
Table~\ref{tab:iv} reports the first-stage F-statistics for the three
instrument groups separately and jointly. The within-market BLP
differentiation instruments deliver a partial $F$ of $1{,}655$ on log
price; the within-fuel-type nested instruments deliver $5{,}144$; and
the battery-cost instruments deliver $15{,}728$. The joint $F$ across
all eleven excluded instruments is $18{,}479$. All three groups
individually and the joint test exceed the conventional Stock--Yogo
weak-instrument threshold of $10$ by orders of magnitude, so price
endogeneity is strongly resolved by the instrument set. The
identifying variation has three economic interpretations matched to
the three IV groups: within-market substitution patterns
($Z_{\mathrm{diff}}$), within-fuel-type substitution patterns
($Z_{\mathrm{nest}}$), and time-series cost shocks transmitted
through battery-input prices ($Z_{\mathrm{bat}}$). The joint power of
the three groups reflects that price variation in the panel is driven
by genuinely distinct sources, none of which is mechanically tied to
the demand residual $\xi_{jm}$.

\input{figures/tab_iv_diagnostics}

We do not report a Hansen $J$ overidentification test.
In a panel with $N\approx 477{,}000$ observations, the asymptotic
$\chi^2(10)$ distribution mechanically rejects almost any non-trivial
moment condition because the test statistic scales linearly with $N$.
Conventional applied-IO practice in large-N panel BLP settings is to
treat Hansen $J$ as uninformative in this regime and to rely on
first-stage strength plus the economic argument for instrument
validity. The economic argument is straightforward in our setting:
$Z_{\mathrm{diff}}$ and $Z_{\mathrm{nest}}$ are functions of competing
products' observable characteristics, which firms take as given in
their own pricing problem; $Z_{\mathrm{bat}}$ is the BNEF battery-pack
price, which is a national time-series shock that no individual product
can manipulate. Conditional on the rich fixed-effect structure and
the demographic interactions in Equation~(\ref{eq:heterogeneity}),
the residual variation in $Z$ is orthogonal to product-specific
demand shocks by construction.

\subsection{Demand Estimation Results}

\begin{table}[htbp]
\centering
\footnotesize
\caption{Demand Estimation Results}
\label{tab:demand}
\begin{tabularx}{\linewidth}{Xrrrrrr}
\toprule
 & \multicolumn{2}{c}{OLS Logit} & \multicolumn{2}{c}{IV Logit} & \multicolumn{2}{c}{Random Logit} \\
\cmidrule(lr){2-3}\cmidrule(lr){4-5}\cmidrule(lr){6-7}
 & Coef. & S.E. & Coef. & S.E. & Coef. & S.E. \\
\midrule
\multicolumn{7}{l}{\textit{Vehicle Attributes}} \\[2pt]
  Log Price & -0.076*** & (0.024) & -4.032*** & (1.458) & -2.925*** & (0.568) \\
  Log Engine Power (GV) & 0.410*** & (0.033) & 0.514 & (1.306) & 0.869** & (0.383) \\
  Log EV Range (km) & 0.716*** & (0.038) & 8.767*** & (1.214) & 8.191*** & (0.338) \\
  EV Range $\times$ Log Density & 0.198* & (0.108) & -1.531*** & (0.571) & -1.428*** & (0.381) \\
[4pt]\multicolumn{7}{l}{\textit{Policy and Market Variables}} \\[2pt]
  EV $\times$ License Restriction & 0.014 & (0.024) & 0.058* & (0.035) & 0.065** & (0.031) \\
  EV $\times$ Pilot City & 0.112*** & (0.021) & 0.058 & (0.038) & 0.040 & (0.038) \\
  EV $\times$ Policy Strength & 0.024 & (0.015) & 0.035 & (0.045) & 0.022 & (0.032) \\
  EV $\times$ Log Income & 7.522*** & (0.861) & 8.774 & (12.601) & 14.129*** & (2.966) \\
  EV $\times$ GDP Per Capita & -0.238*** & (0.088) & -0.416 & (0.544) & -0.559*** & (0.185) \\
  EV $\times$ Log Pop.\ Density & -0.043 & (0.043) & 0.515** & (0.219) & 0.467*** & (0.142) \\
  EV $\times$ Urb.\ Rate & -0.008 & (0.124) & 0.491 & (0.385) & 0.469** & (0.207) \\
  Oil Price & 33.463*** & (1.603) & 81.865*** & (20.788) & 84.148*** & (5.646) \\
  Log Pop.\ Density & -0.059** & (0.028) & -0.056* & (0.034) & -0.037 & (0.032) \\
  GDP Per Capita & 0.387*** & (0.075) & 0.435*** & (0.131) & 0.499*** & (0.082) \\
  Urbanization Rate & -0.026 & (0.052) & -0.052 & (0.067) & -0.078 & (0.055) \\
  Log City Income & 0.013 & (0.047) & -0.006 & (0.195) & -1.165*** & (0.415) \\
\midrule
\multicolumn{7}{l}{\textit{Random Coefficients}} \\[2pt]
  $\pi$(Income $\times$ Price) & & & & & -1.980** & (0.859) \\
\midrule
  Fixed Effects & \multicolumn{2}{c}{\checkmark} & \multicolumn{2}{c}{\checkmark} & \multicolumn{2}{c}{\checkmark} \\
  Observations & \multicolumn{2}{c}{476,787} & \multicolumn{2}{c}{476,787} & \multicolumn{2}{c}{476,787} \\
  $R^2$ & \multicolumn{2}{c}{0.198} & \multicolumn{2}{c}{} & \multicolumn{2}{c}{} \\
  First-Stage $F$ & \multicolumn{2}{c}{} & \multicolumn{2}{c}{3623.7} & \multicolumn{2}{c}{} \\
  GMM Objective & \multicolumn{2}{c}{} & \multicolumn{2}{c}{} & \multicolumn{2}{c}{0.6903} \\
\bottomrule
\end{tabularx}
\\[4pt]
\begin{minipage}{\linewidth}
  \footnotesize\raggedright
  \textit{Notes:} Standard errors clustered by market (OLS and IV)
  or 2-step optimal GMM market-clustered (RC BLP).
  $^{*}p<0.10$, $^{**}p<0.05$, $^{***}p<0.01$.
  All specifications include fuel type, body type, year, firm group,
  and EV~$\times$~year fixed effects.
  Policy variables are standardized to unit variance.
  IV and RC instruments: BLP differentiation IVs on product characteristics,
  within-fuel-type nested IVs, and BNEF battery cost shifters.
\end{minipage}
\end{table}

Table~\ref{tab:demand} reports the demand estimates from the
OLS logit, IV logit, and random-coefficients BLP specifications. The
OLS estimates do not account for the endogeneity of vehicle prices and
EV driving range. As a result, the estimated price coefficient is small
in magnitude and economically implausible. Products with favorable
unobserved demand shocks tend to command higher prices, generating a
positive correlation between prices and the unobserved demand component
$\xi_{jm}$ and biasing the price coefficient toward zero.

Once prices and driving range are instrumented, the estimated price
coefficient increases substantially in magnitude. The IV logit estimate
of the log-price coefficient is $-4.03$, compared with $-0.08$ in the
OLS specification. This large difference indicates that price
endogeneity is quantitatively important in the Chinese automobile
market. The coefficients on vehicle attributes are also substantially
larger in magnitude after instrumenting. In particular, consumers value
both EV driving range and gasoline-vehicle engine power, while higher
fuel prices increase the relative attractiveness of EVs.

The preferred specification is the random-coefficients BLP model
reported in the final two columns. The mean log-price coefficient
remains negative and economically meaningful at $-2.93$. The estimated
income-price interaction parameter is negative and statistically
significant, implying that lower-income households are more sensitive to
vehicle prices than higher-income households. This result is consistent
with the notion that vehicle purchases constitute a larger share of the
budget for lower-income consumers.

The estimates indicate that consumers place substantial value on EV
driving range. The coefficient on log range is positive, while the
interaction between range and population density is negative. Together,
these estimates imply that the marginal value of driving range is lower
in denser cities, where shorter driving distances and greater charging
availability reduce the benefits of additional range.

The demographic interactions reveal substantial heterogeneity in EV
demand across markets. EV demand is stronger in cities with higher
household income, greater population density, and higher urbanization
rates. Rather than relying on an unrestricted EV random coefficient,
the specification captures variation in EV adoption through observable
market characteristics. This approach provides a transparent and
economically interpretable source of heterogeneity in EV preferences
across Chinese cities.

The policy coefficients suggest that non-financial EV incentives also
affect adoption. In particular, EVs receive a significant utility
premium in markets with license-plate restrictions from which they are
exempt. This finding is consistent with evidence that registration
restrictions increase the relative attractiveness of EV ownership in
large urban markets.

\subsection{Demand-Side Fit and Residual Demand}

\begin{figure}[!ht]
\caption{Recovered BLP mean utility $\delta_{jm}$: EV vs.\ GV, 2015--2024}
\label{fig:delta_dist}
\centering
\begin{subfigure}[t]{0.48\linewidth}
  \includegraphics[width=\linewidth]{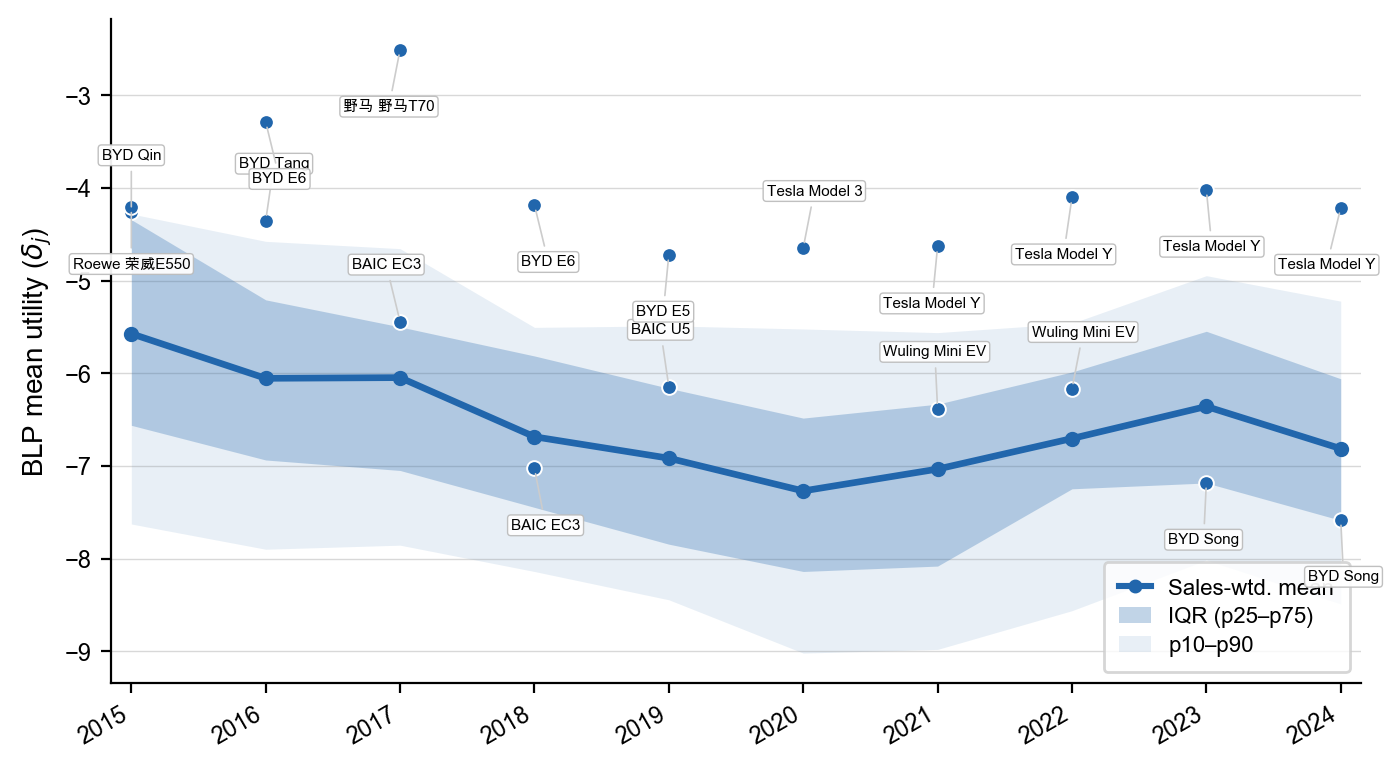}
  \caption{Electric vehicles}
\end{subfigure}
\hfill
\begin{subfigure}[t]{0.48\linewidth}
  \includegraphics[width=\linewidth]{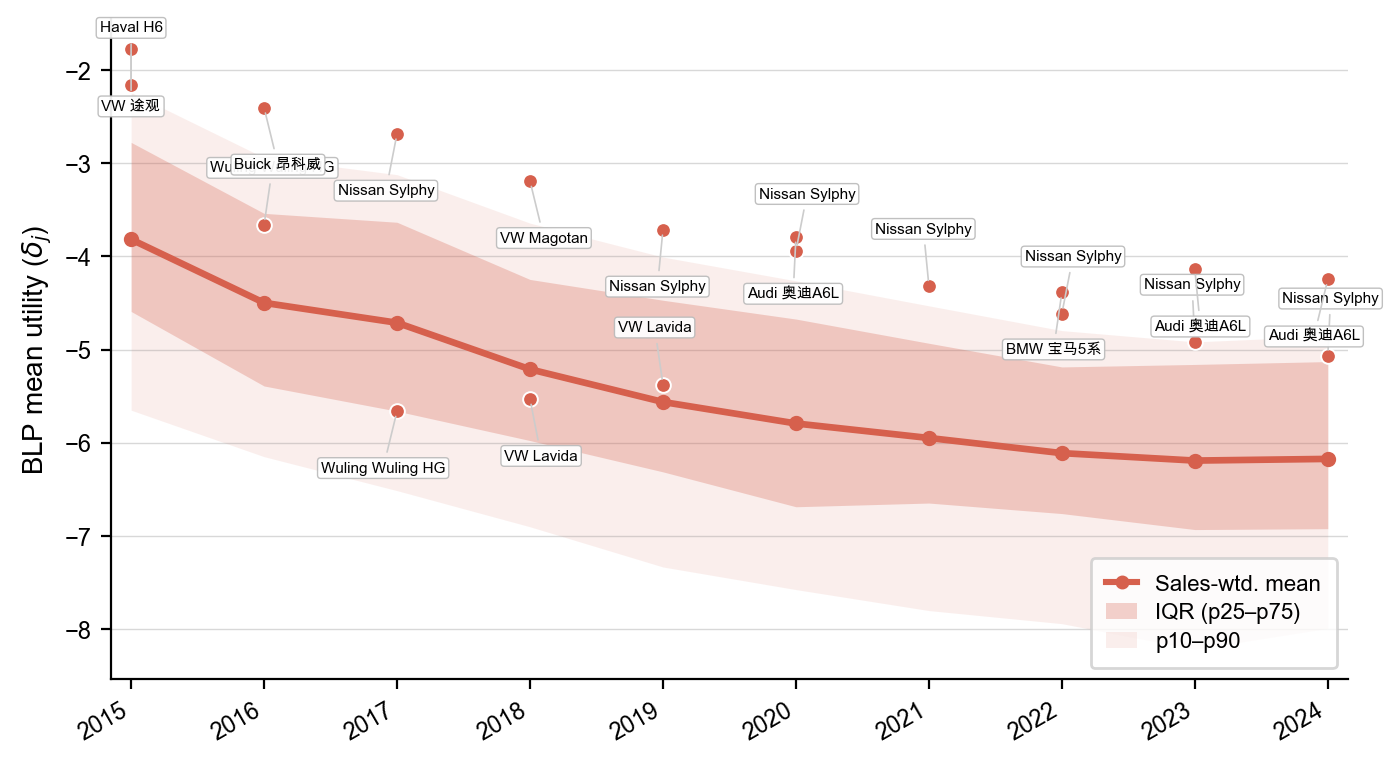}
  \caption{Gasoline vehicles}
\end{subfigure}
\par\smallskip
\flushleft{\footnotesize\textit{Notes:} Line: sales-weighted mean;
dark band: IQR (p25--p75); light band: p10--p90. Selected notable
models are annotated.}
\end{figure}

Figure~\ref{fig:delta_dist} plots the annual distribution of recovered
BLP mean utility for EV and GV products. Panel~(a) reports EV products,
while Panel~(b) reports GV products. The line in each panel is the
sales-weighted mean, the darker band is the interquartile range, and the
lighter band shows the 10th to 90th percentile range.

The figure summarizes how the average market attractiveness of EV and GV
products evolves over time after the BLP inversion. Recovered mean
utility incorporates the product's observed characteristics, price,
fixed effects, and unobserved demand component. It should therefore be
interpreted as the overall mean utility that rationalizes observed
market shares, rather than as the structural demand residual itself. The
comparison between the two panels shows that GV mean utility declines
steadily over the sample period, while EV mean utility displays greater
dispersion and a partial recovery after the early sample years. This
pattern is consistent with a market in which EV products become more
heterogeneous and increasingly competitive with gasoline vehicles over
time.

\begin{figure}[!ht]
\caption{Year-specific mean-utility deviations: EV vs.\ GV}
\label{fig:xi_dist}
\centering
\begin{subfigure}[t]{0.48\linewidth}
  \includegraphics[width=\linewidth]{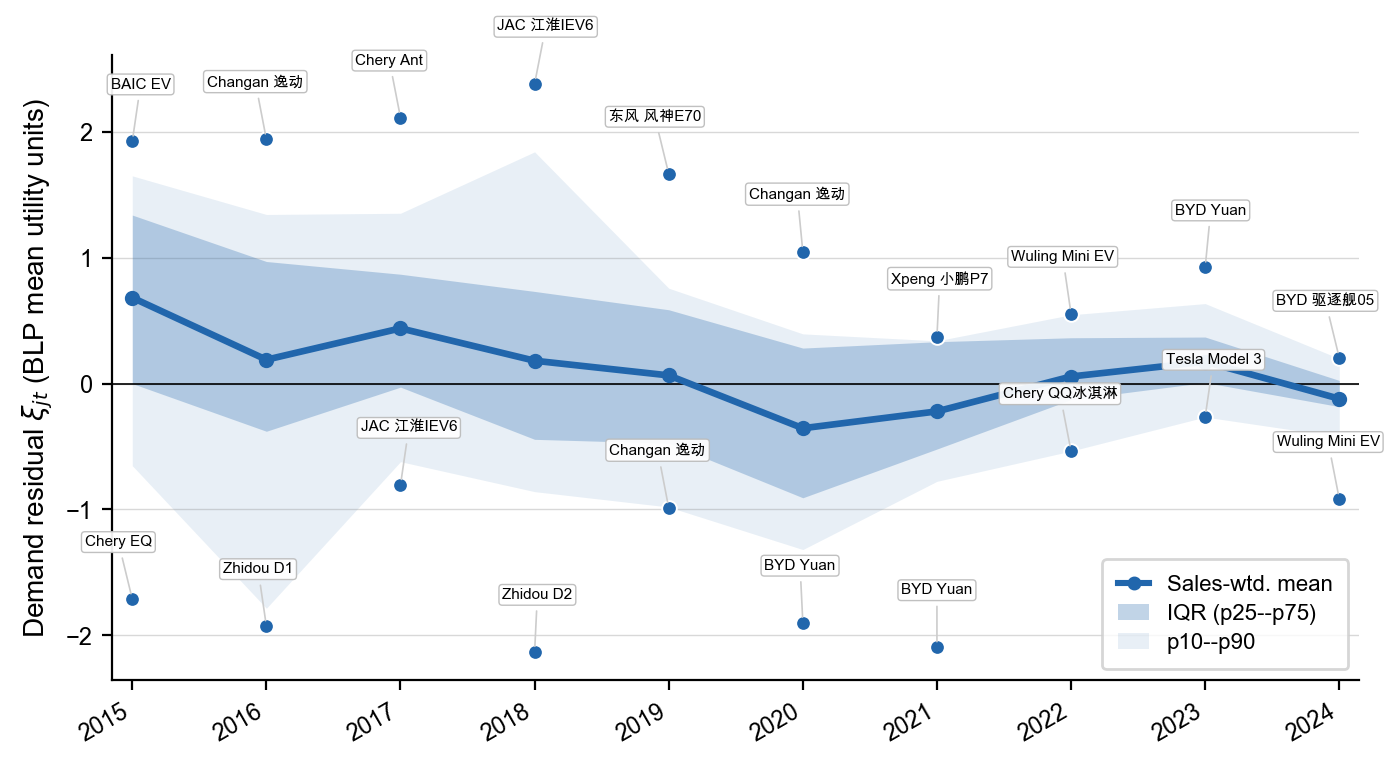}
  \caption{Electric vehicles}
\end{subfigure}
\hfill
\begin{subfigure}[t]{0.48\linewidth}
  \includegraphics[width=\linewidth]{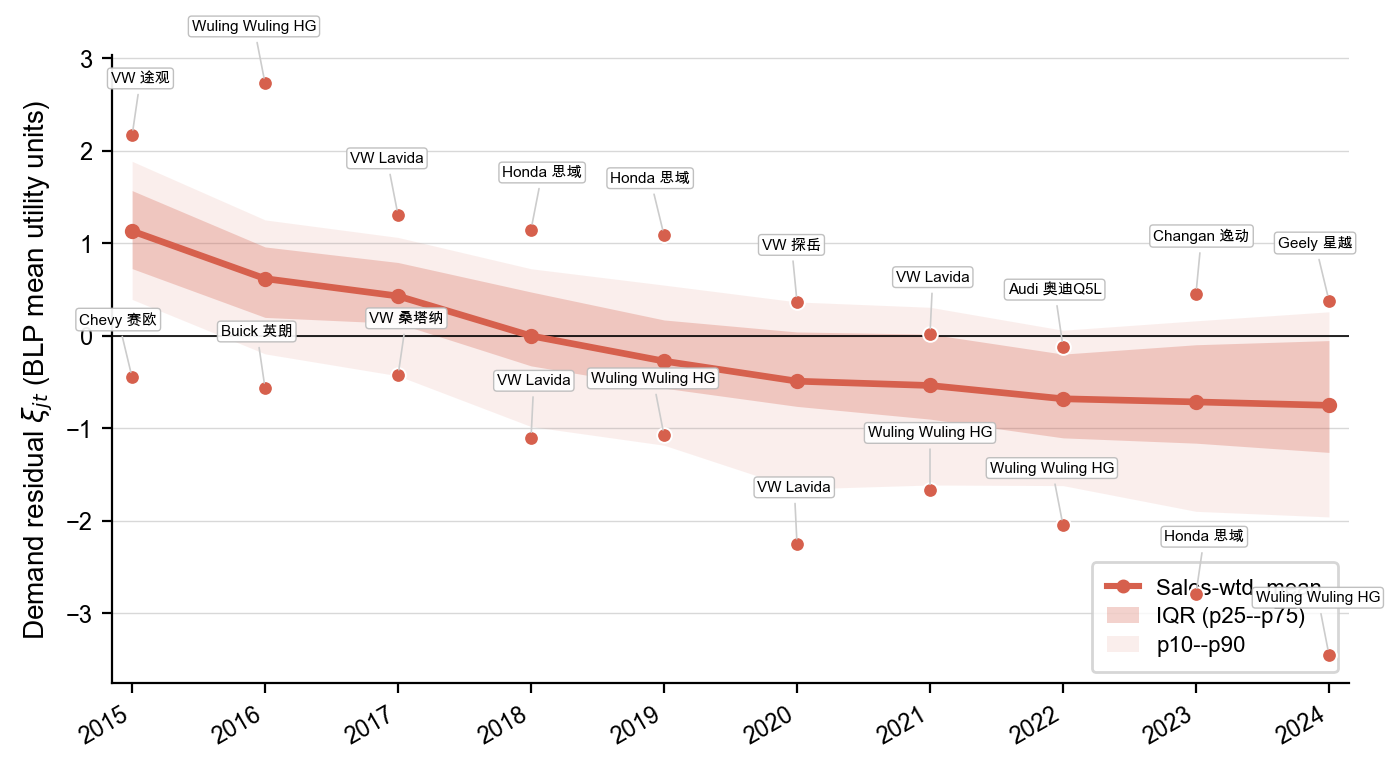}
  \caption{Gasoline vehicles}
\end{subfigure}
\par\smallskip
\flushleft{\footnotesize\textit{Notes:}
$\tilde{\xi}_{jt}=\delta_{jt}-\bar{\delta}_{j}$, where
$\bar{\delta}_{j}$ is product $j$'s sales-weighted mean utility across
all years. Line: sales-weighted mean; dark band: IQR (p25--p75); light
band: p10--p90.}
\end{figure}

Figure~\ref{fig:xi_dist} removes each product's average recovered mean
utility and plots year-specific deviations. This transformation isolates
within-product changes in market attractiveness over time. Panel~(a)
reports EV products, while Panel~(b) reports GV products.

The GV panel shows a clear downward trend in year-specific mean-utility
deviations, indicating that gasoline vehicles become less attractive
relative to their own long-run average over the sample period. The EV
panel is more volatile, with larger dispersion across models and years.
This pattern is consistent with rapid product turnover and changing
consumer perceptions in the EV market. The figure therefore provides
reduced-form evidence that time-varying residual demand shifts play an
important role in the transition from gasoline vehicles to EVs.

\section{Supply-Side Model}

Following \citet{Berry1995}, we recover marginal costs from firms'
static multi-product Bertrand pricing decisions. The demand model is
specified in terms of the consumer net purchase price $p_{jm}$, which
is the price paid by consumers after direct purchase subsidies. Let
$\tau_{jm}$ denote the per-unit purchase subsidy, and let
$\tilde p_{jm}$ denote the corresponding producer price, or the price
received by the firm. These prices are related by
$\tilde p_{jm}
=
p_{jm}
+
\tau_{jm}$. Let
$p_m
=
\left(
p_{1m},
p_{2m},
\ldots,
p_{J_m m}
\right)'$
denote the vector of consumer net prices in market $m$. The demand
system implies market shares $s_{jm}(p_m;\hat{\theta})$, where
$\hat{\theta}$ denotes the estimated demand parameters. In the supply
side, product characteristics, market characteristics, fixed effects,
and demand shocks are held fixed, so market shares are written as
functions of the price vector chosen by firms.

Because the subsidy is taken as fixed by the firm within a market, a
change in the producer price $\tilde p_{jm}$ changes the consumer net
price $p_{jm}$ one-for-one. Firm $f$ owns the set of products
$\mathcal{J}_{fm}$ in market $m$ and chooses prices to maximize
variable profits:
\begin{equation}
\Pi_{fm}
=
\sum_{j\in\mathcal{J}_{fm}}
\left(
\tilde p_{jm}
-
\mathrm{mc}_{jm}
\right)
M_m s_{jm}(p_m),
\label{eq:profit}
\end{equation}
where $\mathrm{mc}_{jm}$ is marginal cost, $M_m$ is market size, and
$s_{jm}(p_m)$ is the market share predicted by the estimated demand
system.

For each product $j\in\mathcal{J}_{fm}$, firm $f$ chooses the producer
price $\tilde p_{jm}$. The first-order condition with respect to
$\tilde p_{jm}$ is
\begin{equation}
\frac{\partial \Pi_{fm}}
{\partial \tilde p_{jm}}
=
M_m s_{jm}(p_m;\hat{\theta})
+
M_m
\sum_{k\in\mathcal{J}_{fm}}
\left(
\tilde p_{km}
-
\mathrm{mc}_{km}
\right)
\frac{\partial s_{km}(p_m;\hat{\theta})}
{\partial \tilde p_{jm}}
=
0.
\label{eq:foc_product_raw}
\end{equation}
Dividing by market size $M_m$ gives
\begin{equation}
s_{jm}(p_m;\hat{\theta})
+
\sum_{k\in\mathcal{J}_{fm}}
\left(
\tilde p_{km}
-
\mathrm{mc}_{km}
\right)
\frac{\partial s_{km}(p_m;\hat{\theta})}
{\partial \tilde p_{jm}}
=
0.
\label{eq:foc_product}
\end{equation}

Since $\tau_{jm}$ is fixed within market $m$, the derivative with
respect to the producer price is equal to the derivative with respect
to the consumer net price, and hence $\frac{\partial s_{km}(p_m)}
{\partial \tilde p_{jm}}
=
\frac{\partial s_{km}(p_m)}
{\partial p_{jm}}$. Also because the demand model is specified in log prices, this derivative is
computed as
$\frac{\partial s_{km}(p_m)}
{\partial p_{jm}}
=
\frac{\partial s_{km}(p_m)}
{\partial \log p_{jm}}
\frac{1}{p_{jm}}$.

Let $\sigma_{ijm}$ denote the individual choice probability of consumer
$i$ for product $j$ in market $m$. The derivative with respect to log
price is implied by the estimated demand system:
\begin{equation}
\frac{\partial s_{km}(p_m)}
{\partial \log p_{jm}}
=
\int
\sigma_{ikm}
\left(
\mathbf{1}\{j=k\}
-
\sigma_{ijm}
\right)
\left(
\beta_p
+
\pi_p \frac{1}{Y_{im}}
\right)
\, dF_m(Y_i).
\label{eq:share_derivative_log_price}
\end{equation}

Stacking the first-order conditions across products within market $m$
gives
\begin{equation}
s_m(p_m)
+
\left(
\Omega_m
\circ
\Delta_m(p_m)
\right)
\left(
\tilde p_m
-
\mathrm{mc}_m
\right)
=
0,
\label{eq:foc_matrix}
\end{equation}
where $\Omega_m$ is the ownership matrix, with element
$\Omega_{jk,m}=1$ if products $j$ and $k$ are owned by the same firm and
zero otherwise. The symbol $\circ$ denotes element-wise multiplication.
The matrix $\Delta_m(p_m)$ is the Jacobian matrix of market shares with
respect to producer prices, with element $\Delta_{jk,m}
=
\frac{\partial s_{km}(p_m)}
{\partial \tilde p_{jm}}$.

Marginal costs are therefore recovered by inverting the first-order
conditions:
\begin{equation}
\mathrm{mc}_m
=
\tilde p_m
+
\left(
\Omega_m
\circ
\Delta_m(p_m)
\right)^{-1}
s_m(p_m).
\label{eq:mc_recovery}
\end{equation}

The recovered marginal costs are used only for products with valid
first-order-condition solutions.

We then parameterize the recovered marginal costs in a single pooled
regression that uses technology-specific interactions to let the
cost-structure of EVs and gasoline vehicles differ:
\begin{equation}
\begin{aligned}
\log \mathrm{mc}_{jm}
&=
\gamma_0
+
\gamma_s
\log \mathrm{size}_{jm}
+
\gamma_h
\log \mathrm{power}_{jm}^{\mathrm{GV}}
+
\gamma_r
\log r_{jm}^{\mathrm{EV}}
+
\gamma_b
\mathrm{BatteryCost}_{jm}^{\mathrm{EV}}
\\
&\quad
+
\lambda_{\mathrm{fuel}(j)}
+
\lambda_{\mathrm{body}(j)}
+
\lambda_{\mathrm{firm}(j)}
+
\lambda_t
+
\omega_{jm}.
\end{aligned}
\label{eq:mc_regression}
\end{equation}

The technology-specific covariates are constructed as masked interactions:
$\log \mathrm{power}_{jm}^{\mathrm{GV}} \equiv (1 - \mathbf{1}[j \in \mathrm{EV}]) \cdot \log \mathrm{power}_{jm}$,
$\log r_{jm}^{\mathrm{EV}} \equiv \mathbf{1}[j \in \mathrm{EV}] \cdot \log r_{jm}$, and
$\mathrm{BatteryCost}_{jm}^{\mathrm{EV}} \equiv \mathbf{1}[j \in \mathrm{EV}] \cdot \mathrm{bnef}_t$,
where $\mathrm{bnef}_t$ is the BNEF annual battery-pack cost ($/$kWh).
This pooled specification is mathematically equivalent to fitting separate
regressions on the EV and GV subsamples for the technology-specific
slopes, but it pools the fixed effects across fuel categories, which
substantially improves the identification of the battery-cost
coefficient by exploiting the EV-versus-GV contrast within each year
rather than relying on residual variation in the time series of battery
prices alone. The fixed effects control for fuel type, body type, firm
group, and year. The coefficient $\gamma_r$ on EV driving range
captures the marginal cost of providing longer range; $\gamma_h$ on
gasoline-vehicle engine power captures the cost of providing higher
engine power; $\gamma_b$ on BatteryCost is the elasticity of EV
marginal cost with respect to the BNEF battery-pack price and is the
structural primitive that drives the Wright's-law channel in the
counterfactual simulations of Section~\ref{sec:no_subsidy_dynamic}.

\subsection{Marginal Cost Estimation Results}
Table~\ref{tab:supply} reports the pooled marginal-cost OLS
estimates from Equation~\eqref{eq:mc_regression}. The technology-specific
covariates (log power for GVs, log range and BatteryCost for EVs) are
constructed as masked interactions so that a single regression delivers
the slope coefficients separately by technology while pooling the FuelType,
BodyType, year, and firm-group fixed effects across all products. The
estimates here are exactly the coefficients used in the decomposition and
the dynamic counterfactual simulations.

\begin{table}[H]
\centering
\footnotesize
\caption{Supply-Side Marginal Cost Estimates (Pooled OLS)}
\label{tab:supply}
\begin{tabularx}{\linewidth}{Xrr}
\toprule
 & Coef. & S.E. \\
\midrule
\multicolumn{3}{l}{\textit{Product Characteristics}} \\[2pt]
  Log Size                            & $+1.5070^{***}$ & (0.0055) \\
  Log Engine Power (GV interaction)   & $+0.9524^{***}$ & (0.0024) \\
  Log EV Range (EV interaction)       & $+0.3967^{***}$ & (0.0034) \\
  BatteryCost\,$\times \mathbf{1}[\mathrm{EV}]$ (USD/kWh) & $+0.0040^{***}$ & (0.0000) \\
\midrule
\multicolumn{3}{l}{\textit{Fuel Type Fixed Effects (base: BEV)}} \\[2pt]
  HEV   & $-1.7490^{***}$ & (0.0228) \\
  ICEV  & $-2.0170^{***}$ & (0.0228) \\
  PHEV  & $+0.6893^{***}$ & (0.0062) \\
  REEV  & $+2.5170^{***}$ & (0.0221) \\
\midrule
\multicolumn{3}{l}{\textit{Year Fixed Effects (base: 2015)}} \\[2pt]
  2016 & $-0.0603^{***}$ & (0.0019) \\
  2017 & $-0.0736^{***}$ & (0.0019) \\
  2018 & $-0.1037^{***}$ & (0.0018) \\
  2019 & $-0.1491^{***}$ & (0.0018) \\
  2020 & $-0.1637^{***}$ & (0.0018) \\
  2021 & $-0.1675^{***}$ & (0.0019) \\
  2022 & $-0.1912^{***}$ & (0.0019) \\
  2023 & $-0.2187^{***}$ & (0.0019) \\
  2024 & $-0.2376^{***}$ & (0.0021) \\
\midrule
  Body Type FE  & \multicolumn{2}{c}{\checkmark} \\
  Firm Group FE & \multicolumn{2}{c}{\checkmark} \\
  Observations  & \multicolumn{2}{c}{476{,}787} \\
  $R^2$         & \multicolumn{2}{c}{0.802} \\
\bottomrule
\end{tabularx}
\\[4pt]
\begin{minipage}{\linewidth}
  \footnotesize\raggedright
  \textit{Notes:} The dependent variable is log marginal cost recovered
  from the multi-product Bertrand first-order conditions using observed
  consumer net prices, subsidies, and the estimated demand system.
  The regression is pooled across all 476{,}787 product--city--year
  observations (BEV, PHEV, REEV, HEV, and ICEV) and the technology-specific
  slopes are identified via masked interactions: Log Engine Power
  is zero for EV rows; Log EV Range and BatteryCost are zero for GV rows.
  Pooling the fixed effects across fuel categories identifies the
  BatteryCost coefficient from the within-year EV-versus-GV contrast,
  which carries non-degenerate variation, rather than from the
  perfectly collinear BNEF time series alone. Standard errors are
  HC1 heteroskedasticity-robust. $^{*}p<0.10$, $^{**}p<0.05$,
  $^{***}p<0.01$.
\end{minipage}
\end{table}

Vehicle size is the largest cost shifter and applies to every product.
With log marginal cost on the left-hand side, $\gamma_s = +1.51$ is a
cost elasticity with respect to exterior vehicle volume: a one-log-point
increase in size (roughly a $170\%$ scale-up) is associated with a
$1.51$-log-point increase in marginal cost. The technology-specific
slopes also have the expected signs. The Log Engine Power coefficient
($+0.95$) on GV rows says that doubling engine power roughly raises
GV marginal cost by $66\%$. The Log EV Range coefficient ($+0.40$)
on EV rows says that doubling driving range raises EV marginal cost by
roughly $28\%$, reflecting the larger battery packs and supporting
electronics needed to provide that range.

The BatteryCost coefficient is the structural primitive that drives
the Wright's-law learning channel of Section~\ref{sec:no_subsidy_dynamic}. We
estimate $\gamma_b = +0.0040$ on the EV-interacted BNEF battery-pack
price (USD/kWh). The interpretation is that, holding fixed product
characteristics, fuel-type, year, body, and firm-group effects, a
\$100/kWh \emph{rise} in the BNEF battery price is associated with a
$0.40$-log-point rise in EV marginal cost relative to gasoline
vehicles in the same year. Equivalently, the realised $258$ USD/kWh
decline in BNEF prices between 2015 and 2024 corresponds to an
EV-versus-GV cost-gap reduction of $0.0040 \times 258 \approx 1.0$
log point, i.e., roughly a $63\%$ contraction in the EV cost premium
attributable to battery-price decline alone. Although $\gamma_b$
appears small in absolute units (USD per kWh is a small denomination
when multiplied through to a log-mc level), the BNEF series itself
spans $258$ USD/kWh between 2015 and 2024, so the implied cumulative
effect is large. The static decomposition's Battery block in
Table~\ref{tab:shapley} reports the equilibrium effect of toggling
the BNEF path from its 2015 value to its 2024 value while holding
cumulative production fixed; the dynamic simulation of
Section~\ref{sec:no_subsidy_dynamic} additionally feeds counterfactual
cumulative production into the BNEF path through Wright's law and
delivers the larger compounded effect.

Identification of $\gamma_b$ requires care. The BNEF battery price is a
single national time series that varies only across years; it would be
perfectly collinear with the EV-specific year fixed effects had we
estimated a regression on EV products alone. Pooling the EV and GV
samples and the fixed effects, while interacting BNEF with the EV
indicator, recovers $\gamma_b$ from the \emph{within-year}
EV-versus-GV cost differential: holding the year fixed, EV products
in a given year inherit the BNEF price shift while GV products do not,
so the EV/GV contrast is what identifies the elasticity. This is the
same coefficient used to construct the
$\mathrm{comp\_learning} = \gamma_b \cdot \mathrm{bnef}_t \cdot
\mathbf{1}[j \in \mathrm{EV}]$ component of the marginal-cost
decomposition in Section~\ref{sec:decomp} and the counterfactual
battery-cost pass-through in Section~\ref{sec:no_subsidy_dynamic}.

The year fixed effects capture the residual time-varying cost
dynamics common to all products after the BatteryCost channel has been
absorbed. Relative to 2015, the year FE decline modestly from
$-0.06$ (2016) to $-0.24$ (2024), corresponding to a roughly $21\%$
decline in pooled marginal costs over the decade attributable to
factors other than battery-price decline (e.g., generic
manufacturing-process improvements, common input-price trends,
exchange-rate effects).

Taken together, the marginal-cost estimates deliver the key economic
pattern needed for the counterfactual analysis. The demand estimates
show that consumers value longer EV range, while the supply estimates
show that range is costly for firms to provide. This joint pattern
creates a meaningful quality-cost tradeoff, rather than treating EV
range as a free demand shifter.

\subsection{Implied Elasticities and Price-Cost Margins}
\begin{figure}[H]
\caption{Own-price elasticities and Lerner indices}
\label{fig:elasticity}
\centering
\begin{subfigure}[t]{0.48\linewidth}
  \includegraphics[width=\linewidth]{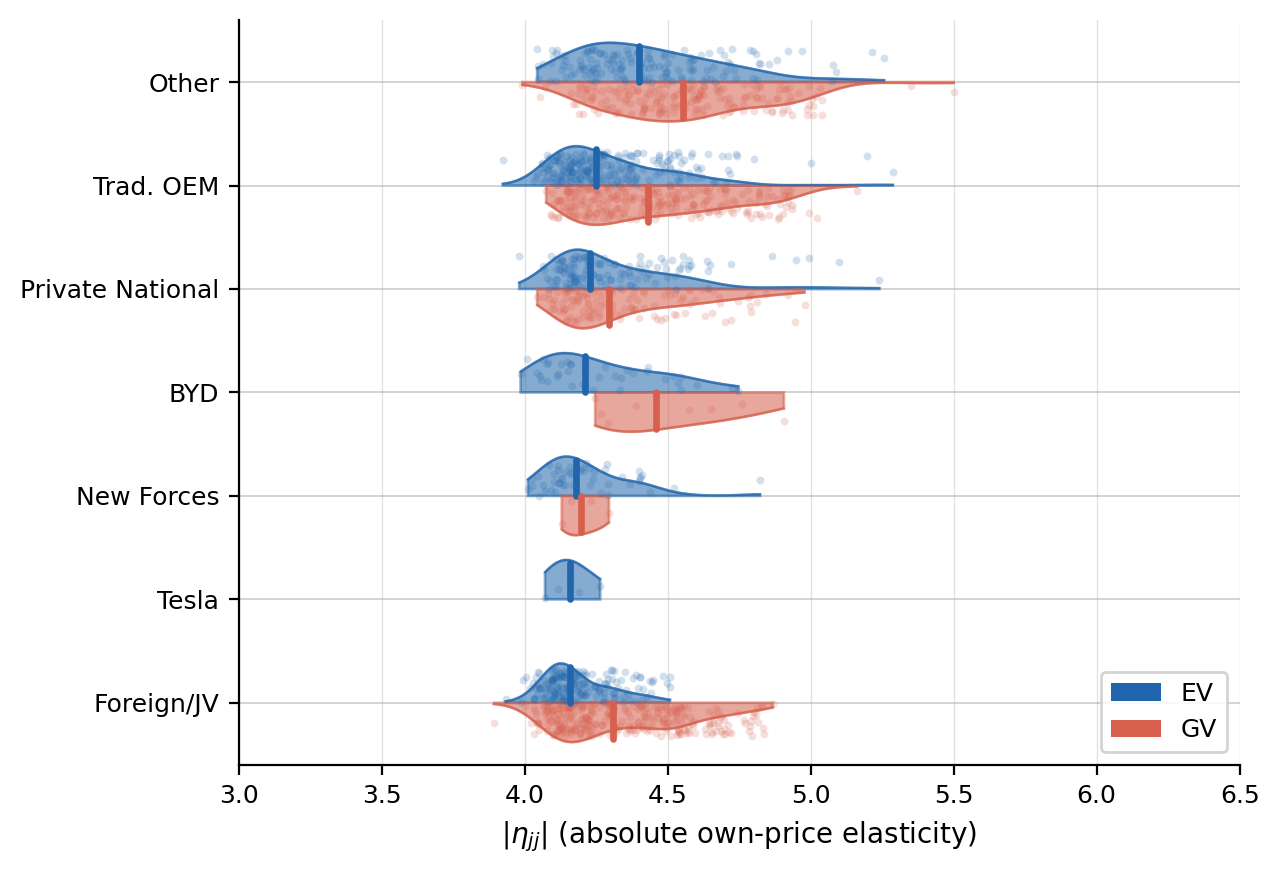}
  \caption{Own-price elasticity $|\eta_{jj}|$}
  \label{fig:own_elasticity}
\end{subfigure}
\hfill
\begin{subfigure}[t]{0.48\linewidth}
  \includegraphics[width=\linewidth]{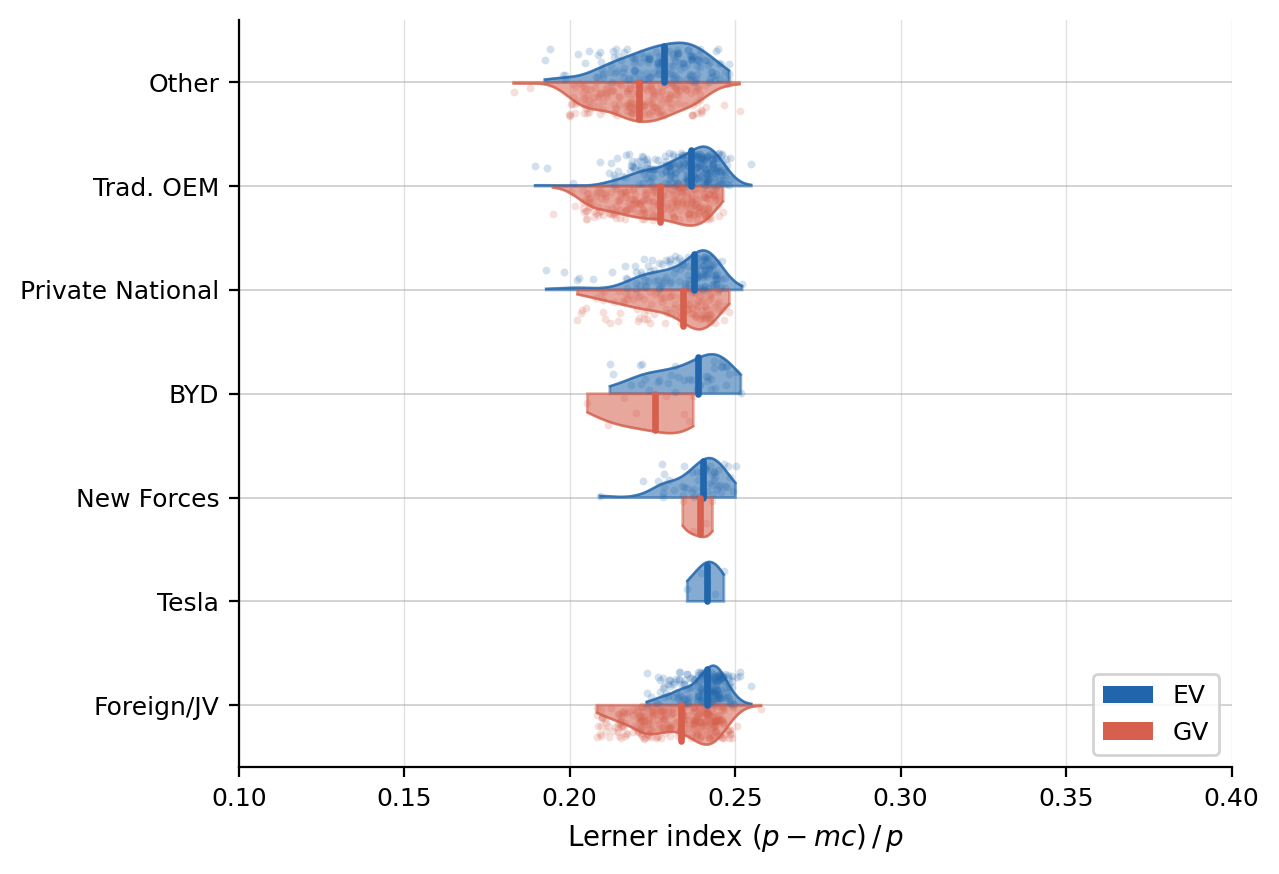}
  \caption{Lerner index $(\tilde p-\mathrm{mc})/\tilde p$}
  \label{fig:lerner_index}
\end{subfigure}
\par\smallskip
\flushleft{\footnotesize\textit{Notes:} Violin half-plots show the
distribution of product-year observations by firm group and fuel type;
dots mark within-group medians. Blue denotes EV products and
red-orange denotes GV products.}
\end{figure}

We combine the recovered marginal costs with the estimated demand system
to evaluate the model's pricing implications. Figure~\ref{fig:elasticity}
reports two implied objects by firm group and fuel type. Panel~\ref{fig:own_elasticity}
shows the distribution of absolute own-price elasticities, defined as the
percentage change in product $j$'s sales induced by a one percent change
in its own consumer net price. Larger absolute values
therefore indicate more price-sensitive demand. These elasticities summarize the substitution patterns implied by
the demand model and enter the supply-side markup inversion through the price-derivative matrix, $\partial s_{km}/\partial \tilde p_{jm}$.
Most elasticities are concentrated between four and five in absolute
value, indicating substantial but plausible price responsiveness. EV and
GV elasticities overlap within most firm groups, suggesting that price
sensitivity is not driven solely by fuel type; instead, groups with
broader product portfolios, such as traditional OEMs, private national
firms, foreign joint ventures, and the residual category,
display wider elasticity distributions, while Tesla and new-force EV brands have more concentrated EV elasticities. Panel~\ref{fig:lerner_index}
reports the corresponding Lerner indices,
$(\tilde p_{jm}-\mathrm{mc}_{jm})/\tilde p_{jm}$, which measure the
share of the producer price kept as a price-cost margin. Higher values
indicate greater pricing power, conditional on marginal cost. The implied
margins are mostly concentrated between 0.20 and 0.30. Differences
between EV and GV margins are modest within many firm groups, but EV
margins are relatively concentrated for the EV-native firms (Tesla,
BYD, and the post-2014 New Forces), whereas the legacy firm groups
(Foreign/JV, traditional state-owned OEMs, the private-national
manufacturers, and the residual ``Other'' category) show wider margin
dispersion. The seven-class firm-group taxonomy used throughout the
paper is defined in
Table~\ref{tab:retention}. Overall, the figure suggests that the
estimated model implies economically reasonable price responsiveness
and price-cost margins while preserving meaningful heterogeneity
across firm groups.

\section{Decomposition of EV Diffusion}
\label{sec:decomp}
Between 2015 and 2024, China's EV market underwent one of the fastest
large-scale technology transitions in recent history. Several forces
moved at the same time: EV products improved rapidly, battery costs
declined, firms expanded EV offerings, purchase subsidies were phased
down, and consumer demand for EVs evolved across markets.

This section decomposes the rise in China's EV market share into
these forces. The exercise is a historical accounting decomposition,
not a policy counterfactual. It asks how much of the 2015--2024
EV-share change can be attributed to product attributes, choice-set
expansion, battery costs, subsidies, residual demand, and demographics.
For each channel, we replace the corresponding primitives with their
endpoint-year values and re-solve for equilibrium prices, quantities,
and market shares.

The estimated demand and supply model is what makes this exercise
possible. Each force operates through both sides of the market:
product attributes and entry shift substitution; battery costs shift
marginal costs and prices; subsidies shift consumer net prices and
firm revenues; demographics shift market size and price sensitivity.
The decomposition therefore captures the full equilibrium effect of
each force, including price responses, markup adjustments, and
reallocation between EVs and GVs.

\subsection{Counterfactual Decomposition Design}
The decomposition proceeds by switching groups of economic primitives from their
2015 values to their endpoint-year values. Let $\mathcal{B}$ denote the set of
decomposition blocks, each representing a potential driver of the EV transition.
For any subset $S\subseteq\mathcal{B}$, let $V_y(S)$ denote the
model-implied national EV share in endpoint year $y$ when the blocks
in $S$ are set to their year-$y$ values and the rest are held at 2015
values. So $V_y(\emptyset)$ is the 2015 baseline share and
$V_y(\mathcal{B})$ is the full-endpoint share. The six blocks are
Quality, Variety, Battery, Subsidy, Residual, and Market; appendix
Table~\ref{tab:blocks} lists the primitives in each. We report the
decomposition at the national aggregate
(Section~\ref{sec:main_results}) and separately for the four city
tiers (Section~\ref{sec:tier_shapley}). The tier-specific
decomposition is the central evidence for the heterogeneous-diffusion
mechanism of the paper title. Each block's contribution varies
systematically across tiers, so ``what drove EV adoption'' has a
different quantitative answer depending on where the city stands on
the diffusion curve.

A simple sequential decomposition introduces blocks one at a time
and assigns each its marginal contribution in a chosen order. But
each block's contribution depends on which other blocks have already
been introduced (the channels are complementary, as discussed in the
introduction). Any single ordering is therefore arbitrary.

To avoid this dependence on an arbitrary ordering, we use the Shapley value from
cooperative game theory \citep{Shapley1953}. Let $\phi_{b,y}$ denote the
contribution of block $b$ to the model-implied change in national EV market
share between 2015 and endpoint year $y$. The Shapley value assigns this
contribution as the average marginal effect of introducing block $b$ across all
possible sequences in which the decomposition blocks could be added:

\begin{equation}
\phi_{b,y}
=
\sum_{S\subseteq \mathcal{B}\setminus\{b\}}
\frac{|S|!\left(|\mathcal{B}|-|S|-1\right)!}
{|\mathcal{B}|!}
\left[
V_y(S\cup\{b\})
-
V_y(S)
\right].
\label{eq:shapley}
\end{equation}

The term $V_y(S\cup\{b\})-V_y(S)$ measures the marginal contribution of block
$b$ when the blocks in $S$ have already been introduced. The weighting term$\frac{|S|!\left(|\mathcal{B}|-|S|-1\right)!}
{|\mathcal{B}|!}$ is the probability that, in a random ordering of all blocks, exactly the blocks
in $S$ appear before block $b$. Thus, the Shapley value averages the marginal
contribution of block $b$ across all possible orderings. With six blocks, this
corresponds to averaging over $6!=720$ possible orderings. The Shapley values satisfy the adding-up property

\begin{equation}
\sum_{b\in\mathcal{B}}
\phi_{b,y}
=
V_y(\mathcal{B})
-
V_y(\emptyset),
\label{eq:shapley_adding_up}
\end{equation}

so the signed sum of the block contributions exactly equals the model-implied
change in national EV market share between 2015 and endpoint year $y$.

For each coalition $S$, we solve for equilibrium prices and
quantities. This equilibrium step matters because the channels
interact through prices and substitution: a battery-cost decline
lowers production costs but also changes equilibrium prices and the
competitive position of EVs against GVs; entry changes the choice
set and competitive pressure across fuel types.

Some coalitions generate very high EV shares, especially when
attribute gains, low battery costs, and favorable residual demand
are introduced together. To prevent the solver from entering
unstable EV-saturation regions, we impose two caps. First, the
product-level counterfactual quantity is capped at twice observed
quantity. Second, the aggregate EV market share is capped at $1.5$
times the observed 2024 EV share. The caps affect only extreme
coalitions.

The saturation regions arise mechanically from the demand elasticity.
Under log utility, the effective own-price elasticity at mean income
is about $4.9$, at the elastic end of the EV-demand literature. At
this elasticity, coalitions that simultaneously improve attributes,
lower marginal cost through battery learning, and turn on favorable
demand residuals drive equilibrium prices toward marginal cost. Small
numerical perturbations then push the fixed-point iteration into a
high-share regime where a few cheap BEVs absorb a disproportionate
share of national demand. The caps truncate these extremes.

The aggregate cap binds in $12$ of the $64$ coalitions at the 2024
endpoint. The binding coalitions sit in the high-V tail; for example,
the full coalition minus Subsidy would push the EV share to roughly
$80\%$ without the cap. The per-product cap binds in a few coalitions
where a single cheap BEV would otherwise capture more than twice its
observed national quantity. The Shapley closure
$\sum_b \phi_b = V(\mathrm{full}) - V(\emptyset)$ remains exact under
the capped $V$, so the caps shift individual $\phi_b$ magnitudes but
not the decomposition identity.

Before interpreting the block contributions, we verify that the
model matches the aggregate EV-share transition. The block
contributions sum to the model-implied endpoint change
$V(\mathrm{full})-V(\emptyset)$ by construction; they explain the
observed change only to the extent that the model-implied change is
close to the raw data. Under the 2015 environment, the
model-implied EV share is $1.03\%$. Under 2024 values, it is
$44.66\%$, an aggregate change of $43.64\%$. The observed aggregate
change is $44.30\%$, so the endpoint gap is $-0.75\%$. Appendix
Table~\ref{tab:endpoints} reports the same diagnostic for each
intermediate endpoint year. The gaps are small, so the decomposition
closely tracks the observed transition. The remaining $-0.75\%$ gap
reflects the BLP solver's incomplete reproduction of observed 2024
shares, not a defect of the decomposition. The waterfall and
cumulative figures use the model-implied change as the reference
object so that the Shapley values close the endpoint exactly.
\subsection{Order Sensitivity and Shapley Averaging}

\begin{figure}[!ht]
\caption{Order sensitivity of sequential decompositions}
\label{fig:ordering}
\centering

\begin{subfigure}[t]{0.32\linewidth}
  \includegraphics[width=\linewidth]{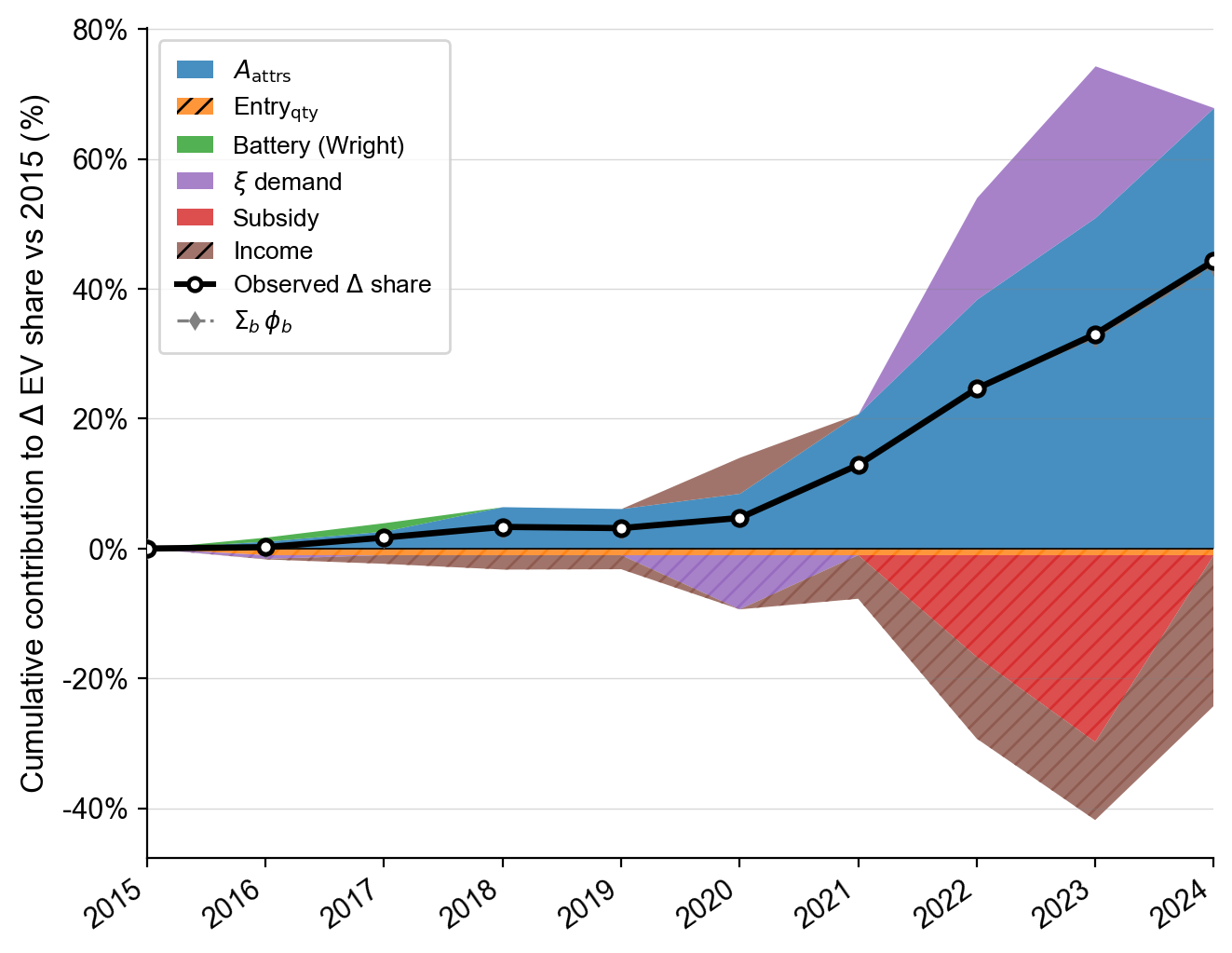}
  \caption{Sequence A: Variety first}
  \label{fig:ordering_A}
\end{subfigure}
\hfill
\begin{subfigure}[t]{0.32\linewidth}
  \includegraphics[width=\linewidth]{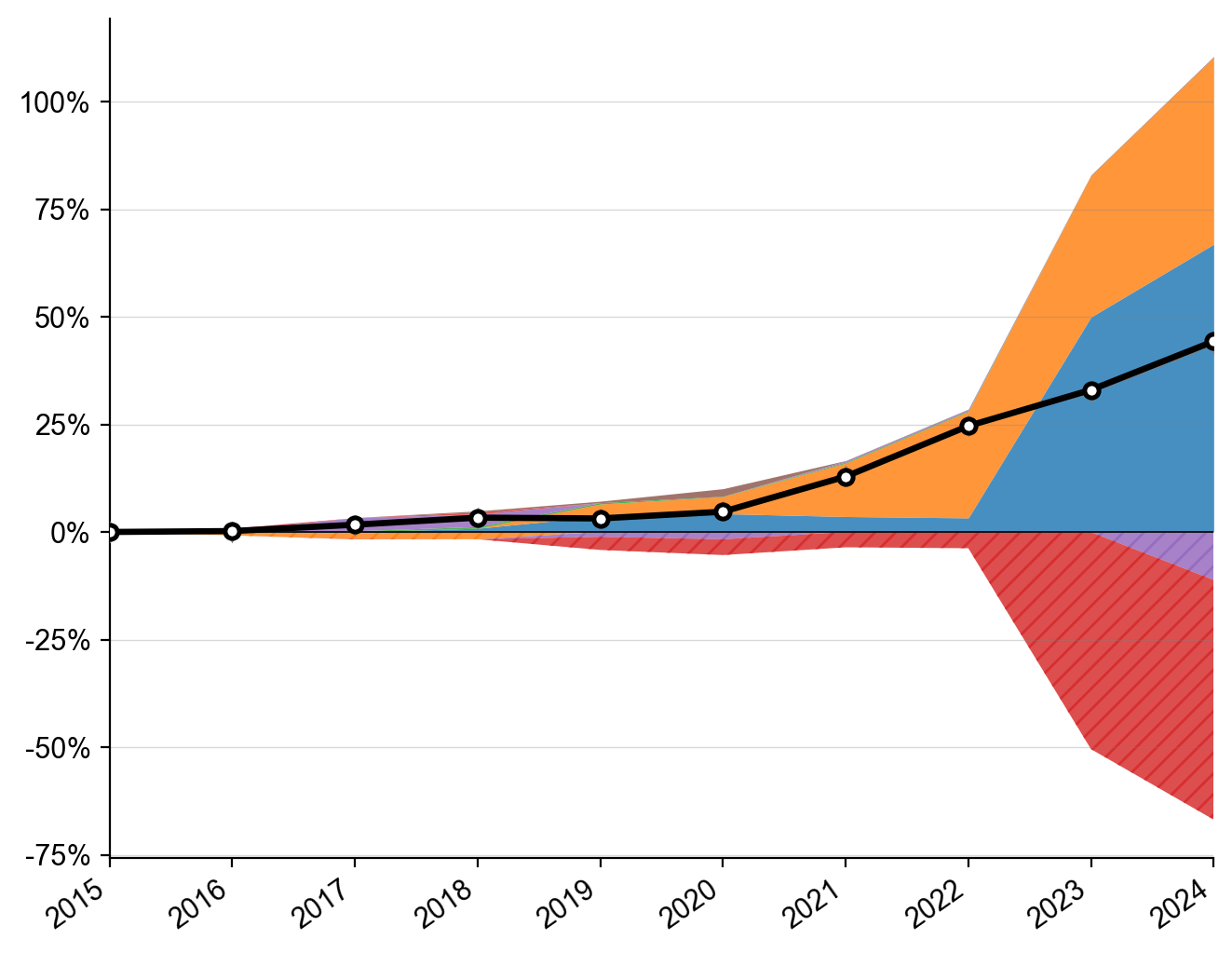}
  \caption{Sequence B: Variety last}
  \label{fig:ordering_B}
\end{subfigure}
\hfill
\begin{subfigure}[t]{0.32\linewidth}
  \includegraphics[width=\linewidth]{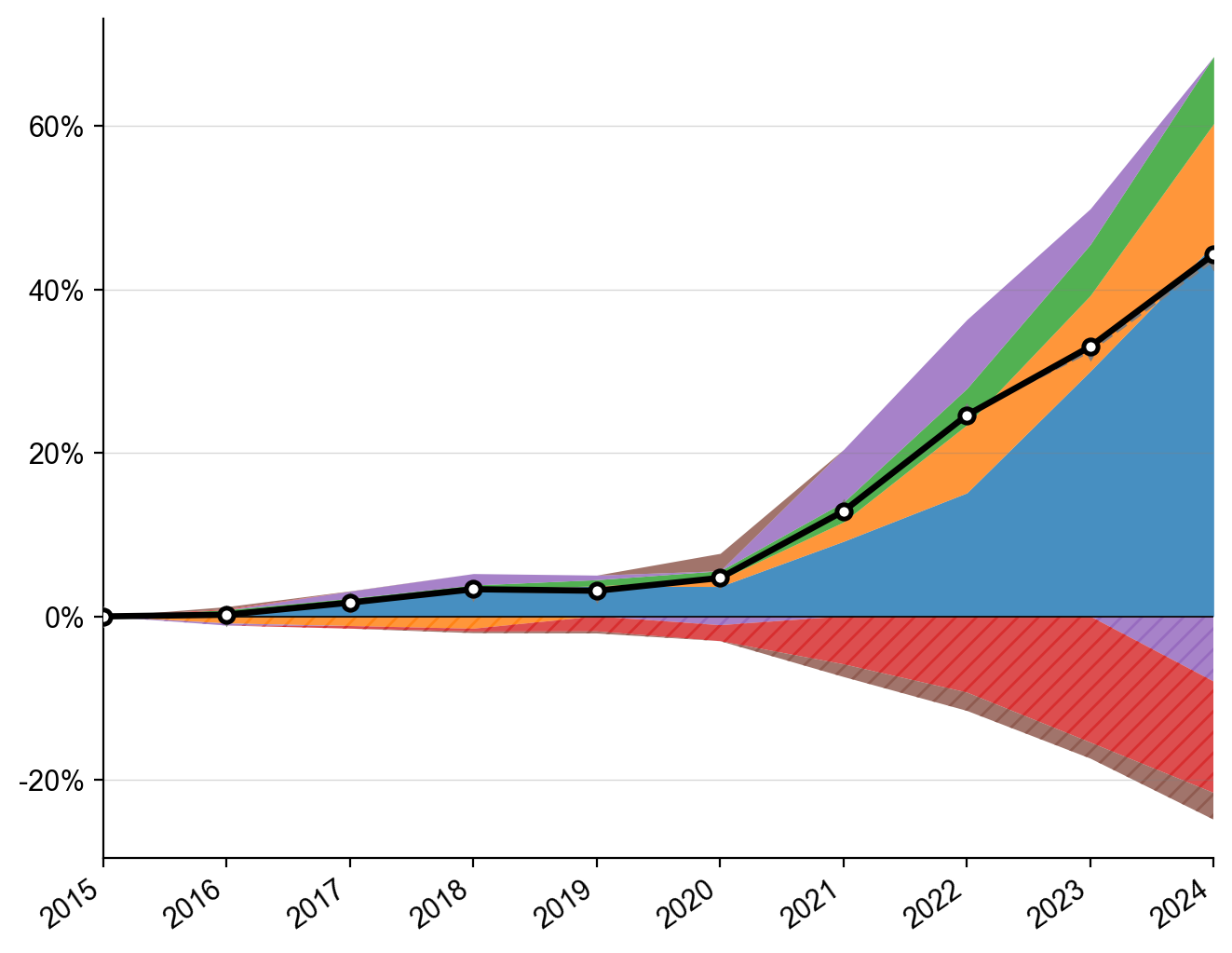}
  \caption{Shapley value}
  \label{fig:ordering_shapley}
\end{subfigure}

\par\smallskip
\flushleft{\footnotesize\textit{Notes:} The figure decomposes the
model-implied change in national EV market share from 2015 to 2024.
Panels~\subref{fig:ordering_A}
and~\subref{fig:ordering_B} report sequential decompositions
under two different orderings of the same six blocks.
Panel~\subref{fig:ordering_shapley} reports the Shapley value,
which averages marginal contributions over all 720 possible orderings.
The black line is the observed aggregate change in EV share, and the
gray dashed line is the model-implied change,
$\sum_{b\in\mathcal{B}}\phi_{b,2024}
= V_{2024}(\mathcal{B}) - V_{2024}(\emptyset)$.
Hatched areas indicate negative contributions. All changes are reported
in percentage units of aggregate EV share.}
\end{figure}

Figure~\ref{fig:ordering} illustrates why averaging over
orderings is important in this setting. The figure focuses on the main
2015--2024 transition.
Panels~\subref{fig:ordering_A}
and~\subref{fig:ordering_B} report two sequential decompositions that use the same six blocks but introduce them in
different orders. Because the blocks interact, the contribution assigned
to a given block depends on the order in which the counterfactual
primitives are switched on.

This order dependence is especially clear for the Variety block, which
captures changes in the set of EV products. In
Panel~\subref{fig:ordering_A}, Variety is introduced first
and receives little credit because new EV products are added to an
otherwise 2015-like environment. In
Panel~\subref{fig:ordering_B}, the same block is introduced
last and receives much more credit because the market has already been
transformed by improved EV quality, lower battery costs, changed
subsidies, residual demand shifts, and market conditions. Thus, a
sequential decomposition can assign very different importance to the same
economic force simply because of the chosen ordering.

Panel~\subref{fig:ordering_shapley} reports the Shapley
decomposition. Rather than choosing one sequence, the Shapley value
averages each block's marginal contribution over all $6! = 720$ possible
orderings. This produces an order-invariant allocation of the
model-implied 2015--2024 change across the six blocks. The figure shows
that the EV transition is highly nonseparable: the effects of Quality,
Variety, Battery, Subsidy, Residual, and Market conditions depend on
which other primitives have already changed.

\subsection{Main Decomposition Results}
\label{sec:main_results}
\begin{figure}[!ht]
\caption{Shapley waterfall: 2015--2024 EV-share change}
\label{fig:waterfall}
\centering
\includegraphics[width=0.95\linewidth]{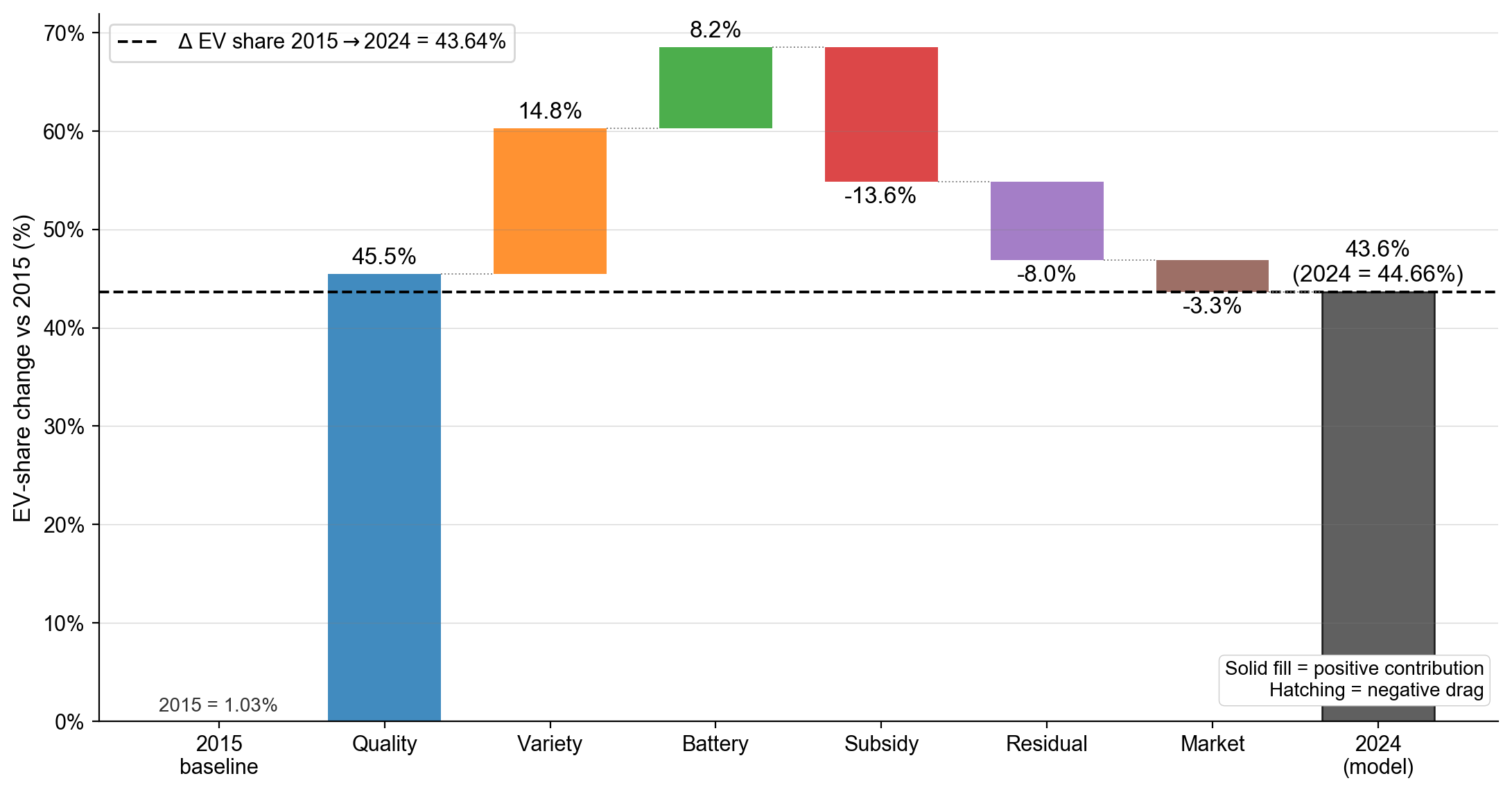}
\par\smallskip
\flushleft{\footnotesize\textit{Notes:} Each bar is the Shapley value
$\phi_b$. The dashed line marks the model-implied endpoint difference.
The signed sum $\sum_b \phi_b=43.64\%$ closes the endpoint exactly.
Solid bars indicate positive contributions; hatched bars indicate
negative contributions. All values are reported in EV-share units.}
\end{figure}

We then turn to the main decomposition for the terminal endpoint year,
2024. Table~\ref{tab:shapley} and Figure~\ref{fig:waterfall} report the
Shapley contributions to the model-implied change in national EV market
share from 2015 to 2024. The EV share rises from $1.03\%$ in the 2015
baseline to $44.66\%$ in the model-implied 2024 endpoint, so the endpoint
difference is $43.64\%$.

\begin{table}[!ht]
\centering
\caption{Shapley decomposition of the 2015--2024 EV-share change}
\label{tab:shapley}
\small
\begin{tabular}{lrl}
\toprule
Block & Contribution & Interpretation \\
\midrule
Quality & $45.49\%$ & Product-quality and attribute-related channel \\
Variety & $14.81\%$ & Larger and more EV-oriented product set \\
Battery & $8.20\%$ & Battery-related cost reductions \\
Subsidy & $-13.63\%$ & Direct subsidy phase-out relative to 2015 \\
Residual & $-7.96\%$ & Residual demand shifts after observed controls \\
Market & $-3.27\%$ & Income, market size, and demographic changes \\
\midrule
Signed sum & $43.64\%$ & Model-implied EV-share change \\
\bottomrule
\end{tabular}
\begin{flushleft}
\footnotesize
\textit{Notes:} Contributions are Shapley values for the 2015--2024
change in national EV market share. The signed sum equals
$V_{2024}(\mathcal{B})-V_{2024}(\emptyset)$. All values are reported in
EV-share units. The reported signed sum uses unrounded Shapley values.
\end{flushleft}
\end{table}

The dominant contribution comes from the Quality block. This block
toggles the per-product observable attributes that vary between 2015
and 2024 for products present in both endpoint sets: log driving
range, log engine power on GV products, log vehicle size, and the
range $\times$ density interaction. It excludes the FuelType,
BodyType, and SizeSegment fixed effects (category-level intercepts
that do not vary by year and would be double-counted with Variety).
It also excludes prices, policy variables, demand residuals, and
demographic interactions, which sit in other blocks. The $45.49\%$
contribution therefore measures the equilibrium effect of the
within-product attribute trajectory, not any single attribute and not
the changing product set.

Variety is the second-largest force at $14.81\%$. Consumers in 2024
face a much richer set of EV options than in 2015. The interpretation
is close to the new-product channel in differentiated-products demand
\citep{Petrin2002}, where new products change the substitution
environment and expand consumer choice. Variety is treated as an
accounting block rather than modeled as a dynamic firm decision.

Battery-related cost changes contribute $8.20\%$. This block measures
the equilibrium effect of switching the battery-cost path from its
2015 value to its 2024 value, holding cumulative EV production at the
observed trajectory. The larger dynamic effect of Wright's-law
learning — where subsidy-induced production feeds back into the
battery-cost path — is reported separately in
Section~\ref{sec:no_subsidy_dynamic}. The static contribution is
smaller than Quality and Variety, but economically important as a
complement: lower battery costs matter most when firms can embody
them in attractive EV products.

The Subsidy block contributes $-13.63\%$. This does not mean that
subsidies reduce EV demand. The historical comparison switches the
subsidy environment from 2015 to 2024, a period during which direct
purchase subsidies were phased down. The Subsidy block therefore
measures policy phase-out, not the marginal value of subsidy support
in the 2024 market. That marginal value is reported separately in
Section~\ref{sec:no_subsidy_dynamic}, and the static-vs-dynamic
reconciliation is in
Section~\ref{sec:policy_relation_decomposition}.

Residual and Market are also negative in the average-order
decomposition. These negative contributions should be interpreted
conditionally. After accounting for Quality, Variety, Battery, and
Subsidy, the remaining residual demand shifts and market-level changes do
not by themselves explain the rise in EV share. In particular, the
Market block combines several forces, including changes in income
distributions, population, market size, and price sensitivity; its
negative sign is a net effect in EV-share terms, not evidence that higher
income reduces EV demand mechanically.

Figure~\ref{fig:waterfall} visualizes the same decomposition. The large
Quality contribution is partially offset by the negative Subsidy,
Residual, and Market blocks. The final gray bar shows that the positive
forces dominate, yielding a model-implied endpoint difference of
$43.64\%$. 

\subsection{Heterogeneity Across City Tiers}
\label{sec:tier_shapley}

The aggregate decomposition masks heterogeneity across city markets.
China's EV transition did not diffuse uniformly: large, high-income
cities adopted EVs earlier; lower-tier cities became more important
later. We therefore apply the same Shapley decomposition to city
tiers.

\begin{figure}[H]
\centering
\caption{Tier-specific Shapley decomposition of the 2015--2024 EV-share change}
\label{fig:tier_shapley}
\includegraphics[width=0.92\linewidth]{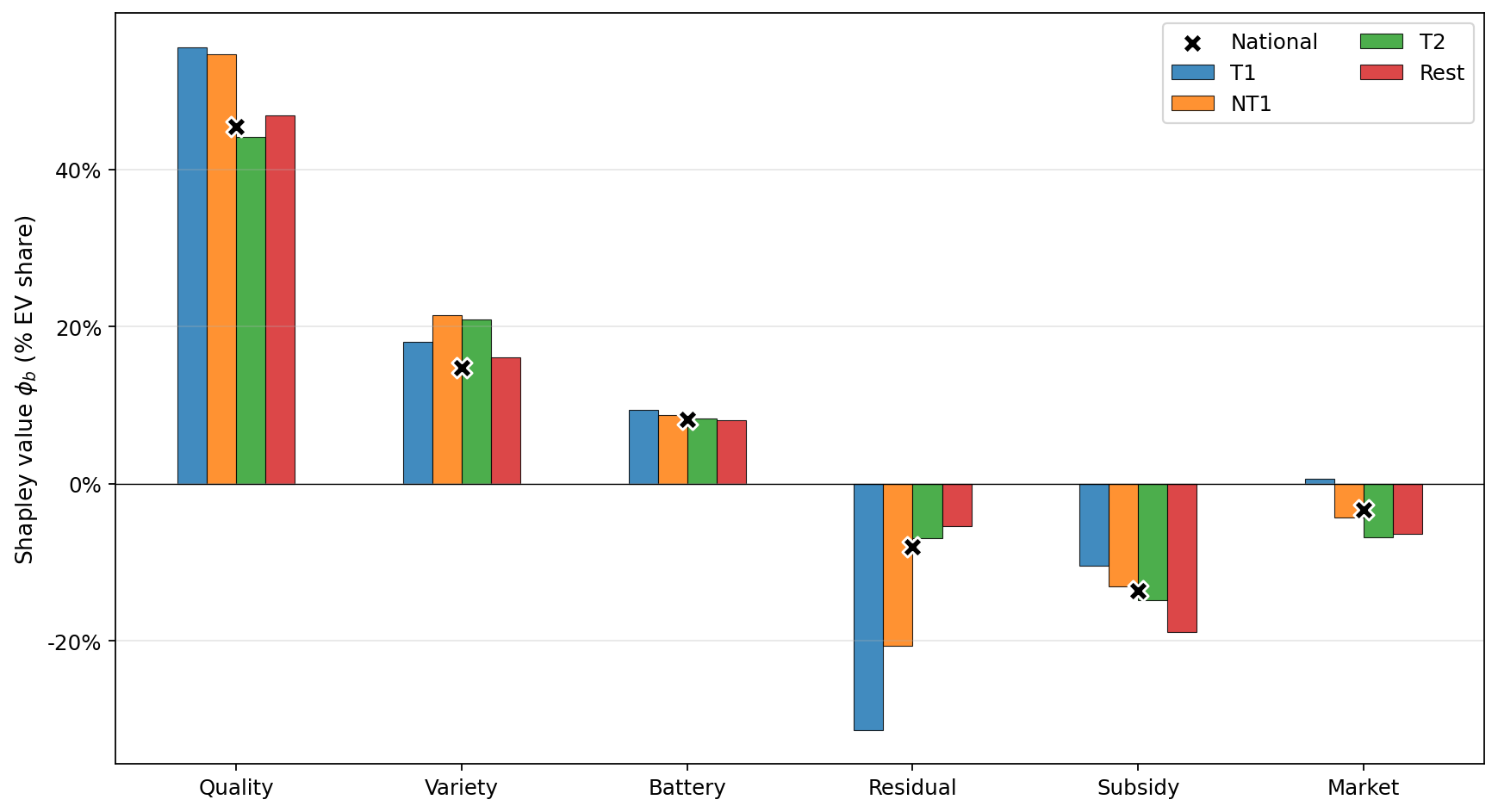}
\par\smallskip
\flushleft{\footnotesize\textit{Notes:} The figure reports Shapley
values by block for each city tier and for the national aggregate. For
each coalition, equilibrium prices and quantities are solved using the
national product-market equilibrium. Tier-specific values are then
computed by aggregating the resulting city-level EV shares within each
tier. The black X marks the national Shapley value from
Table~\ref{tab:shapley}. All values are reported in EV-share units.}
\end{figure}

Figure~\ref{fig:tier_shapley} reports the result; appendix
Table~\ref{tab:tier_shapley} reports the numerical values. The
exercise uses the same six blocks and the same coalition equilibria
as the national decomposition. For each coalition $S$, equilibrium
prices and quantities are solved at the national product-market
level. We then construct a tier-specific value function
$V_{t,2024}(S)$ by aggregating the resulting city-level EV shares
within tier $t$. The adding-up condition holds exactly within each
tier:
\[
\sum_{b\in\mathcal{B}}\phi_{b,t,2024}
=
V_{t,2024}(\mathcal{B})-V_{t,2024}(\emptyset).
\]

Three patterns stand out. First, Quality is the largest positive block
in every tier, but its contribution is larger in higher-tier city
markets. The Quality contribution is especially large in Tier 1 and New
Tier 1, consistent with EV product-quality improvements being especially
valuable in larger and richer urban markets, where consumers were better
positioned to substitute toward improved EV products.

Second, the Subsidy block is more negative in lower-tier cities. As in
the national decomposition, this block measures the effect of moving from
the 2015 subsidy environment to the 2024 subsidy environment. A more
negative Subsidy contribution therefore means that the historical
phase-down of direct purchase subsidies reduced model-implied EV share
more strongly in that tier. In EV-share terms, the phase-down was more
costly for lower-tier markets.

Third, the Residual block is much more negative in Tier 1 than in
the Rest tier. After accounting for Quality, Variety, Battery,
Subsidy, and Market, the remaining residual component does not add
to Tier 1 EV growth. This does not mean Tier 1 EV demand weakened
mechanically. The observable channels already account for most of
the Tier 1 transition, leaving a negative average-order residual.

The tier results reinforce the heterogeneous-diffusion interpretation of
the paper. The same national technology and policy transition generates
different EV-share contributions across city tiers because cities differ
in market size, income, product exposure, and stage of diffusion.

\subsection{Dynamics of the Transition}

\begin{figure}[!ht]
\caption{Cumulative Shapley decomposition from the 2015 baseline}
\label{fig:cumulative}
\centering

\begin{subfigure}[t]{0.48\linewidth}
  \includegraphics[width=\linewidth]{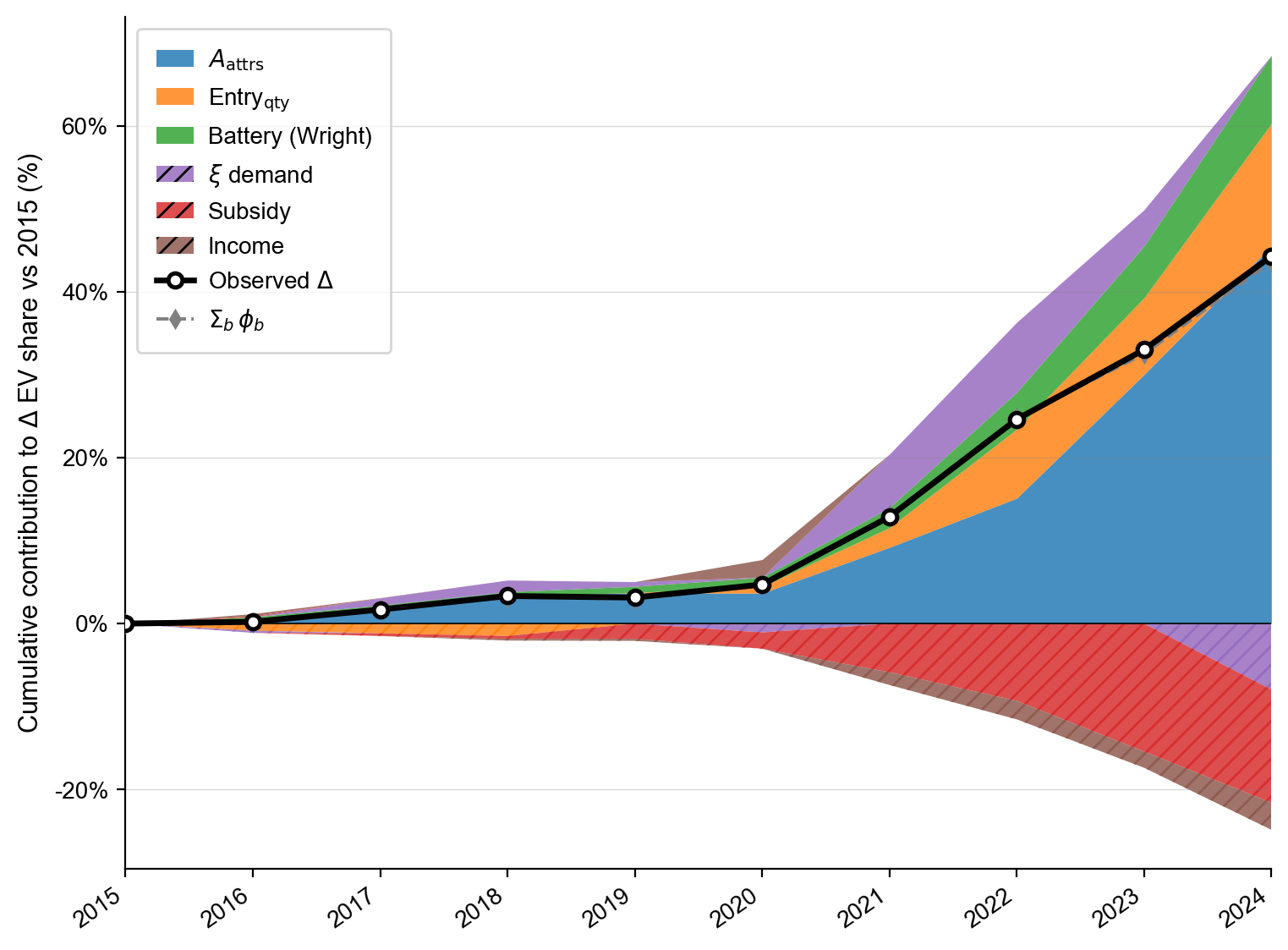}
  \caption{EV market share}
  \label{fig:cumulative_share}
\end{subfigure}
\hfill
\begin{subfigure}[t]{0.48\linewidth}
  \includegraphics[width=\linewidth]{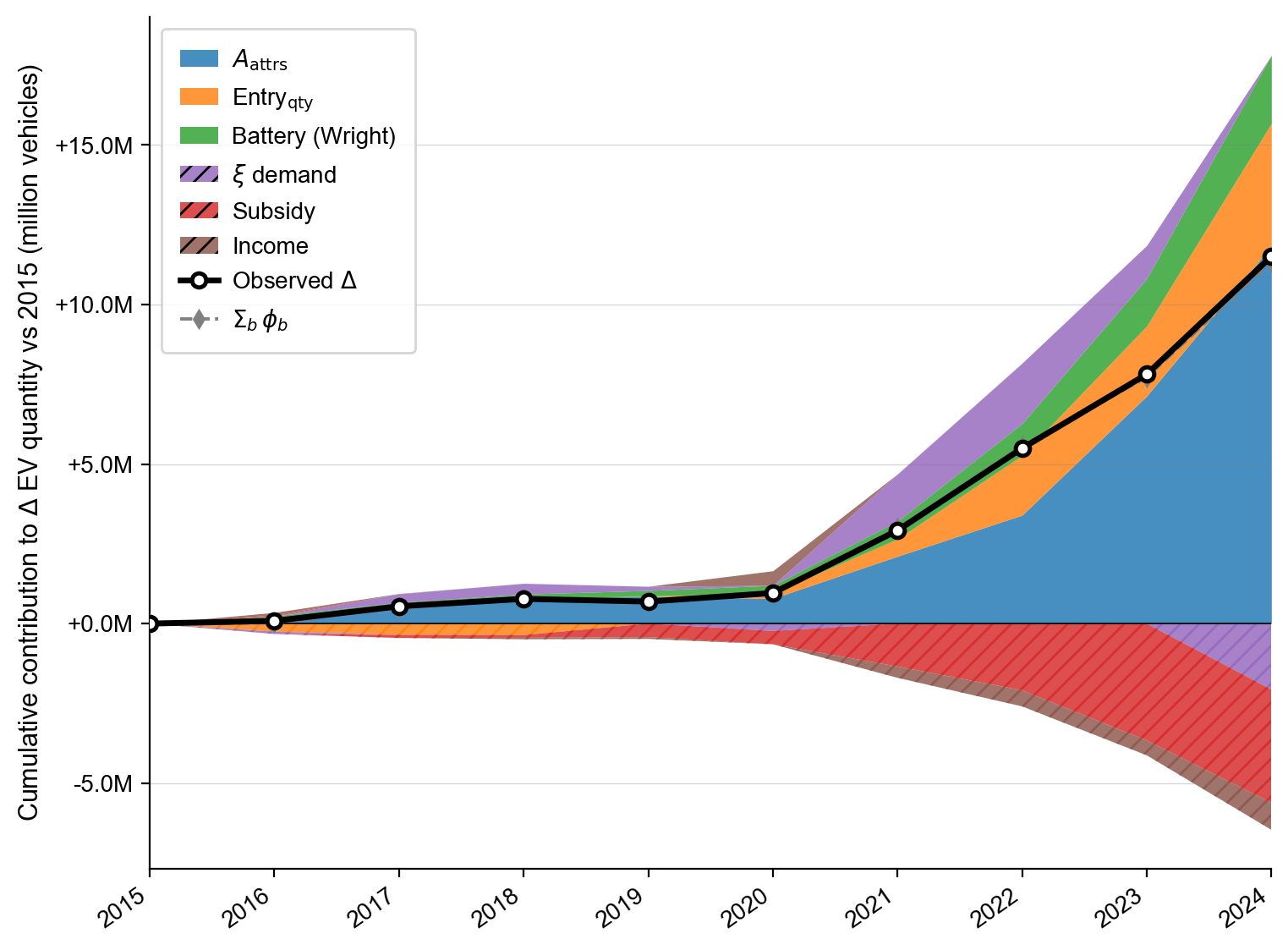}
  \caption{EV quantity}
  \label{fig:cumulative_qty}
\end{subfigure}

\par\smallskip
\flushleft{\footnotesize\textit{Notes:} Stacked areas show positive
blocks with solid fill and negative blocks with hatched fill. The black
line is the observed change relative to 2015; the gray dashed line is
the model-implied Shapley sum $\sum_b\phi_b$. The market-share panel is
reported in EV-share units, while the quantity panel is reported in
million vehicles.}
\end{figure}

Figure~\ref{fig:cumulative} extends the decomposition year by year from
the 2015 baseline. Panel~\subref{fig:cumulative_share} reports the
cumulative contribution to EV market share, while
Panel~\subref{fig:cumulative_qty} reports the corresponding contribution
to EV quantities. The black line is the observed change, and the gray
dashed line is the model-implied Shapley sum.

The figure shows that EV diffusion accelerates sharply after 2020. In
the early years, all blocks have relatively small contributions because
EVs remain a thin segment of the market. After 2020, the contribution of
Quality rises rapidly, reflecting the improvement in EV attributes,
range, and overall product appeal. Variety also becomes increasingly
important as the EV choice set expands. Battery-related cost changes add
a meaningful contribution, especially in later years when cost reductions
can be embodied in a broader set of EV products.

The negative Subsidy block grows over time because the historical subsidy
schedule becomes less generous relative to the 2015 baseline. The
negative Residual and Market components partly offset the positive
product and cost channels, but they are not large enough to reverse the
overall transition. Panel~\subref{fig:cumulative_qty} shows that the
same mechanisms drive not only EV-share growth but also the large
increase in the number of EVs sold.

\subsection{Order Sensitivity and Interpretation}
The Shapley values provide an order-invariant average contribution, but
the range of sequential contributions is itself informative. It shows
that China's EV transition was driven by complementary forces rather than
by a single isolated channel. Quality matters more when battery costs are
low and EV variety has expanded. Variety matters more when the available
EV products are high quality. Battery-cost reductions matter more when
firms have attractive EV products through which to pass cost reductions
into prices and quality.

Table~\ref{tab:sensitivity} reports this order-sensitivity diagnostic
for the 2024 decomposition. For each block, the table shows the Shapley
value, the minimum and maximum marginal contribution across coalitions,
the range, and the equal-weighted mean of marginal contributions.

\begin{table}[!ht]
\centering
\caption{Order-sensitivity diagnostic for the 2024 Shapley values}
\label{tab:sensitivity}
\small
\begin{tabular}{lrrrrr}
\toprule
Block & Shapley & Min & Max & Range & Equal-weighted mean \\
\midrule
Quality & $45.49\%$ & $0.00\%$ & $67.89\%$ & $67.89\%$ & $40.85\%$ \\
Variety & $14.81\%$ & $-20.08\%$ & $66.85\%$ & $86.94\%$ & $12.14\%$ \\
Battery & $8.20\%$ & $0.00\%$ & $55.41\%$ & $55.41\%$ & $3.06\%$ \\
Subsidy & $-13.63\%$ & $-56.12\%$ & $0.00\%$ & $56.12\%$ & $-13.95\%$ \\
Residual & $-7.96\%$ & $-57.89\%$ & $19.06\%$ & $76.95\%$ & $-11.08\%$ \\
Market & $-3.27\%$ & $-23.32\%$ & $31.39\%$ & $54.70\%$ & $0.55\%$ \\
\bottomrule
\end{tabular}
\begin{flushleft}
\footnotesize
\textit{Notes:} The table reports the distribution of marginal
contributions across coalitions in the 2024 Shapley decomposition. All
values are reported in EV-share units.
\end{flushleft}
\end{table}

The large ranges reinforce the main conclusion. The EV transition is not
well described as the sum of independent effects. Instead, EV diffusion
reflects the interaction of product-quality improvements, variety
expansion, battery-cost reductions, policy changes, and market-level
demand conditions. The robust qualitative message is that Quality is
central, Variety and Battery are important complements, and the negative
Subsidy contribution reflects historical policy phase-out rather than
evidence that subsidies were ineffective.

\section{Policy Counterfactuals and Incidence}
\label{sec:policy_counterfactuals}

The decomposition above explains the historical EV transition from 2015
to 2024. The policy counterfactuals in this section ask a different
question: how would EV adoption and subsidy incidence change under
alternative subsidy regimes? This distinction is important. The Subsidy
block in the Shapley decomposition compares the observed 2024 subsidy
environment with the 2015 subsidy environment. By contrast, the
counterfactuals below compare market outcomes with and without subsidy
support under specified economic environments.

\subsection{No-Subsidy Dynamic Counterfactual}
\label{sec:no_subsidy_dynamic}

\begin{figure}[H]
\caption{No-subsidy counterfactual: dynamic simulation}
\label{fig:dynamic}
\centering

\begin{subfigure}[t]{0.58\linewidth}
  \includegraphics[width=\linewidth]{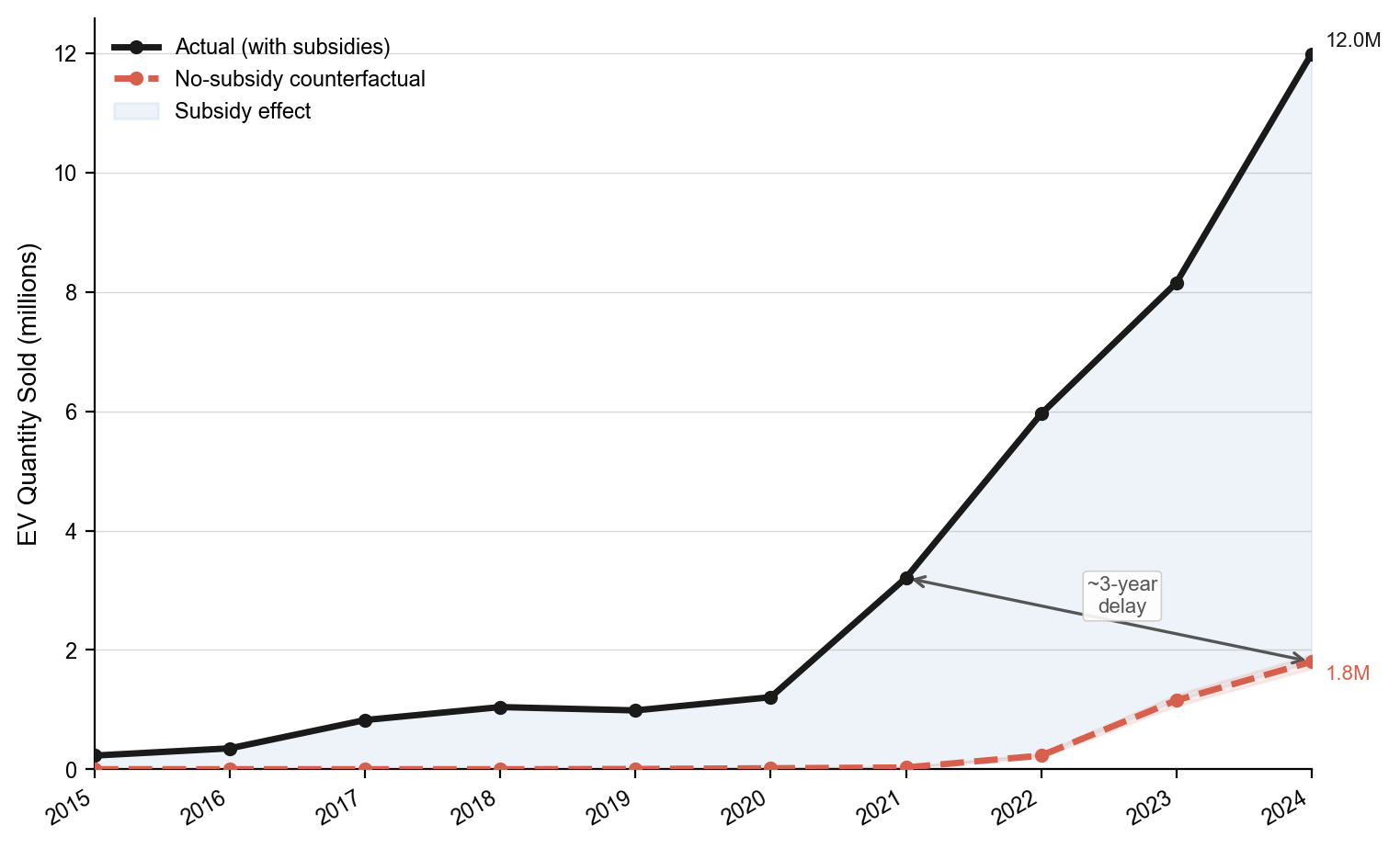}
  \caption{EV quantity path, 2015--2024}
  \label{fig:dynamic_path}
\end{subfigure}
\hfill
\begin{subfigure}[t]{0.38\linewidth}
  \includegraphics[width=\linewidth]{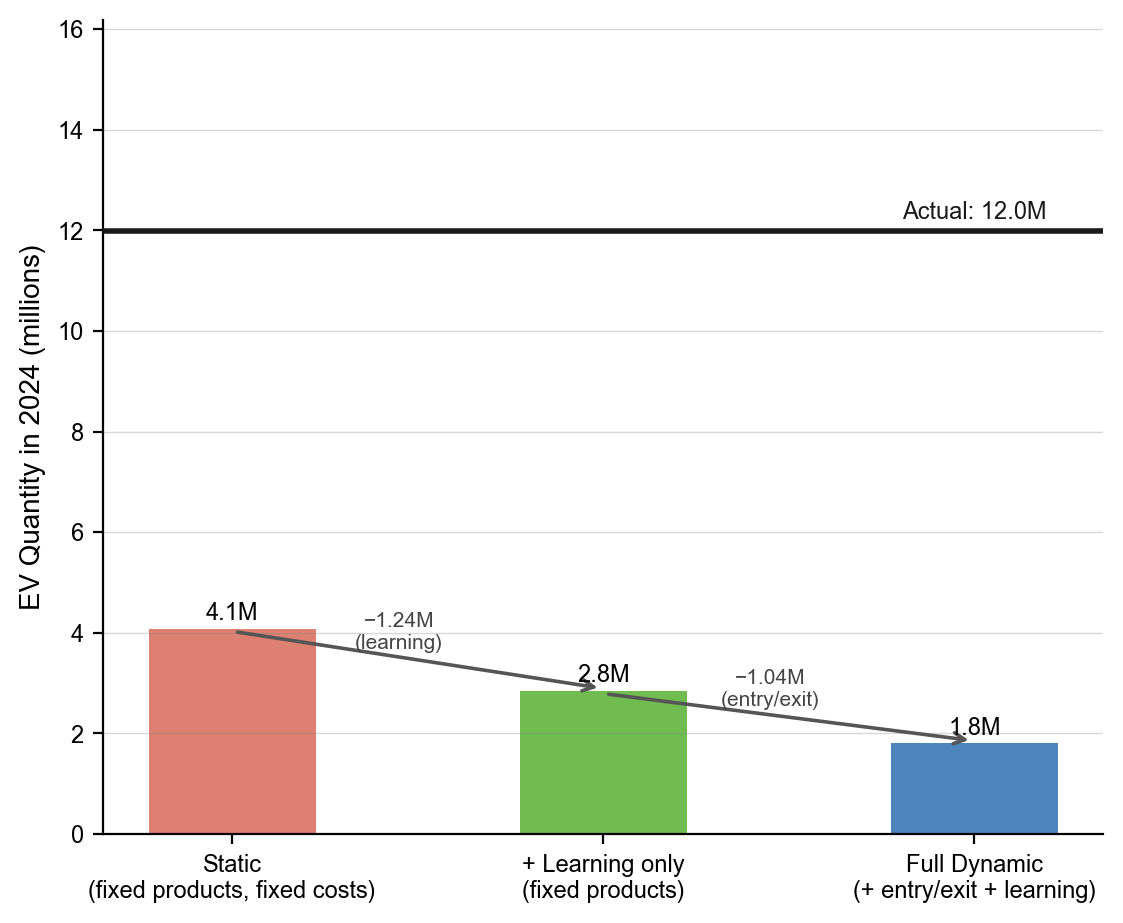}
  \caption{2024 quantity by dynamic regime}
  \label{fig:dynamic_channels}
\end{subfigure}

\par\smallskip
\flushleft{\footnotesize\textit{Notes:}
Panel~\subref{fig:dynamic_path} compares the actual EV quantity
path with the no-subsidy counterfactual under the full dynamic regime
with endogenous entry and Wright's Law battery-cost learning. Shading
shows Monte Carlo uncertainty across 50 draws. The arrow marks the
approximately three-year adoption delay attributable to subsidies.
Panel~\subref{fig:dynamic_channels} reports 2024 no-subsidy EV
quantity under three regimes: Static, Learning only, and Full dynamic.
The arrows quantify the learning and entry/exit channels. Quantities are
reported in million vehicles.}
\end{figure}

Figure~\ref{fig:dynamic} studies the role of subsidies in EV
market development using a dynamic no-subsidy counterfactual. The
simulation compares the observed subsidy path with a counterfactual path
in which the national purchase subsidy is set to zero over the sample
period. The exercise incorporates two dynamic mechanisms: battery-cost
learning through Wright's Law at a calibrated rate of $18\%$ per
doubling of cumulative production\footnote{We fit the Wright curve
$\log c_t = \log A - b \log Q_t$ to the 2013--2024 BNEF battery-cost
panel paired with global (CN + ROW) cumulative EV stocks from the IEA
Outlook, obtaining $\hat b = 0.373$ ($s.e. = 0.044$) and an implied
learning rate of $22.8\%$ per doubling ($s.e. = 2.3\%$). The
literature median over the broader lithium-ion learning-rate range is
$18\%$ (\citealp{NewellJaffeStavins1999}; \citealp{Popp2002}). We
report the simulation at the literature median because it is the more
conservative of the two; a sensitivity exercise at the data-fit rate
of $22.8\%$ shifts the 2024 no-subsidy EV share by less than $0.5\%$
relative to the headline numbers.} and endogenous product availability through
entry and exit.

The dynamic simulation does not model
forward-looking expectations on either side of the market. Within each
year, consumers choose myopically among the available products
conditional on observed prices and attributes, and firms set prices
through a per-city Bertrand--Nash equilibrium without internalizing
the cumulative-production-to-battery-cost feedback when choosing
quantities. The Wright's-law channel therefore operates entirely
through state evolution between years rather than through optimizing
behavior within years. A fully forward-looking model with rational
expectations of the battery-cost path is a substantively different
exercise that we leave to future work; the current results should be
read as a sequence of one-shot static equilibria linked by state
transitions, not as a dynamic programming solution.

Panel~\subref{fig:dynamic_path} reports the EV quantity path
under the full dynamic regime. The actual path reaches $12.0$ million EVs
in 2024 (the observed national NEV registration count was approximately
$11.15$ million, so the simulation overshoots observed quantity by
about $7.6\%$; the gap is small relative to the magnitude of the
no-subsidy effect), while the no-subsidy path reaches only $1.8$
million EVs. The
counterfactual path remains far below the actual path throughout the
diffusion period, and the figure indicates an approximately three-year
delay in EV adoption. Thus, the subsidy does not merely shift purchases
within a given static market; it accelerates the transition by supporting
early demand, cumulative learning, and subsequent product availability.

Panel~\subref{fig:dynamic_channels} decomposes the 2024
no-subsidy quantity across three modeling regimes. In the Static regime,
with product availability and costs held fixed, the no-subsidy market
would still sell $4.1$ million EVs. Allowing battery costs to respond
through cumulative learning reduces the no-subsidy quantity to
$2.8$ million. Allowing both learning and entry/exit reduces it further
to $1.8$ million. The reduction from $4.1$ million to $2.8$ million
captures the learning channel, while the additional reduction from
$2.8$ million to $1.8$ million captures the entry and exit channel.
These dynamic responses amplify the direct demand effect of subsidy
removal: weaker early demand slows battery-cost declines and reduces the
future variety of EV products. Appendix Table~\ref{tab:regimes} reports the
corresponding 2024 quantities by dynamic regime and makes explicit the
learning and entry/exit amplification channels.

\paragraph{Robustness to the demand elasticity.} The
no-subsidy effects depend on the demand model's price elasticity. Our
canonical estimate gives an effective $|\alpha| \approx 4.9$ at mean
income, which is at the elastic end of the EV-BLP literature. To
bracket the result we re-run the dynamic simulation with $|\alpha|$
rescaled to $2.5$, the approximate median of estimates in the
EV-demand literature (\citealp{Springel2021}; \citealp{BollingerGillingham2019};
\citealp{Li2017}). The same subsidy-removal experiment then implies
a 2024 EV quantity of $5.4$ million (versus $4.1$ million under the
canonical $|\alpha|$) in the Static regime,
$4.7$ million (versus $2.8$ million) under Learning only, and
$3.9 \pm 0.2$ million (versus $1.8 \pm 0.07$ million) under the Full
dynamic regime. The corresponding subsidy-removal effects on EV share
are $-13.6\,$pp, $-18.9\,$pp, and $-23.4 \pm 1.4\,$pp respectively, or
roughly $30\%$ smaller in magnitude than the canonical numbers. The
qualitative ranking across regimes (entry/exit amplifies learning,
which amplifies direct demand effects) is preserved. The robustness
check confirms that the central message about subsidy support and
dynamic amplification is not driven by the upper-bound demand
elasticity; a less-elastic demand model produces quantitatively
smaller but qualitatively identical results.

\subsection{Subsidy Incidence Across City Income Tiers}
\label{sec:subsidy_incidence}

\begin{figure}[!ht]
\centering
\caption{Cross-tier EV diffusion lag, 2015--2024}
\label{fig:lag}
\includegraphics[width=0.85\linewidth]{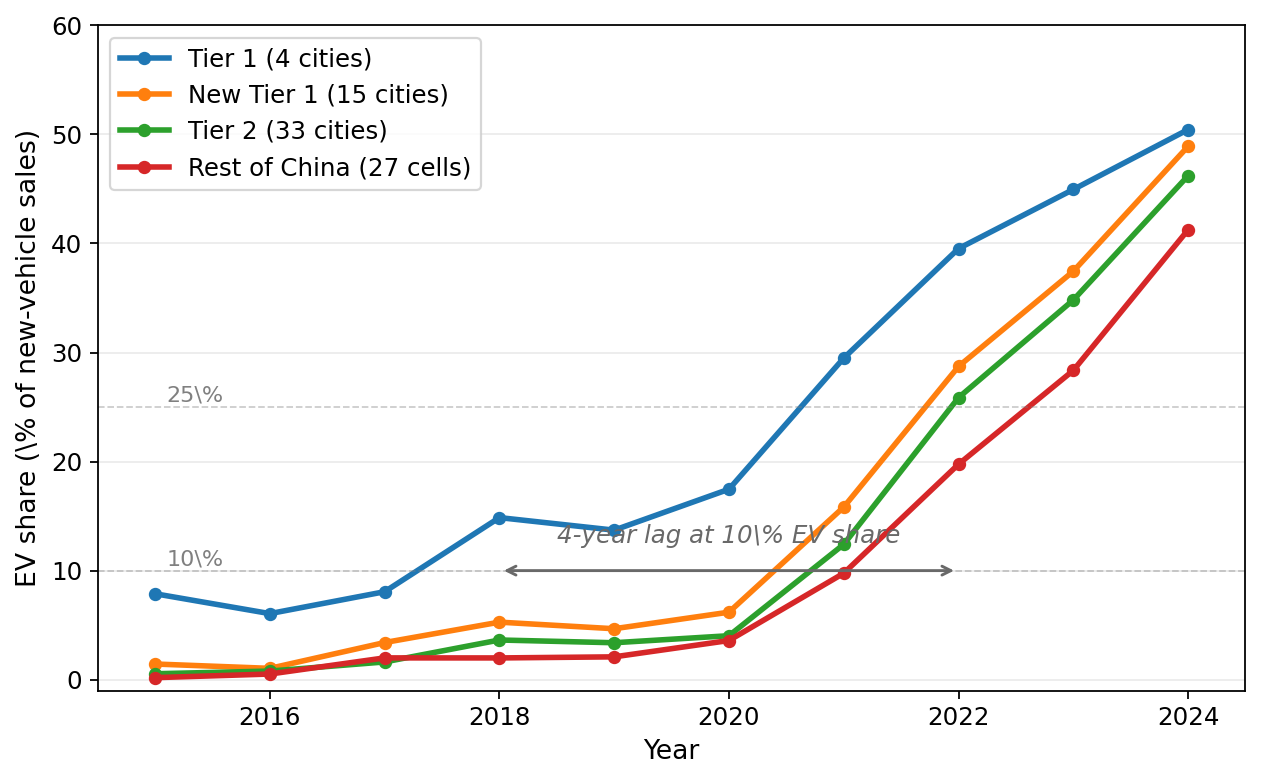}
\par\smallskip
\flushleft{\footnotesize\textit{Notes:} EV share is BEV $+$ PHEV $+$
REEV as a fraction of new-vehicle sales in each tier. Tier 1 covers
Beijing, Shanghai, Guangzhou, and Shenzhen; New Tier 1 covers 15 new
first-tier cities; Tier 2 covers 33 remaining second-tier cities in
the panel; Rest covers the 27 small-city residual cells. The
double-headed arrow marks the 4-year lag between Tier 1 and Rest
crossing the 10\% EV-share threshold.}
\end{figure}

Before discussing the incidence of subsidy benefits,
Figure~\ref{fig:lag} establishes a stylized fact about heterogeneous
EV diffusion across city tiers. The national aggregate EV-share series
hides material differences in adoption timing: Tier 1 cities (Beijing,
Shanghai, Guangzhou, Shenzhen) crossed a 10\% EV share in 2018,
whereas the Rest tier (small-city cells) only crossed the same
threshold in 2022, a 4-year lag. The lag compresses as the technology
matures, narrowing to roughly 2 years at the 25\% threshold and to
about 1 year at the 40\% threshold (Table~\ref{tab:lag}). This timing
pattern is the empirical content of the ``heterogeneous diffusion''
framing of the paper: the same technology arrives in different cities
at different points in the diffusion curve, so the same policy
instrument applied to the same national market produces very different
contemporaneous effects depending on each city's stage of diffusion.
The decomposition of historical subsidy benefits below is most
naturally read in this light.

\input{figures/tab_diffusion_lag}

\begin{figure}[!ht]
\caption{EV subsidy incidence by city income tier}
\label{fig:incidence}
\centering

\begin{subfigure}[t]{0.54\linewidth}
  \includegraphics[width=\linewidth]{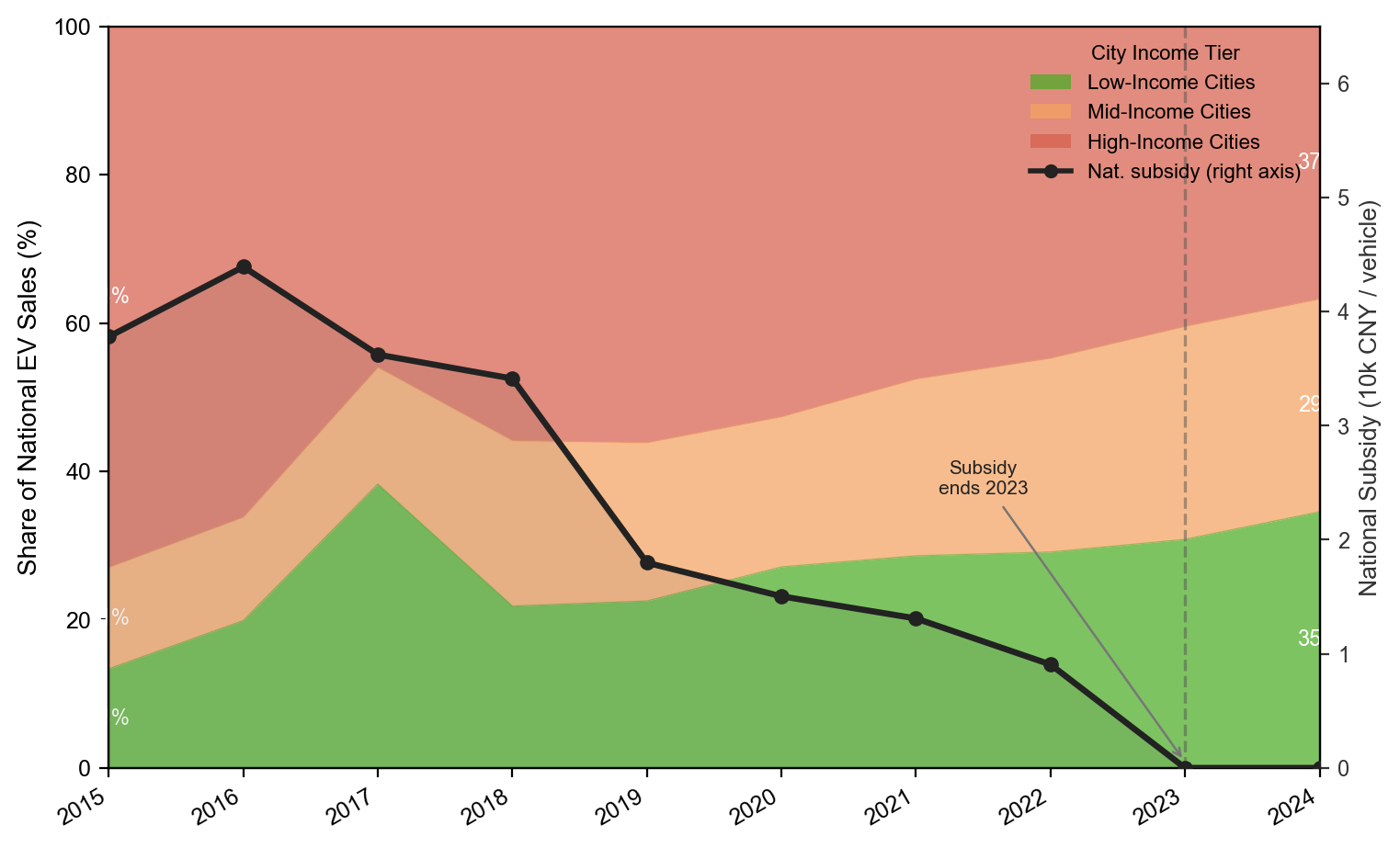}
  \caption{EV adoption share by city income tier, 2015--2024}
  \label{fig:incidence_share}
\end{subfigure}
\hfill
\begin{subfigure}[t]{0.42\linewidth}
  \includegraphics[width=\linewidth]{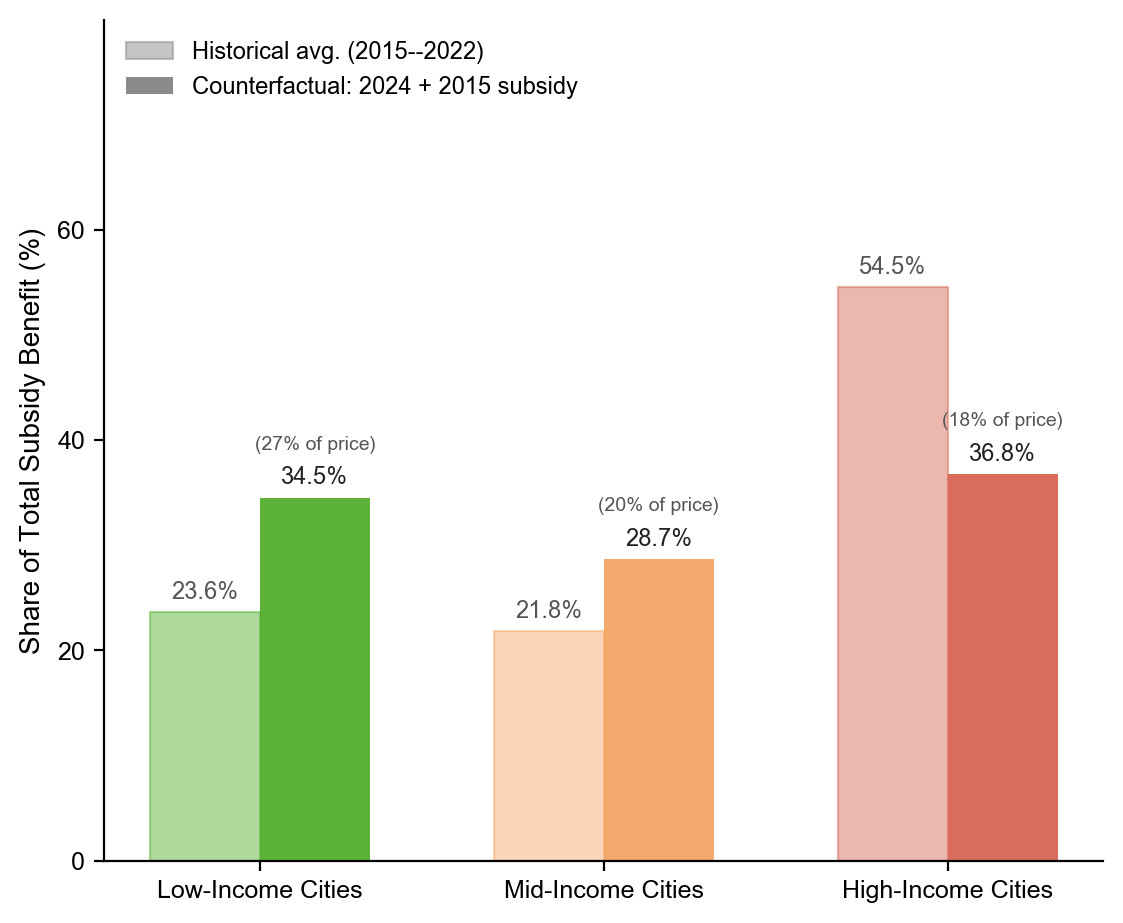}
  \caption{Subsidy benefit: historical vs.\ counterfactual}
  \label{fig:incidence_benefit}
\end{subfigure}

\par\smallskip
\flushleft{\footnotesize\textit{Notes:}
Panel~\subref{fig:incidence_share} reports the share of national EV
sales accounted for by Low-, Mid-, and High-income cities, where income
tiers are defined by terciles of 2020 GDP per capita. The black line
reports the national per-vehicle subsidy on the right axis.
Panel~\subref{fig:incidence_benefit} reports the share of cumulative
subsidy benefits accruing to each income tier. The historical bars use
the 2015--2022 subsidy schedule. The counterfactual bars apply the 2015
subsidy level to the 2024 market structure, holding quantities and
product mix fixed. Annotations report the effective subsidy rate as a
share of the average vehicle price in each tier.}
\end{figure}

Figure~\ref{fig:incidence} examines how subsidy benefits are
distributed across city income tiers. Cities are divided into Low-, Mid-,
and High-income groups using terciles of 2020 GDP per capita. The figure
then compares the historical allocation of EV purchases and subsidy
benefits with a counterfactual that applies the 2015 subsidy level to the
2024 market.

Panel~\subref{fig:incidence_share} shows a substantial shift in the
geography of EV adoption. In 2015, High-income cities accounted for
$73\%$ of national EV sales, while Low-income cities accounted for only
$13\%$. By 2024, EV adoption had spread much further down the income
distribution: Low-income cities accounted for $35\%$ of EV sales, while
High-income cities accounted for $37\%$. This diffusion occurred just as
the national purchase subsidy was phased out. As a result, the subsidy
program was most active when EV adoption was concentrated in richer
markets, but ended when adoption had become more evenly distributed.

Panel~\subref{fig:incidence_benefit} makes this timing issue clear.
Historically, High-income cities received $54.5\%$ of cumulative subsidy
benefits, compared with $23.6\%$ for Low-income cities and $21.8\%$ for
Mid-income cities. In the counterfactual that applies the 2015 subsidy
level to the 2024 market, the distribution becomes much more balanced:
Low-income cities receive $34.5\%$, Mid-income cities receive $28.7\%$,
and High-income cities receive $36.8\%$. The same nominal subsidy is also
larger relative to vehicle prices in lower-income cities, so the
effective subsidy rate is highest where affordability constraints are
likely to be most relevant.

Taken together, the two panels show a misalignment in the timing of
subsidy support. During the active subsidy years, benefits were skewed
toward High-income cities because those cities adopted EVs first. By the
time EV demand had spread to lower-income cities, the subsidy had been
withdrawn. The subsidy therefore had a regressive historical incidence,
even though applying the same subsidy schedule to the later, more mature
EV market would have produced a more balanced distribution of benefits. Appendix Tables~\ref{tab:welfare}--\ref{tab:retention}
report the corresponding private-surplus accounting, consumer-surplus
incidence by city tier, and firm-group exposure under subsidy removal.

\paragraph{Direct vs.\ indirect channels of the consumer-welfare loss.}
The CS reduction under subsidy removal travels through two distinct
mechanisms. The \emph{direct} channel is the within-year price-elasticity
response: removing the cash subsidy and purchase-tax exemption raises the
consumer's net price by the full subsidy amount at the observed marginal
cost, and demand contracts along the BLP elasticity. The \emph{indirect}
channel runs through the Wright's-law feedback: lower counterfactual EV
quantities in earlier periods reduce cumulative production, which
raises the counterfactual battery-cost path, which raises EV marginal
cost, which raises the equilibrium price at which consumers actually
transact in 2024. The two channels are conceptually distinct because
the direct channel reflects the contemporaneous incidence of the cash
transfer, while the indirect channel reflects the loss of compounded
learning benefits.

Table~\ref{tab:channels} reports the decomposition
by city income tier. Two patterns stand out. First, the indirect
Wright's-law channel accounts for roughly half of the total CS
reduction at the national level ($48.0\%$), so a structural welfare
analysis that ignores cumulative learning would understate the
consumer cost of subsidy removal by approximately a factor of two.
Second, the indirect share is moderately larger in T1, NT1, and T2
cities ($49\%$--$50\%$) than in the residual ``Rest'' category
($44\%$). The intuition is that the Wright channel transmits to
welfare through EV prices, and EV market share is concentrated in
the higher-tier cities, so a given equilibrium-price perturbation
moves more consumer surplus in those tiers.

The consumer-surplus magnitudes reported in
Table~\ref{tab:channels} are evaluated under the BLP inclusive-value
formula $\mathrm{CS}_i = (1/|\alpha_i|)\log\sum_j \exp(u_{ij})$ and
inherit the units of the demand model. Because $|\alpha_i|$ enters
the denominator, the absolute scale of CS depends on the demand-elasticity
specification: a less elastic model yields a numerically larger CS
loss for the same change in equilibrium prices and quantities. The
$50/50$ split between direct and indirect channels, the ordering of
per-capita losses across tiers, and the $80\%$ pass-through of
Table~\ref{tab:passthrough} are invariant to the absolute CS scale.
The aggregate magnitudes in Table~\ref{tab:welfare} (e.g., $6.0$
billion yuan CS gain from the 2024 subsidy) are therefore best read
as ratios to the fiscal cost and as comparisons across tiers, not as
standalone monetary estimates.

\begin{table}[!ht]
\centering
\caption{Direct and indirect channels of the CS loss by city tier}
\label{tab:channels}
\footnotesize
\begin{tabular}{lrrrrr}
\toprule
City tier & Pop.\ (M) & CS loss (bn) & Direct & Indirect & Per cap.\ (yuan) \\
\midrule
Tier 1 & 83.6 & 24.4 & 50.7\% & 49.3\% & 292 \\
New Tier 1 & 200.5 & 50.0 & 49.9\% & 50.1\% & 250 \\
Tier 2 & 242.8 & 49.6 & 50.9\% & 49.1\% & 204 \\
Rest of China & 796.2 & 48.9 & 55.8\% & 44.2\% & 62 \\
\midrule
National & 1{,}323.1 & 172.9 & 52.0\% & 48.0\% & 131 \\
\bottomrule
\end{tabular}
\begin{flushleft}
\footnotesize
\textit{Notes:} Decomposition of the 2024 consumer-surplus loss under
subsidy removal into a direct cash-transfer channel and an indirect
Wright's-law channel. The direct channel removes the subsidy at
observed marginal cost. The indirect channel additionally feeds the
counterfactual EV-quantity reduction through cumulative production
into the Wright's-law battery-cost path, raising counterfactual EV
marginal costs and equilibrium prices. Wright-counterfactual prices
are drawn from the deterministic exogenous-entry endogenous-range
regime of the dynamic simulation. Direct and indirect shares are the
contributions of each channel to the total CS loss in each tier.
CS values are evaluated under the log-price BLP specification
that drives the main decomposition; absolute magnitudes therefore
differ in scale from the linear-price reporting in
Table~\ref{tab:welfare}, but the channel decomposition
is invariant to that scale.
\end{flushleft}
\end{table}

The decomposition also has a policy-design implication. The direct
channel is what a cash subsidy on a fixed product set buys. The
indirect channel is the subsidy's contribution to scale and to the
compounded learning that flows from accumulated production. Policy
instruments that secure the production-scale margin without
distributing cash transfers, such as production-targeted investment
incentives or charging-infrastructure provision, would capture a
portion of the indirect channel without the full direct
cash-transfer cost. The decomposition cannot rank such instruments,
but it identifies that roughly half of the realized consumer-welfare
benefit of the 2024 EV subsidy is accounted for by the
production-scale margin rather than by the cash transfer alone.

\paragraph{Joint distributional reading.} Tables~\ref{tab:channels},
\ref{tab:passthrough}, and \ref{tab:retention} together provide a
distributional account of the 2024 subsidy that the aggregate
no-subsidy effect alone would miss. On the consumer side, per-capita
consumer-surplus loss ranges from $292$ yuan in Tier 1 cities to
$62$ yuan in the Rest tier, a roughly $5$-fold gradient. On the firm side, the
EV business retained under subsidy removal varies from $27\%$ for
Tesla and $18\%$ for the post-2014 New Forces down to $11\%$ for
traditional state-owned manufacturers. On the policy-mechanism side,
pass-through of the nominal subsidy into consumer surplus is $80\%$
on average and $84\%$ in Tier 1, indicating that a substantial fraction
of the subsidy disbursement reaches the buyer rather than being
absorbed into firm markups. The same nominal subsidy therefore
distributes its benefit unevenly across cities, across firms, and
between consumers and producers. These three margins of unevenness
are the empirical content of the heterogeneous-diffusion framing: the
2024 EV market is at a stage of diffusion where a single policy
instrument has very different effects on Tier 1 versus Rest tier
consumers, on EV-native firms versus legacy OEMs, and on direct cash
incidence versus indirect learning incidence. A complete welfare
appraisal of EV industrial policy in China cannot reduce to a single
``effect of subsidy'' number; it must allow the effect to vary across
these three dimensions simultaneously.

\subsection{Relation to the Decomposition Results}
\label{sec:policy_relation_decomposition}

These policy counterfactuals should be read separately from the Subsidy
block in the Shapley decomposition. The Shapley result asks how the
historical change in the subsidy environment from 2015 to 2024
contributed to the observed EV-share transition. Because direct purchase
subsidies were phased down over this period, the Subsidy block is
negative in the decomposition.

The no-subsidy counterfactual asks a different question: how much did
subsidy support matter relative to a world without the subsidy? In that
exercise, subsidies increase EV adoption by lowering consumer net prices,
supporting cumulative learning, and sustaining product availability.
There is therefore no contradiction between a negative historical
Subsidy contribution in the decomposition and a positive role for
subsidies in the policy counterfactual. The two exercises answer
different questions: the first explains the historical transition, while
the second evaluates the market consequences of alternative subsidy
regimes.

To help reconcile the three subsidy magnitudes that
appear in the paper: the static Subsidy Shapley of $-13.63\%$
measures the marginal effect of switching the subsidy schedule from
its 2015 value to its 2024 value (a phase-down) while holding all
other channels fixed; the dynamic full-removal effects of
$-22.8$ to $-32.8\%$ across the three regimes measure the marginal
effect of removing the 2024 subsidy entirely while allowing
Wright's-law learning and entry to respond endogenously; and the
demand-elasticity robustness estimate of $-23.4\%$ under
$|\alpha|=2.5$ measures the same dynamic removal under a less elastic
calibration. The three numbers measure different counterfactuals; we
report all three because each addresses a different empirical
question.

\paragraph{Externalities not in the welfare arithmetic.}
The welfare accounting in this section is private surplus only:
consumer surplus, producer surplus, and the government's fiscal cost
of the cash and tax-exemption components. It does not include the
external benefit of substituting EVs for ICE vehicles in tailpipe CO$_2$
emissions, in local air-quality pollutants (PM$_{2.5}$, NO$_x$), or in
externality-relevant industrial-policy spillovers (national-champion
build-up in battery and supply-chain industries). Standard estimates of
the marginal social cost of CO$_2$ in the range of $\$50$--$\$200$ per
ton, applied to the $\sim 11$ million annual EV substitutions implied
by the actual scenario, would deliver an externality benefit on the
order of tens of billions of yuan per year. A complete cost-benefit
treatment of the subsidy would add this externality benefit to the
private-surplus accounting reported here, which would shift the
verdict on the policy's net welfare effect. We do not attempt that externality calculation in this paper because
it depends on assumptions about the marginal CO$_2$ price, the ICE
counterfactual fleet, and the regional grid mix that are outside the
structural model. \citet{HollandMansurMullerYates2016} show that the
sign and magnitude of EV environmental benefits in the U.S.\ depend
strongly on the local generation mix, and
\citet{HollandMansurMullerYates2019} trace the distributional
consequences across regions; the same logic applies to China's
spatially heterogeneous grid.

The two results have a joint reading for policy redesign. The
negative historical Shapley contribution is mechanical: the 2024
schedule is less generous than the 2015 schedule. The positive
counterfactual contribution is structural: some subsidy support
delivers consumer-side gains in the 2024 market. These are not in
tension. Together they imply that the timing and targeting of the
schedule are themselves policy choices.

The 2015--2024 schedule front-loaded benefits onto early high-income
adopters and phased out before EV adoption reached lower-tier cities.
The realized incidence pattern in Table~\ref{tab:channels} therefore
reflects the schedule, not just the underlying technology. A schedule
that linked eligibility to city income tier or to the marginal
buyer's purchase-price quintile would shift incidence toward
lower-tier cities at the same aggregate fiscal cost. A schedule that
combined cash transfers with production-side incentives that capture
the indirect Wright channel would deliver similar consumer-side gains
at lower fiscal cost. We do not solve a mechanism-design problem
here. We report that the historical schedule was front-loaded, that
the realized incidence is regressive in a quantifiable way, and that
the decomposition supplies the magnitudes a designer would need to
evaluate any specific redesign.

\section{Conclusion}

The revised evidence supports a more credible and more nuanced account
of China's EV diffusion. The transition was not driven by a single
force. It combined improved EV product quality, falling battery costs,
entry of new products and firms, and stronger urban EV demand. Direct
subsidies helped create and sustain the market, but their observed
2015--2024 time-series contribution is negative because the direct
subsidy schedule was phased out.

The contribution of the paper is not a neat ranking of mechanisms.
It is showing why such a ranking is intrinsically difficult when
quality, cost, entry, and policy are complements. The decomposition
organizes a complicated transition into transparent blocks and
reports how sensitive those blocks are to ordering. Endpoint fit and
sign discipline make the results informative; wide Shapley ranges
define the limits of precision.

The heterogeneous-diffusion framing is not decorative. The Shapley
decomposition, the no-subsidy counterfactual, and the cross-tier
incidence accounting agree on the main qualitative features. Product
improvement is the largest single driver of aggregate diffusion. But
its welfare-relevant manifestation is concentrated in high-income
cities and in EV-native firms. Two policy implications follow.
First, about half of the consumer-welfare benefit flows through the
indirect Wright channel rather than the direct cash transfer, so
production-targeted instruments could capture much of the indirect
channel at lower fiscal cost. Second, the regressive incidence is a
feature of the schedule, not of the technology. Eligibility tied to
city income tier or to marginal-buyer purchase-price quintile is
within reach without changing the structural model.

Three further limitations are worth flagging.
\emph{(i)} The demand and supply models cover the new-vehicle market
only. The 2020s development of a secondary EV market is not in the
estimation sample. Used-EV substitution would attenuate the
no-subsidy effects measured at the new-vehicle margin because some
buyers priced out of the new-EV market would substitute toward used
EVs rather than to new ICE vehicles. \citet{XingLeardLi2021} quantify
the EV-replaces-ICE margin in the U.S.\ on the new-vehicle side; the
analogous Chinese used-vehicle margin remains an open question. \emph{(ii)} The
sample covers $79$ cities accounting for approximately $42\%$ of
national new-vehicle sales; the BLP-sample-implied 2024 EV quantity of
$0.42$ million scales to the national NEV registration count of
roughly $11$ million under a market-size scaling factor of about
$28$. The per-capita welfare ratios, channel decompositions, and tier
incidence ratios reported in Section~\ref{sec:policy_counterfactuals}
are invariant to this scaling, but absolute fiscal-cost-to-welfare
ratios should be read in BLP-sample units. \emph{(iii)} Charging
infrastructure enters our model only through the city-level fixed
effects and the EV-$\times$-density interaction. A specification that
modeled the charger-installed-base as a state variable jointly
co-evolving with EV adoption would close a feedback loop that we
treat as fixed; this is a natural extension.

\newpage
\bibliographystyle{plainnat}
\bibliography{references}

\newpage
\appendix
\section{Appendix Figures}

\begin{figure}[!ht]
\caption{BEV driving-range trajectories by product, 2015--2024.}
\label{fig:range_traj}
\centering
\includegraphics[width=0.88\linewidth]{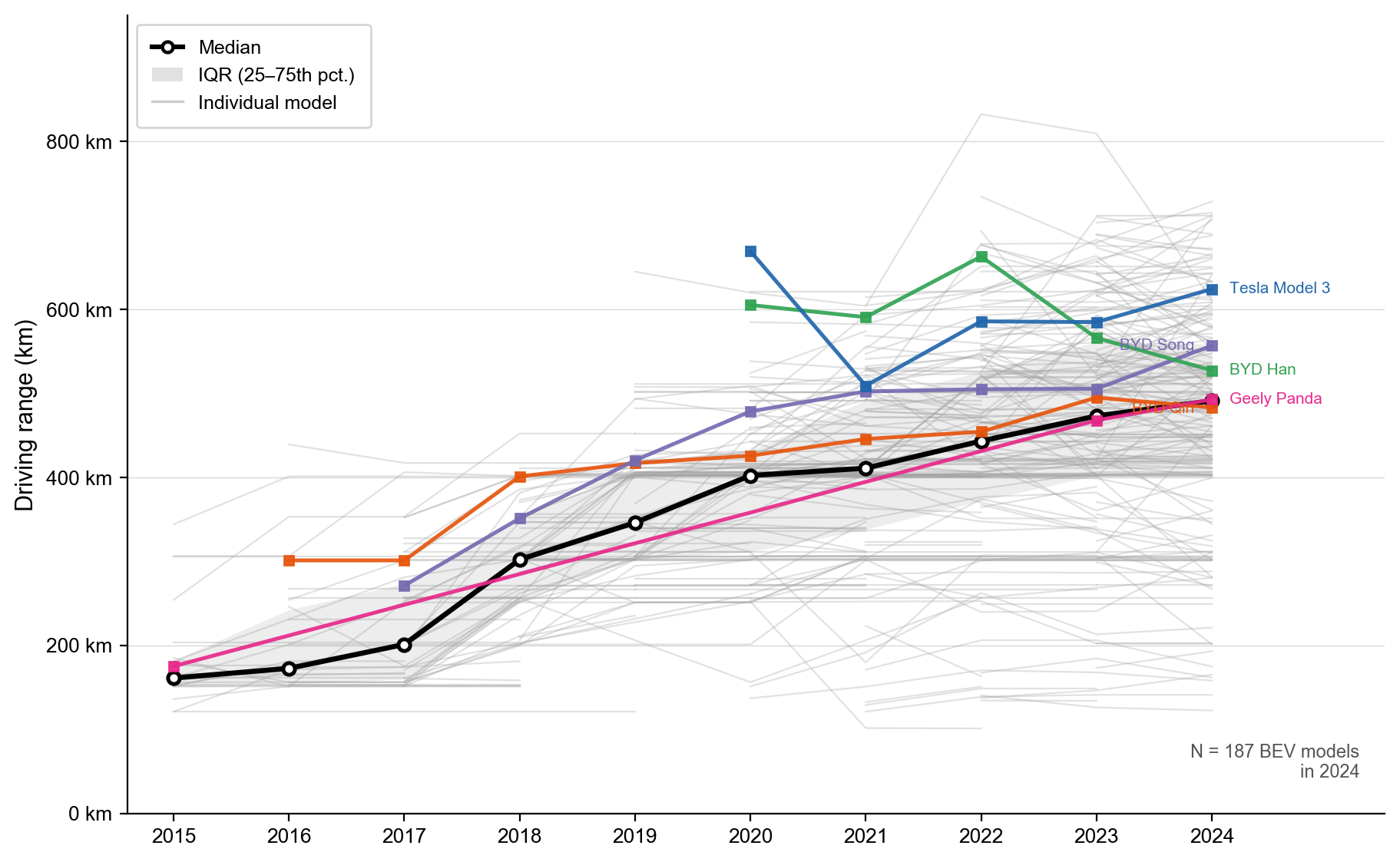}
\par\smallskip
\flushleft{\footnotesize\textit{Notes:} Each thin gray line traces one BEV model's range (km) over the years it was sold. The black line is the cross-model median; the shaded band is the interquartile range (25th--75th percentile). Five representative models are highlighted. Range upgrades are product-specific in timing and magnitude, motivating the flexible per-product, per-year range treatment in demand estimation.}
\end{figure}

\begin{figure}[!ht]
\caption{Distribution of recovered marginal costs, EV vs.\ GV}
\label{fig:mc_dist}
\centering
\includegraphics[width=0.88\linewidth]{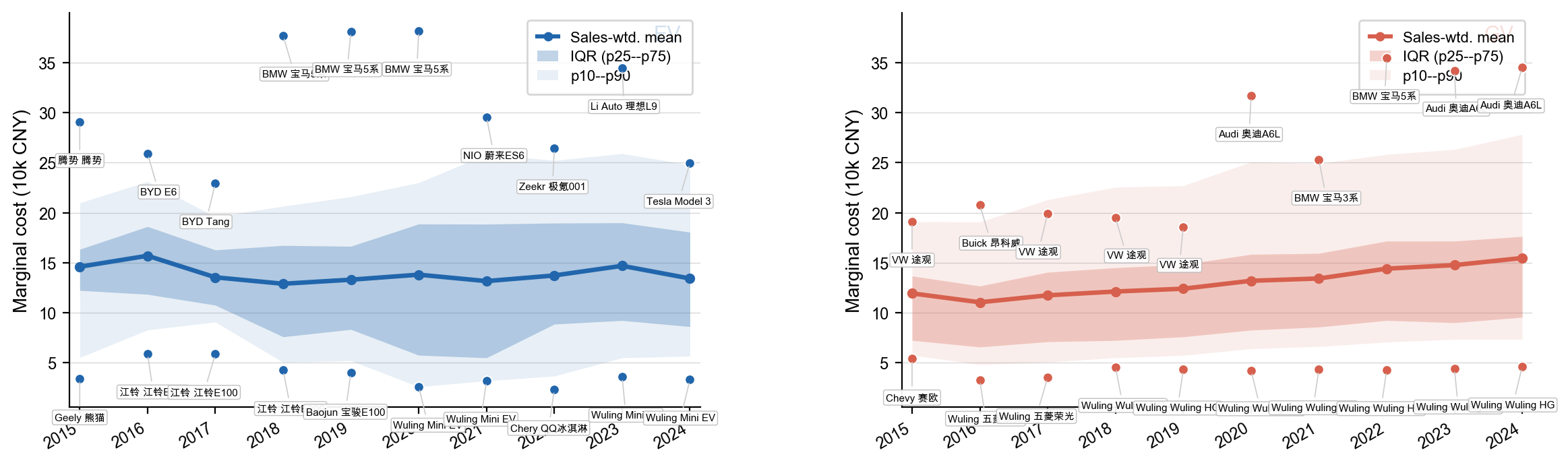}
\par\smallskip
\flushleft{\footnotesize\textit{Notes:} Distribution of the
product-year-city marginal costs recovered from the multi-product
Bertrand first-order conditions, plotted separately for EV
(BEV + PHEV + REEV) and GV (ICE + HEV) products. The EV distribution
is wider than the GV distribution, reflecting heterogeneity in battery
size, electric drivetrain configuration, and the steeper trajectory of
EV cost decline over 2015--2024 that is identified by the Year and
EV $\times$ Year effects in Equation~(\ref{eq:mc_regression}). The right
tail of the EV distribution corresponds to large-battery long-range
BEVs and premium PHEVs whose marginal cost levels remain comparable
to mid-size GV products throughout the sample.}
\end{figure}

\begin{figure}[!ht]
\caption{Distribution of unobserved marginal-cost residual $\omega_{jm}$}
\label{fig:mc_unobs}
\centering
\includegraphics[width=0.88\linewidth]{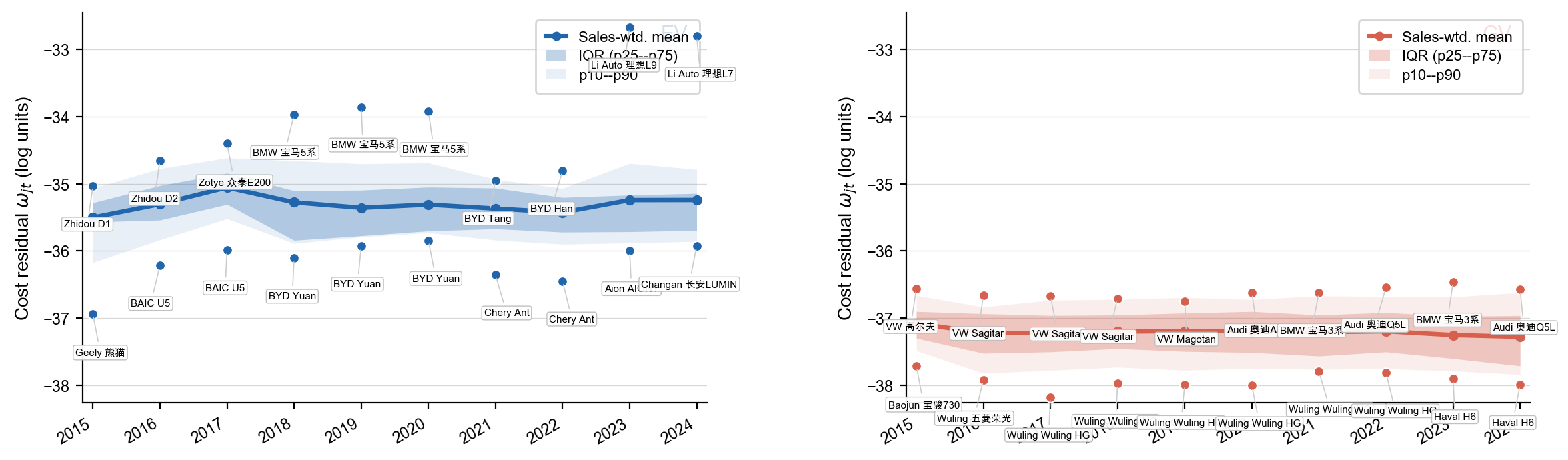}
\par\smallskip
\flushleft{\footnotesize\textit{Notes:} Distribution of the unobserved
supply-side residual $\omega_{jm}$, defined as the part of recovered log
marginal cost not explained by the pooled OLS specification in
Equation~(\ref{eq:mc_regression}). The residual is mean-zero by
construction and tightly concentrated for both EVs and GVs, indicating
that the observable characteristics (vehicle size, engine power, EV
driving range, BNEF battery cost, FuelType, BodyType, firm-group, and
year fixed effects) absorb most of the cross-product variation in
recovered marginal costs. The wider EV right tail reflects residual
variation in battery chemistry and supply-chain procurement that is
not captured by the BNEF aggregate-cost time series.}
\end{figure}

\newpage
\section{Appendix Tables}
\begin{table}[H]
\centering
\caption{Blocks in the Shapley Decomposition}
\label{tab:blocks}
\small
\begin{tabularx}{\textwidth}{p{0.16\textwidth}p{0.18\textwidth}X}
\toprule
Figure label & Block name & Economic content \\
\midrule

Quality
&
Product attributes
&
Within-product values of log driving range (EV products), log engine
power (GV products), log vehicle size, and the range $\times$ density
interaction; toggled between 2015 and 2024 values for products
present in both endpoint sets. Excludes fixed effects (which are
year-invariant and handled within Variety), prices, policy variables,
and demand residuals.
\\[4pt]

Variety
&
Entry and product variety
&
Changes in the available product set, including the expansion and
reshaping of EV and GV product offerings.
\\[4pt]

Battery
&
Battery costs
&
Battery-related cost changes that affect EV marginal costs and the cost
of providing longer driving range.
\\[4pt]

Subsidy
&
Subsidies and tax incentives
&
Changes in direct purchase subsidies, local subsidy components, and tax
exemptions between 2015 and the endpoint year.
\\[4pt]

Residual
&
Residual demand and cost
&
Residual demand and cost components not explained by observed product
attributes, prices, policies, demographics, or fixed effects.
\\[4pt]

Market
&
Income and market size
&
Changes in household income, population, market size, and demographic
variables that affect demand and price sensitivity.
\\

\bottomrule
\end{tabularx}
\begin{minipage}{\textwidth}
\footnotesize
\textit{Notes:} The first column reports the short labels used in the
decomposition figures. Each block is switched from its 2015 value to its
endpoint-year value while the remaining blocks are held fixed at their
2015 values. For each coalition of blocks, equilibrium prices,
quantities, and market shares are recomputed before calculating the
Shapley contribution.
\end{minipage}
\end{table}

\begin{table}[H]
\centering
\caption{Endpoint fit of the year-by-year decomposition}
\label{tab:endpoints}
\small
\begin{tabular}{lrrrrr}
\toprule
Endpoint year & $V(\emptyset)$ & $V(\mathrm{full})$ & Observed $\Delta$ & $\sum_b\phi_b$ & Gap \\
\midrule
2016 & $1.03\%$ & $1.10\%$ & $0.21\%$ & $0.08\%$ & $-0.13\%$ \\
2017 & $1.03\%$ & $2.65\%$ & $1.69\%$ & $1.63\%$ & $-0.06\%$ \\
2018 & $1.03\%$ & $4.22\%$ & $3.33\%$ & $3.20\%$ & $-0.13\%$ \\
2019 & $1.03\%$ & $4.00\%$ & $3.14\%$ & $2.97\%$ & $-0.17\%$ \\
2020 & $1.03\%$ & $5.71\%$ & $4.71\%$ & $4.69\%$ & $-0.02\%$ \\
2021 & $1.03\%$ & $14.05\%$ & $12.88\%$ & $13.02\%$ & $0.15\%$ \\
2022 & $1.03\%$ & $25.80\%$ & $24.63\%$ & $24.78\%$ & $0.15\%$ \\
2023 & $1.03\%$ & $33.53\%$ & $33.02\%$ & $32.50\%$ & $-0.51\%$ \\
2024 & $1.03\%$ & $44.66\%$ & $44.30\%$ & $43.64\%$ & $-0.75\%$ \\
\bottomrule
\end{tabular}
\begin{flushleft}
\footnotesize
\textit{Notes:} The baseline environment is 2015. In each row, 
$V(\emptyset)$ is the model-implied EV share under the 2015 environment, 
while $V(\mathrm{full})$ is the model-implied EV share when all blocks are 
set to the listed endpoint-year values. The Shapley values satisfy 
$\sum_b \phi_b = V(\mathrm{full}) - V(\emptyset)$ by construction. 
The gap is $\sum_b \phi_b$ minus the observed aggregate EV-share change. 
All values are reported in percentage units of aggregate EV share.
\end{flushleft}
\end{table}

\input{tables/tab_shapley_by_tier}

\newpage
\section{Additional Policy Counterfactual Results}
\label{app:policy_counterfactuals}

\subsection{Additional Dynamic Counterfactual Results}
\label{app:dynamic_counterfactuals}

Table~\ref{tab:regimes} accompanies the discussion of dynamic
amplification channels in Section~\ref{sec:no_subsidy_dynamic} by
reporting the no-subsidy 2024 EV quantity under each of the three
regimes alongside the corresponding shortfall relative to actual.
The reduction from $4.1$ to $2.8$ million between the Static and
Learning regimes isolates the contribution of Wright's-law battery-cost
learning to subsidy efficacy; the further reduction from $2.8$ to
$1.8$ million between the Learning and Full dynamic regimes isolates
the contribution of endogenous entry and exit. The two channels
deliver roughly comparable additional reductions of $1.2$ and $1.0$
million vehicles, indicating that neither channel dominates the
dynamic-amplification mechanism.

\begin{table}[H]
\centering
\caption{No-subsidy EV quantity by dynamic regime}
\label{tab:regimes}
\small
\begin{tabular}{lrrr}
\toprule
Regime & No-subsidy EV quantity & Shortfall from actual & Incremental reduction \\
\midrule
Static & 4.1 million & 7.9 million & -- \\
Learning only & 2.8 million & 9.2 million & 1.2 million \\
Full dynamic & 1.8 million & 10.2 million & 1.0 million \\
\bottomrule
\end{tabular}
\begin{flushleft}
\footnotesize
\textit{Notes:} The actual 2024 EV quantity is $12.0$ million. The
Static regime holds product availability and battery costs fixed. The
Learning only regime allows battery costs to respond through cumulative
production while holding product availability fixed. The Full dynamic
regime allows both battery-cost learning and entry/exit responses. The
incremental reduction is measured relative to the previous row.
\end{flushleft}
\end{table}

\subsection{Additional Welfare and Incidence Accounting}
\label{app:welfare_incidence}

Tables~\ref{tab:welfare}, \ref{tab:passthrough}, and~\ref{tab:retention}
report the welfare-accounting components, the city-tier subsidy
pass-through ratios, and the firm-group EV-quantity-retained shares
that underlie the discussion in Section~\ref{sec:subsidy_incidence}.
The three tables share a common scenario: the 2024 actual subsidy
package is removed under the canonical endo\_endo regime of the
dynamic simulation, and the resulting consumer-surplus, producer-surplus,
and quantity changes are decomposed across consumer tiers and producer
groups. Together they document that the same nominal subsidy distributes
its benefit unevenly across cities (Table~\ref{tab:passthrough}),
across firm types (Table~\ref{tab:retention}), and between consumers
and producers (Table~\ref{tab:welfare}).

\begin{table}[!ht]
\centering
\caption{Subsidy-removal welfare accounting and incidence}
\label{tab:welfare}
\small
\begin{tabular}{lr}
\toprule
Object & Estimate \\
\midrule
2024 EV subsidy spend & 209.6 bn yuan/year \\
Producer surplus gain from subsidy & 536.0 bn yuan/year \\
Consumer surplus gain from subsidy & 172.9 bn yuan/year \\
Total private surplus gain (CS+PS) & 708.9 bn yuan/year \\
Total surplus per yuan of subsidy & 3.38 \\
PS share of private surplus gain & 76\% \\
Mean subsidy per EV & 17,813 yuan \\
Average per-capita CS loss under removal & 131 yuan/year \\
\midrule
\textit{Memo:} Implicit Tier 1 license-plate subsidy (excluded) & 4.0 bn yuan/year \\
\midrule
Tier-1 per-capita CS loss under removal & 292 yuan/year \\
New Tier-1 per-capita CS loss under removal & 250 yuan/year \\
Tier-2 per-capita CS loss under removal & 204 yuan/year \\
Rest-of-China per-capita CS loss under removal & 62 yuan/year \\
\bottomrule
\end{tabular}
\begin{flushleft}
\footnotesize
\textit{Notes:} The table reports the 2024 subsidy-removal accounting,
all values on a common national scale. Subsidy spend is the fiscal
cost of the 2024 NEV cash schedule plus the $10\%$ purchase-tax
exemption. Consumer surplus is computed from the BLP inclusive-value
formula on the $79$-city panel and scaled to the national NEV
registration count by multiplying by a market-size factor of $28.6$
(the ratio of national 2024 NEV registrations $\approx 11.15$ million
to BLP-sample EV quantity $\approx 0.39$ million). Producer surplus
is computed at the national level from the recovered Bertrand markup
and the model-implied national EV quantity. The total surplus per
yuan of subsidy ($3.38$) and the within-PS ranking across firm
groups in Table~\ref{tab:retention} are the headline distributional
takeaways. The table does not include the implicit value of the EV
license-plate priority in Tier 1 cities, which is reported separately
in Table~\ref{tab:passthrough} and discussed in
Section~\ref{sec:policy_background}.
\end{flushleft}
\end{table}

\begin{table}[!ht]
\centering
\caption{Subsidy pass-through to consumer surplus by city tier, 2024}
\label{tab:passthrough}
\small
\begin{tabular}{lrrrrr}
\toprule
City tier & Cities & EV qty (M) & Subsidy received (bn) & CS gain (bn) & Pass-through \\
\midrule
Tier 1 & 4 & 1.29 & 29.1 & 24.4 & 83.8\% \\
New Tier 1 & 15 & 3.15 & 63.6 & 50.0 & 78.6\% \\
Tier 2 & 33 & 3.26 & 60.5 & 49.6 & 82.0\% \\
Rest of China & 27 & 4.43 & 62.5 & 48.9 & 78.3\% \\
\midrule
National & 79 & 12.13 & 215.7 & 172.9 & 80.2\% \\
\bottomrule
\end{tabular}
\begin{flushleft}
\footnotesize
\textit{Notes:} Pass-through ratio is the fraction of nominal subsidy
disbursement to EV buyers in each tier that translates into a
consumer-surplus gain. EV qty is the BLP-sample EV quantity in the
2024 panel (a within-sample share-of-agent-pool measure that scales
proportionally to the national NEV registration count). Subsidy
received aggregates the per-product cash subsidy plus 10\% purchase-tax
exemption over the tier's 2024 EV products. CS gain is the actual-minus
no-subsidy consumer-surplus change under the full-dynamic regime of
Table~\ref{tab:channels}. Pass-through is below
$100\%$ in all tiers because Bertrand--Nash firms capture part of the
subsidy in higher markups, and pass-through is moderately higher in
Tier 1 and Tier 2 (premium-product mix is less elastic to net price
than the small-vehicle EV mix in NT1 and Rest tiers).
\end{flushleft}
\end{table}

\begin{table}[!ht]
\centering
\caption{Firm-group EV quantity retained under subsidy removal}
\label{tab:retention}
\small
\begin{tabular}{lrrrr}
\toprule
Firm group & Actual quantity & No-subsidy quantity & Retained share \\
\midrule
BYD              & 4.26 million & 0.69 $\pm$ 0.07 million & 16.1\% \\
Tesla            & 0.74 million & 0.20 $\pm$ 0.05 million & 26.7\% \\
Foreign/JV       & 0.70 million & 0.10 $\pm$ 0.02 million & 14.7\% \\
Trad.\ OEM       & 2.69 million & 0.31 $\pm$ 0.05 million & 11.4\% \\
New Forces       & 1.53 million & 0.27 $\pm$ 0.05 million & 17.9\% \\
Private National & 1.98 million & 0.25 $\pm$ 0.03 million & 12.6\% \\
Other            & 0.08 million & 0.00 $\pm$ 0.00 million & \phantom{0}4.7\% \\
\midrule
\textbf{Total} & \textbf{12.00 million} & \textbf{1.82 million} & \textbf{15.2\%} \\
\bottomrule
\end{tabular}
\begin{flushleft}
\footnotesize
\textit{Notes:} 2024 EV quantity by firm group under the actual-subsidy
scenario and the endo\_endo no-subsidy counterfactual (50 Monte Carlo
draws of the stochastic Stage~D entry, reported as cross-draw mean
$\pm$ one standard deviation). Retained share is the no-subsidy mean
EV quantity divided by the actual quantity for each firm group. Firm
groups follow the seven-class manufacturer classification described in
the supply-side appendix: BYD is reported separately given its dominant
market share; Tesla is reported separately for its premium-import
exposure; Foreign/JV pools international joint ventures; Trad.\ OEM
pools the state-owned legacy manufacturers; New Forces pools the
post-2014 EV-only entrants (Nio, XPeng, Li Auto, etc.); Private
National pools private-domestic firms; Other pools the remaining
small-share manufacturers. The Tesla and New Forces groups retain the
largest share of their 2024 EV business under subsidy removal, while
small-share \textit{Other} manufacturers and traditional OEMs are the
most exposed.
\end{flushleft}
\end{table}

\subsection{Supply-side and Robustness Diagnostics}
\label{app:supply_robustness}

Table~\ref{tab:lerner_byyear} reports the quantity-weighted mean Lerner
index recovered from the multi-product Bertrand-Nash first-order
conditions, broken down by fuel type and calendar year. The Lerner
index measures the price-cost margin as a share of the producer
price. Two patterns stand out. First, all five fuel-type margins sit
in the $0.20$ to $0.25$ band throughout the sample, indicating that
the demand and supply specifications together imply economically
reasonable pricing power that is similar in magnitude to U.S.\ and
European automobile-BLP estimates. Second, the BEV and ICE Lerner
ratios rise by about $3\%$ between 2015 and 2024,
consistent with consolidation and product-line specialization over
the diffusion window. The PHEV margin declines slightly over the
same window, reflecting the entry of small-battery PHEVs at lower
markups. The REEV class appears in 2021 and stabilizes near $0.25$.

\begin{table}[!ht]
\centering
\caption{Q-weighted Lerner index by fuel type and year}
\label{tab:lerner_byyear}
\small
\begin{tabular}{lrrrrr}
\toprule
Year & BEV & HEV & ICE & PHEV & REEV \\
\midrule
2015 & 0.213 & 0.219 & 0.205 & 0.234 & --- \\
2016 & 0.220 & 0.221 & 0.213 & 0.237 & --- \\
2017 & 0.217 & 0.225 & 0.215 & 0.235 & --- \\
2018 & 0.221 & 0.229 & 0.219 & 0.238 & --- \\
2019 & 0.232 & 0.236 & 0.227 & 0.241 & --- \\
2020 & 0.236 & 0.240 & 0.232 & 0.245 & --- \\
2021 & 0.238 & 0.242 & 0.235 & 0.244 & 0.249 \\
2022 & 0.240 & 0.243 & 0.239 & 0.243 & 0.248 \\
2023 & 0.243 & 0.244 & 0.240 & 0.241 & 0.247 \\
2024 & 0.242 & 0.246 & 0.242 & 0.239 & 0.247 \\
\bottomrule
\end{tabular}
\begin{flushleft}
\footnotesize
\textit{Notes:} The Lerner index is computed at the product-year-city
level from the recovered marginal costs and observed consumer net
prices and then aggregated by quantity-weighted mean within each
(fuel type, year) cell. ICE is ICEV.
\end{flushleft}
\end{table}

Table~\ref{tab:alpha_robust} reports the canonical and lit-median
robustness exercises side by side. The canonical column uses the
estimated demand-elasticity $|\alpha| \approx 4.9$ from the demand model
of Section~3.
The robustness column rescales $|\alpha|$ to the EV-demand
literature median of $2.5$ before re-running the dynamic simulation.
The structural ranking across the three dynamic regimes is preserved
in both calibrations: the Static effect is smaller in magnitude
than the Learning-only effect, which is in turn smaller than the
Full-dynamic effect. The robustness column delivers subsidy effects
roughly $30\%$ smaller in magnitude than the canonical column, but
the qualitative finding that subsidies remove a substantial share of
2024 EV adoption is preserved.

\begin{table}[!ht]
\centering
\caption{Sensitivity of subsidy-removal effects to the demand elasticity}
\label{tab:alpha_robust}
\small
\begin{tabular}{lrrr}
\toprule
Regime & $|\alpha|=4.9$ (canonical) & $|\alpha|=2.5$ (lit median) & Ratio \\
\midrule
Static (exog\_exog) & $-22.75\%$ & $-13.60\%$ & 0.60 \\
Learning (exog\_endo) & $-28.91\%$ & $-18.85\%$ & 0.65 \\
Full dynamic (endo\_endo) & $-32.76 \pm 0.65\%$ & $-23.40 \pm 1.35\%$ & 0.71 \\
\bottomrule
\end{tabular}
\begin{flushleft}
\footnotesize
\textit{Notes:} 2024 subsidy-removal effect on EV share under the
canonical demand-elasticity calibration ($|\alpha|=4.9$) and the
literature-median calibration ($|\alpha|=2.5$, obtained by scaling
both $\beta_{\log p}$ and $\pi_p$ by the same factor). The robustness
column is from a fresh dynamic-simulation run that uses identical
data, BLP demand attributes, and Wright's-law parameters as the
canonical run, but with the rescaled price coefficients. Ratio is the
robustness-column estimate divided by the canonical estimate. The
endo\_endo standard deviations are across $50$ stochastic-entry draws.
\end{flushleft}
\end{table}

The $|\alpha|=2.5$ rescaling is applied only to the dynamic
simulation. We do not re-estimate the static Shapley decomposition at
the rescaled elasticity because the Shapley equilibrium-solver inputs
include both the demand parameters and the recovered marginal costs,
and a consistent re-derivation would require re-running the
marginal-cost backout under the rescaled demand parameters before
re-solving the coalition equilibria. Such a re-derivation is a
substantive recomputation rather than a sensitivity exercise, and we
leave it to future work that targets static Shapley robustness as a
primary research question. The dynamic-sim robustness reported in
Table~\ref{tab:alpha_robust} provides a calibrated bound on how the
no-subsidy effects scale with $|\alpha|$ and is the most policy-relevant
sensitivity diagnostic that the present specification supports.

\begin{table}[!ht]
\centering
\caption{City tier definitions}
\label{tab:tier_definitions}
\small
\renewcommand{\arraystretch}{1.15}
\begin{tabular}{lrp{0.68\linewidth}}
\toprule
Tier & Markets & Member markets \\
\midrule
Tier 1 & 4 & Beijing, Shanghai, Guangzhou, Shenzhen. \\
\addlinespace[4pt]
New Tier 1 & 15 & Chengdu, Hangzhou, Wuhan, Chongqing, Nanjing,
Tianjin, Suzhou, Xi'an, Changsha, Shenyang, Qingdao, Zhengzhou,
Dalian, Dongguan, Ningbo. \\
\addlinespace[4pt]
Tier 2 & 33 & Foshan, Wuxi, Hefei, Fuzhou, Kunming, Xiamen, Jinan,
Harbin, Changchun, Shijiazhuang, Quanzhou, Lanzhou, Urumqi, Guiyang,
Nanning, Nanchang, Taiyuan, Haikou, Zhuhai, Wenzhou, Shaoxing,
Jiaxing, Taizhou (Zhejiang), Huizhou, Zhongshan, Changzhou, Nantong,
Tangshan, Xuzhou, Linyi, Weifang, Yantai, Jinhua. \\
\addlinespace[4pt]
Rest & 27 & Province-level residual cells, each labeled
``\textit{Province} -- Other'' and aggregating all prefectures in
that province not separately identified in the BLP panel: Anhui,
Fujian, Gansu, Guangdong, Guangxi, Guizhou, Hainan, Hebei,
Heilongjiang, Henan, Hubei, Hunan, Inner Mongolia, Jiangsu, Jiangxi,
Jilin, Liaoning, Ningxia, Qinghai, Shaanxi, Shandong, Shanxi,
Sichuan, Tibet, Xinjiang, Yunnan, Zhejiang. \\
\midrule
Total & 79 & \\
\bottomrule
\end{tabular}
\begin{flushleft}
\footnotesize
\textit{Notes:} The Tier~1 / New Tier~1 / Tier~2 classification
follows the China Business Network (CBN) annual city ranking, which
is the standard tier definition in applied China-economics work. The
Tier~2 row lists all prefecture-level cities that appear as their own
market in the BLP panel but are not in the CBN Tier~1 or New-Tier~1
lists. The Rest row covers the 27 province-level residual cells used
to absorb prefectures that the panel does not identify separately.
Membership is fixed across all calendar years in the sample. The same
79 markets appear in every year from 2015 to 2024, yielding
$79 \times 10 = 790$ market-year observations.
\end{flushleft}
\end{table}

\begin{table}[!ht]
\centering
\caption{Firm-group definitions}
\label{tab:firm_group_definitions}
\small
\renewcommand{\arraystretch}{1.15}
\begin{tabular}{lrrp{0.50\linewidth}}
\toprule
Firm group & Brands & 2024 EV \% & Member brands \\
\midrule
BYD & 4 & $100\%$ &
BYD, Denza, Yangwang, Fang Cheng Bao. \\
\addlinespace[4pt]
Tesla & 1 & $100\%$ &
Tesla (Shanghai Gigafactory). \\
\addlinespace[4pt]
Foreign / JV & 28 & $7\%$ &
Foreign brands sold in China (often via joint ventures with Chinese
SOEs): VW, Audi, Skoda, BMW, Mercedes-Benz, Mini, Smart; Toyota,
Honda, Nissan, Mazda, Mitsubishi, Lexus, Acura, Infiniti; Buick,
Chevrolet, Cadillac, Ford, Lincoln; Hyundai, Kia; Peugeot, Citro\"en,
Fiat; MG, Jaguar, Land Rover, Lotus, Volvo, Polestar. \\
\addlinespace[4pt]
Trad.\ OEM (SOE) & 28 & $50\%$ &
Brands under the seven traditional state-owned auto groups: SAIC
(Roewe, MG, Maxus, IM, Rising), FAW (Hongqi, Bestune, Besturn),
Dongfeng (Dongfeng, Voyah, Aeolus, M-Hero), Changan (Changan,
Deepal, Avatr, Kaicene), GAC (Trumpchi, Aion, Hyper), BAIC (Beijing,
ARCFOX, Stelato), Brilliance (Jinbei, Zhonghua). \\
\addlinespace[4pt]
New Forces & 8 & $100\%$ &
EV-native startups founded after 2014: NIO, Li Auto, XPeng,
Leapmotor, Nezha, Xiaomi, WM Motor, HiPhi. \\
\addlinespace[4pt]
Private National & 19 & $39\%$ &
Brands under the five non-SOE Chinese auto groups: Geely (Geely,
Geometry, Zeekr, Galaxy, Lynk \& Co, Lotus, Polestar, Smart\,$^\dagger$),
Great Wall (Haval, WEY, Tank, ORA), Chery (Chery, Exeed, Jetour,
iCAR), JAC, Seres. \\
\addlinespace[4pt]
Other & 23 & $77\%$ &
Niche manufacturers and exits: Skywell, Leapmotor-spinoffs, Dorcen,
Hanteng, Bestune-spinoffs, JMC, Haima, Yundu, Zotye, smaller
LEV-segment producers. \\
\midrule
Total & 111 & $43\%$ & \\
\bottomrule
\end{tabular}
\begin{flushleft}
\footnotesize
\textit{Notes:} ``Brands'' counts distinct brand-level entries in the
2024 BLP panel. ``2024 EV \%'' is the share of 2024 product-rows in
the group that have FuelType $\in \{$BEV, PHEV, REEV$\}$, weighted
equally across products. Group assignment follows
\texttt{code/\_brand\_groups.py}. Foreign brands are classified by
the consumer-perceived brand name, not by the manufacturing JV
parent: e.g., ``Volkswagen'' is Foreign/JV even though it is produced
by SAIC-Volkswagen and FAW-Volkswagen. $^\dagger$Smart, Polestar,
and Lotus appear under Private National because Geely holds the
majority stake; their consumer-perceived brand identity remains
foreign and would alternatively place them in Foreign/JV. ``Other''
collects manufacturers that do not map to any of the five Chinese
national parent groups and are not in Tesla, BYD, or the New Forces
list; this group is dominated by exiting or niche players and is
included to keep the row totals consistent with the BLP panel.
\end{flushleft}
\end{table}

\end{document}

%% file: figures/tab_iv_diagnostics.tex
\begin{table}[!ht]
\centering
\small
\caption{First-stage F-statistics for the price endogeneity instruments}
\label{tab:iv}
\begin{tabular}{lrr}
\toprule
Instrument group & \# excluded IVs & First-stage $F$ \\
\midrule
Within-market diff IVs (3 cols) & 3 & 1,655.0 \\
Within-fuel-type diff IVs (3 cols) & 3 & 5,143.7 \\
Battery-cost IVs (5 cols) & 5 & 15,728.1 \\
\midrule
\textbf{All excluded IVs (joint)} & 11 & \textbf{18,479.3} \\
\bottomrule
\end{tabular}

\begin{flushleft}
\footnotesize
\textit{Notes:} The dependent variable is log consumer net price. The three IV groups are the within-market BLP differentiation instruments $Z_{\mathrm{diff}}$, the within-fuel-type nested differentiation instruments $Z_{\mathrm{nest}}$, and the battery-cost shifters $Z_{\mathrm{bat}} = \mathrm{BNEF}_t \times \{1[\mathrm{BEV}], 1[\mathrm{PHEV}], 1[\mathrm{REEV}], \log\mathrm{batcap}, \mathrm{BNEF}_t \times \log\mathrm{batcap}\}$. The per-group $F$ statistic tests joint significance of the group conditional on the exogenous X-controls and the other two groups. The overall first-stage $F$ tests joint significance of all $11$ excluded IVs. All groups individually and jointly exceed the conventional Stock--Yogo threshold of $10$ by orders of magnitude, indicating that price is strongly identified.
\end{flushleft}
\end{table}

%% file: figures/tab_diffusion_lag.tex
\begin{table}[!ht]
\centering
\small
\caption{Year each tier first crosses an EV-share threshold}
\label{tab:lag}
\begin{tabular}{lrrrr}
\toprule
City tier & 5\% threshold & 10\% threshold & 25\% threshold & 40\% threshold \\
\midrule
T1 & 2015 & 2018 & 2021 & 2023 \\
NT1 & 2018 & 2021 & 2022 & 2024 \\
T2 & 2021 & 2021 & 2022 & 2024 \\
Rest of China & 2021 & 2022 & 2023 & 2024 \\
\midrule
\emph{Lag vs.\ Tier 1} & 6\,yr (Rest) & 4\,yr (Rest) & 2\,yr (Rest) & 1\,yr (Rest) \\
\bottomrule
\end{tabular}

\begin{flushleft}
\footnotesize
\textit{Notes:} The table reports the first calendar year in which each city tier's Q-weighted EV share (BEV $+$ PHEV $+$ REEV as a fraction of all new-vehicle sales in the tier) reaches each threshold. T1 = \{Beijing, Shanghai, Guangzhou, Shenzhen\}; NT1 = 15 new first-tier cities (Chengdu, Hangzhou, Wuhan, Chongqing, Nanjing, Tianjin, Suzhou, Xi'an, Changsha, Shenyang, Qingdao, Zhengzhou, Dalian, Dongguan, Ningbo); T2 = 33 second-tier cities in the panel; Rest of China = the 27 small-city residual cells. The bottom row reports the diffusion lag of the Rest tier relative to Tier 1.
\end{flushleft}
\end{table}

%% file: tables/tab_shapley_by_tier.tex
\begin{table}[H]
\centering
\small
\caption{Tier-specific Shapley decomposition of the 2015--2024 EV-share change}
\label{tab:tier_shapley}
\begin{tabular}{lrrrrr}
\toprule
Block & Tier 1 & New Tier 1 & Tier 2 & Rest & National \\
\midrule
Quality & $55.59\%$ & $54.66\%$ & $44.12\%$ & $46.89\%$ & $45.49\%$ \\
Variety & $18.05\%$ & $21.41\%$ & $20.91\%$ & $16.13\%$ & $14.81\%$ \\
Battery & $9.41\%$ & $8.72\%$ & $8.34\%$ & $8.11\%$ & $8.20\%$ \\
Subsidy & $-10.38\%$ & $-13.05\%$ & $-14.81\%$ & $-18.90\%$ & $-13.63\%$ \\
Residual & $-31.33\%$ & $-20.63\%$ & $-6.93\%$ & $-5.43\%$ & $-7.96\%$ \\
Market & $0.69\%$ & $-4.33\%$ & $-6.78\%$ & $-6.39\%$ & $-3.27\%$ \\
\midrule
$\sum_b\phi_b$ & $42.02\%$ & $46.79\%$ & $44.85\%$ & $40.42\%$ & $43.64\%$ \\
$V(\emptyset)$ & $8.71\%$ & $1.45\%$ & $0.55\%$ & $0.19\%$ & $1.03\%$ \\
$V(\mathcal{B})$ & $50.73\%$ & $48.24\%$ & $45.40\%$ & $40.61\%$ & $44.66\%$ \\
\bottomrule
\end{tabular}

\begin{flushleft}
\footnotesize
\textit{Notes:} The table reports tier-specific Shapley decompositions
of the model-implied 2015--2024 EV-share change at the 2024 endpoint.
For each coalition, equilibrium prices and quantities are solved at the
national product-market level. Tier-specific value functions are then
computed by aggregating the resulting city-level EV shares within each
tier. Each column closes within its own value function:
$\sum_b\phi_b=V(\mathcal{B})-V(\emptyset)$. Comparisons across tier
columns show how the same channel contributes to EV-share changes within
different city groups, rather than a mechanical allocation of the
national Shapley value across tiers. Tier 1 includes Beijing, Shanghai,
Guangzhou, and Shenzhen; New Tier 1 includes 15 new first-tier cities;
Tier 2 includes 33 second-tier cities in the panel; Rest includes the
remaining small-city cells. All values are reported in EV-share units.
\end{flushleft}
\end{table}